\documentclass[twocolumn]{aastex62}

\usepackage{amsmath}
\usepackage{float}
\usepackage{soul}
\setstcolor{magenta}
\usepackage{lineno}
\newcommand{\mjup}{$M_{\rm Jup}$}

\newcommand{\kms}{km\,s$^{-1}$}
\newcommand{\msun}{$M_{\odot}$}
\newcommand{\Msun}{$M_{\odot}$}

\newcommand{\masyr}{mas\,yr$^{-1}$}
\newcommand{\ebv}{E($B$-$V$)}
\newcommand{\rv}{$v_{r}$}

\received{...}
\revised{...}
\accepted{...}

\submitjournal{AJ}

\shorttitle{The debris disk of HR 8799}
\shortauthors{Faramaz et al.}

\begin{document}

%%%%% Title %%%%%%

\title{A detailed characterization of HR 8799's debris disk with ALMA in Band 7}

%%%%% Authors %%%%%%

\correspondingauthor{Virginie Faramaz}
\email{virginie.c.faramaz@jpl.nasa.gov}

\author{Virginie Faramaz}
\affiliation{Jet Propulsion Laboratory, California Institute of Technology, 4800 Oak Grove drive, Pasadena CA 91109, USA.}

\author{Sebastian Marino}
%\affiliation{Max Planck Institute for Astronomy, K\"onigstuhl 17, D-69117 Heidelberg, Germany.}
\affiliation{Institute of Astronomy, University of Cambridge, Madingley Road, Cambridge CB3 0HA, UK}
\affiliation{Jesus College, University of Cambridge, Jesus Lane, Cambridge CB5 8BL, UK}

\author{Mark Booth}
\affiliation{Astrophysikalisches Institut und Universit{\"a}tssternwarte, Friedrich-Schiller-Universit{\"a}t Jena, Schillerg{\"a}{\ss}chen 2-3, 07745 Jena, Germany.}

\author{Luca Matr\`a}
\affiliation{Centre for Astronomy, School of Physics, National University of Ireland Galway, University Road, Galway, Ireland}

\author{Eric E. Mamajek}
\affiliation{Jet Propulsion Laboratory, California Institute of Technology, 4800 Oak Grove drive, Pasadena CA 91109, USA.}

\author{Geoffrey Bryden}
\affiliation{Jet Propulsion Laboratory, California Institute of Technology, 4800 Oak Grove drive, Pasadena CA 91109, USA.}

\author{Karl R. Stapelfeldt}
\affiliation{Jet Propulsion Laboratory, California Institute of Technology, 4800 Oak Grove drive, Pasadena CA 91109, USA.}

\author{Simon Casassus}
\affiliation{Departamento de Astronomia, Universidad de Chile, Casilla 36-D, Santiago, Chile.}

\author{Jorge Cuadra}
\affiliation{Departamento de Ciencias, Facultad de Artes Liberales, Universidad Adolfo Ib\'a\~nez, Av.\ Padre Hurtado 750, Vi\~na del Mar, Chile.}
\affiliation{N\'ucleo Milenio de Formaci\'on Planetaria - NPF, Chile}

\author{Antonio S. Hales}
\affiliation{Joint ALMA Observatory, Alonso de C\'ordova 3107, Vitacura 763-0355, Santiago, Chile.}
\affiliation{National Radio Astronomy Observatory, 520 Edgemont Road, Charlottesville, Virginia, 22903-2475, USA.}

\author{Alice Zurlo}
\affiliation{N\'ucleo de Astronom\'ia, Facultad de Ingenier\'ia y Ciencias, Universidad Diego Portales, Av. Ejercito 441, Santiago, Chile}
\affiliation{Escuela de Ingenier\'ia Industrial, Facultad de Ingenier\'ia y Ciencias, Universidad Diego Portales, Av. Ejercito 441, Santiago, Chile}

%%%%% Abstract %%%%%%

\begin{abstract}
The exoplanetary system of HR 8799 is one of the rare systems in which multiple planets have been directly imaged. Its architecture is strikingly similar to that of the Solar System, with the four imaged giant planets surrounding a warm dust belt analogous to the Asteroid Belt, and themselves being surrounded by a cold dust belt analogue to the Kuiper Belt. Previous observations of this cold belt with ALMA in Band 6 (1.3 mm) revealed its inner edge, but analyses of the data differ on its precise location. It was therefore unclear whether the outermost planet HR 8799 b was dynamically sculpting it or not. We present here new ALMA observations of this debris disk in Band 7 (340 GHz, 880 \micron). 
These are the most detailed observations of this disk obtained so far, with a resolution of 1" (40~au) and sensitivity of $9.8\,\mu\mathrm{Jy\,beam^{-1}}$, which allowed us to recover the disk structure with high confidence. In order to constrain the disk morphology, we fit its emission using radiative transfer models combined with a MCMC procedure. We find that this disk cannot be adequately represented by a single power law with sharp edges. It exhibits a smoothly rising inner edge and smoothly falling outer edge, with a peak in between, as expected from a disk that contains a high eccentricity component, hence confirming previous findings. Whether this excited population and inner edge shape stem from the presence of an additional planet remains, however, an open question.
\end{abstract}


%%%%% Body %%%%%%

\section{Introduction} \label{sec:intro}

Many exoplanet-host stars also host debris disks \citep[at least 20\% of FGK stars,][]{Marshall2014,Yelverton2020}, which are revealed by a far-infrared excess due to the presence of dust grains, a by-product of the collisional activity among km-sized solid bodies \citep{Aumann1984}. Just as in the Solar System, the debris disks and planets in extrasolar systems are thought to interact throughout their evolution \citep{Wyatt2008,Krivov2010}, leading to the dynamical formation of structures in disks in response to the influence of the planets \citep{MoroMartin2013}. For instance the Asteroid belt and the Kuiper belt have been partially shaped by interactions with the planets of the Solar System as seen by their limited spatial extent and resonant structures. The detailed study of these interactions has permitted the dynamical history of the Solar System to be reconstructed \citep{Malhotra1995,Levison2008}. Therefore, in addition to informing us on the dynamical history of an exoplanetary system, the structure of debris disks can give valuable insights on the planetary companions that might create it and that might not be directly observable \citep[see e.g., ][]{Mouillet1997,Wyatt1999,Kalas2005,Booth2017,Marino2018,Marino2019,Marino2020,Faramaz2019,Matra2019}.

Situated at a distance of $41.3\,$pc\footnote{Note that most previous studies used the \textit{Hipparcos} distance of 39.4\,pc, which means that previous results on the distance to the star in au are affected by a $\sim 5\%$ difference, which we deemed low enough that is does not influence the comparison.} \citep{GaiaCollaboration2016,GaiaCollaboration2018,Bailer-Jones2018}, HR 8799 is a young $\lambda$ Boo-type star \citep[classified F0V kA5mA5 $\lambda$ Boo;][]{Gray2014} with an age that is estimated to be between 30 and 160 Myr \citep{Marois2008,Geiler2019}. Further detail about the star's characteristics can be found in Appendix ~\ref{sec:stellar}.

For a decade, this star was the only star around which multiple planets have been observed through the direct imaging method \citep[until the recent discovery of two giant planets on wide orbits around TYC 8998-760-1 earlier this year;][]{Bohn2020}.  
Discovered by \citet{Marois2008,Marois2010}, four giant planets orbit HR 8799 with projected orbital radii extending from 15 to 68 au, and masses estimated to be $\sim 7\,$ Jupiter masses (\mjup) for each of the three innermost planets -- HR 8799 c,d, and e -- and $\sim 6\,$\mjup\, for the outermost HR 8799 b \citep{WangJason2018}. In addition, this system contains multiple debris disk components.

The debris disk of HR 8799 was first detected with \textit{IRAS} \citep{Sadakane1986}. A detailed study of this disk, combining \textit{IRAS}, \textit{ISO}, \textit{Spitzer} and JCMT/SCUBA flux measurements allowed \citet{Su2009} to infer the presence of three debris components: an extended halo of small dust grains, a warm debris disk, and a colder debris disk, with an inner edge located at  $\sim 90$ au, then presumed to be sculpted by the outermost planet HR 8799 b. Furthermore, the \textit{Herschel} observations of \citet{Matthews2014} inferred a disk inner edge location of $100\,\pm\,10$\,au, which is compatible with the findings of \citet{Su2009}.

Therefore, the system of HR 8799 follows an architecture strikingly similar to that of the Solar System, but rescaled at wider separations: the four giant planets surround a warm belt analogous to the Main Asteroid Belt, while being themselves surrounded by a cold debris disk, analogous to the Kuiper Belt (which will be the focus of this paper). Simply put, the system of HR 8799 is a younger, broader, and more massive version of the Solar System. 
It is thus not surprising that this system has been monitored over the last decade with many different facilities, not only to observe the planets themselves \citep{Soummer2011,Maire2015,Zurlo2016,Wertz2017,WangJi2018,WangJason2018,Petit2020}, but also its cold debris disk (references above and therein).

While observations of this disk at wavelengths shorter than those of \textit{Spitzer} are difficult because its surface brightness is too low at these wavelengths \citep[which led for instance to a non-detection with \textit{HST;}][]{Gerard2016}, it has nevertheless been abundantly resolved at longer wavelengths, in particular using sub-millimetre and millimetre observations: with the Caltech Submillimeter Observatory (CSO) at 350 microns \citep{Patience2011}, with the SubMillimeter Array (SMA) at 880 microns \citep{Hughes2011}, with the James Clerk Maxwell Telescope (JCMT) at 450 and 850 microns \citep{Williams2006,Holland2017}, with the Atacama Large (sub)Millimeter Array (ALMA) at 1.3 mm \citep{Booth2016} and with the SubMillimeter Array (SMA) at 1.3 mm \citep{Wilner2018}. 

The first resolved sub-mm observations of this debris disk by \citet{Patience2011} with CSO suggested the presence of a brightness asymmetry and a clumpy structure, but it was difficult to be formally conclusive as the signal-to-noise ratio (SNR) of these observations was low. The observations of \citet{Hughes2011} with the SMA were the first interferometric observations of the disk. Whilst the SNR of the observations was poor as well, using the visibilities in combination with the SED, \citet{Hughes2011} demonstrated that both the CSO and SMA data can be adequately fit by an axisymmetric model with no significant evidence for clumps, and found a best fit for the disk inner edge of about 150 au. Using higher angular resolution (yet still low SNR) ALMA 1.3 mm observations, \citet{Booth2016} place the ring inner edge at 145 au with an accuracy of 10\%, consistent with the results of \citet{Hughes2011}.

This is significantly further out than in \textit{Herschel} observations ($100\,\pm\,10\,$au), and this position seems incompatible with the scenario where the outermost planet HR 8799 b sculpts the inner edge of the disk. Indeed, using the upper limits available at that time on the semi-major axis and mass of planet b, which maximizes the theoretical radial extent of the zone cleared by this planet, \citet{Booth2016} concluded that any debris disk exterior to this planet should possess an inner edge located at $\sim 100\,$au. This value was later confirmed by \citet{Read2018} through dynamical modeling and N-body simulations, as well as theoretically by \citet{WangJason2018}, who used the most up-to-date constraints and orbital fit for planet b to derive the expected debris disk inner edge location at $93^{+3}_{-2}\,$au. Consequently, an inner edge observed much further out at 145 au suggests the possible presence of an extra perturbing planet orbiting beyond planet b. 
However, when reanalysing the ALMA 1.3 mm observations, \citet{Wilner2018} found instead a smaller best fit inner edge of $115^{+16}_{-17}\,$au and, when combining with their SMA observations at the same wavelength, an inner edge of $104^{+8}_{-12}\,$au, consistent with \textit{Herschel} derived constraints. It is unclear as of now where the difference between the results of \citet{Wilner2018} and \citet{Booth2016} comes from. 

Following the findings of \citet{Booth2016}, a dynamical study by \citet{Read2018} finds that it is possible for an extra planet to be present exterior to planet b, and which can carve an inner edge at 145 au \citep{Booth2016}, all the while not disrupting the equilibrium of the whole planetary system. They find that the most suitable candidates almost exclusively fall below a Jovian mass -- and therefore below the direct imaging detection limit of 1.25 \mjup\ \citep{Maire2015} -- with a putative planet of 0.1\,\mjup\, orbiting at 138 au dynamically shaping the planetesimal belt in a way that best matches the radial profile observed with ALMA in Band 6 (1.3 mm). Note however that they compare their synthetic observations to the ALMA restored profile, which depends on the underlying best fit model of \citet{Booth2016}, but is nevertheless the best that can be done with such low sensitivity observations. 

Note as well, that \citet{Read2018} did not explore the possibility of a fifth planet in mean-motion resonance with planet b, which could provide additional stability and allow for this fifth planet to be more massive than a Jovian mass. A subsequent dynamical study by \citet{Gozdziewski2018} confirmed this possibility and  found indeed that in that case, a fifth planet could reach masses up to 3\,\mjup. They also confirmed that the four known planets are in a double Laplace resonance configuration, and that a significant proportion of stable orbits for a fifth planet exterior to planet b actually involve yet another resonant configuration, extending the existing chain. 

\citet{Gozdziewski2018} subsequently investigated the carving of the inner edge of the debris disk by an additional fifth planet, and found that this inner edge should exhibit a complex shape, with potentially large resonant clumps. The formation of resonant clumps by a fifth planet was also suggested by \citet{Read2018}.

A possible hint that such resonant trapping has indeed happened could be present in the findings of \citet{Geiler2019}, who used a fully collisional model of the debris disk to find a model that consistently fits both the ALMA 1.3 mm and Herschel data. They found that these two datasets could not be reconciled unless the disk was made of two distinct populations, a quiescent low-eccentricity one on one hand, and an excited one on the other hand, exhibiting larger eccentricities. This is analogous to the populations found in our own Kuiper belt, with the excited population stemming from scattering by Neptune as it migrates outwards in the Kuiper belt, all the while trapping material in resonance \citep[see, e.g.,][and references therein]{Morbidelli2020}.

On the other hand, \citet{Gozdziewski2018} found that it was possible to find a scenario to explain the location of the inner edge at 145 au, which does not imply the presence of an additional planet. Indeed, they found that is was possible to find appropriate initial wider configurations for the four known planets and migration rates to lead them to reach their final double Laplace resonant state, all the while leaving a debris disk which will be made of a scarce population between 90 and 150\,au, followed by a main disk starting at $\sim 150$\,au. Their work suggests that in that case, a complex asymmetric inner edge shape and resonant trapping may occur.
An extension of this work in \citet{Gozdziewski2020} led them to depict more precisely that while the ring inner edge location would agree with that of \citet{Wilner2018}, a ring of high-eccentricity planetesimals should also be expected at 140-160 au, that is, near the inner edge reported by \citet{Booth2016}. This high eccentricity component would induce a higher dust production rate and stronger disk emission, such that the disk radial intensity profile would not be adequately fit by a single power-law, as both \citet{Booth2016} and \citet{Wilner2018} assumed. 

There are thus a number of questions that remain open about this exoplanetary system and its cold debris disk. What is the location of the inner edge of this debris disk? Is an extra planet required to sculpt it? What is the shape of this inner edge? Is it symmetric, or does it exhibit a complex asymmetric shape and resonant clumps? Is the disk adequately fit by a single power-law profile or is there a peak emission due to local velocity dispersion and high eccentricity planetesimals? In order to answer these questions, we need better constraints in the sub-mm on the geometry of the debris ring of HR 8799, and in particular on the shape of the disk inner edge and on the presence of clumps. These are crucial for assessing the dynamical history of this system, and for making conclusions on the potential presence of an additional planet. This motivated us to observe this debris disk at high angular resolution with ALMA in Band 7 (870 microns), and to push those to resolve the disk at sufficient sensitivity. At this wavelength -- shorter than Band 6 (1.3 mm) -- we can expect the cold disk to be brighter, while still dominating the observations, as the warm belt and the halo of small grains seen in far-IR are not expected to emit significantly at mm wavelengths. 

We present our observations in Section \ref{sec:obs}, and their modelling in Section \ref{sec:model}. We then discuss our results and summarize our conclusions in Sections \ref{sec:discussion} and \ref{sec:conclusion}, respectively.

%------------------------------------------------------------------------------------------------
%NOTES:

%Patience 2011: emission found consistent with disk extending from 100-300 au, but with clumpy structure, could it be that they saw the BG source? The system moves slowly, it should be there in the observations. HOW COME NO ONE SAW THAT CLUMP IN PREVIOUS OBSERVATIONS AT THE SAME LAMBDA?
%Hughes 2011: clump as well, but this time the clump is on the other side of the star than seen with CSO observations. However, they say that a smooth axisymmetric disk fits well the observations (adding CARMA obs to the dataset). based on visibilities and SED modelling, show the disk should be broad. using the averaged visibilities show a best fit for the inner edge semi-major axis of around 150 au
%------------------------------------------------------------------------------------------------

\section{ALMA Observations} \label{sec:obs}

We present here ALMA observations of the debris disk of HR 8799 in Band 7 (340 GHz, 880 microns), under the project 2016.1.00907.S (PI: V. Faramaz). They comprise 12m-array observations carried out from 2018 May 13 to 2018 June 1 (6 separate observations), and observations carried out with the Atacama Compact Array (ACA) from 2016 October 26 to 2017 July 7 (28 separate observations). Tables~\ref{tab:obslog12m} and \ref{tab:obslogACA}, summarize the 12m Array and the Atacama Compact Array (ACA) observations, respectively.

ACA data were taken using baselines ranging from 8.9 to 48.9 m, which corresponds to angular scales of 20\farcs1 and 3\farcs7, respectively, while 12m-array data were taken using baselines ranging from 15 to 314 m, corresponding to angular scales of 12\arcsec\ and 0\farcs6. Given the distance of the star ($d\,=\,41.3\,$pc), these angular resolutions translate to spatial scales that were probed with the ACA ranging from 150 to 830\,au, and those probed with the 12m-array ranging from 24 to 490\,au.

%New parallax measured by Gaia is 24.218 mas with an error of 0.088 mas. Gives a distance of 41.3 pc pm 0.15 pc

The spectral setup consisted of four spectral windows, each 2 GHz wide. For both the 12m array and ACA observations, three were centered on 334, 336, and 348 GHz, and divided into 128 channels of width 15.625 MHz $\approx$ 13.6\,\kms. As CO gas emission was reported in previous ALMA observations \citep{Booth2016}, we used the fourth spectral window to target the CO J=3-2 emission line. Therefore, the fourth spectral window was centered on 346 GHz, with a large number of finer channels (4096 for the ACA observations and 3840 for the 12m array observations), leading to a spectral resolution of 0.5 MHz $\approx$ 0.4\,\kms. Analysis of the CO gas line emission can be found in Appendix \ref{sec:app_gas}.
%delta v= c x delta nu / nu values.
The total time on source was 19.6 hours with the ACA and 4.5 hours with the 12m array.

%%%%%%%%%%%%%%%
%We averaged the data according to the prescription of Equations 3.192 and 3.194 for bandwidth and time smearing.. They tell us which spectral and time averaging can be performed without compromising the data and losing information. The reference to know how much can be averaged in time and channels is Essential Radio Astronomy (online at https://www.cv.nrao.edu/~sransom/web/Ch3.html#S7.SS3.p2 )
%%%%%%%%%

%%% TABLE 12m 

\begin{deluxetable*}{cc|cccc|ccc}[htbp] 
\tablecaption{Summary of our ALMA/12m array observations at 870 microns (Band 7). \label{tab:obslog12m}}
%\tablenum{2}
\tablewidth{0pt}
\tablehead{
\colhead{Date\tablenotemark{a}} &
\colhead{Time\tablenotemark{a}} &
\colhead{On source} & \colhead{$\mathrm{N_{Ant.}}$} & \colhead{PWV} & \colhead{Elevation} & \multicolumn{3}{c}{Calibrators} \\
\colhead{(YYYY-mm-dd)} & \colhead{(UTC)} &
\colhead{(min)} & \colhead{} & \colhead{(mm)} & \colhead{(deg)} & \colhead{Flux} & \colhead{Bandpass} & \colhead{Phase}
}
\startdata
2018-05-13 & 10:44:13.4 & 44.8 & 43 & 0.36-0.44 & 42.0-46.0 & J2148+0657 & J2253+1608 & J2253+1608  \\[2pt]
2018-05-21 & 10:39:38.8 & 45.3 & 48 & 0.65-0.77 & 44.5-46.0 & J2148+0657 & J2253+1608 & J2253+1608  \\[2pt]
2018-05-24 & 10:09:52.7  & 45.3 & 46 & 0.23-0.35  & 43.0-46.0 & J2148+0657 & J2253+1608 & J2253+1608  \\[2pt]
2018-05-29 & 09:22:59.8 & 45.3 & 45 & 0.51-0.63 & 41.0-45.5 & Titan & J2253+1608 & J2253+1608  \\[2pt]
2018-05-29 & 10:47:57.1 & 45.3 & 45 & 0.63-0.70 & 44.0-46.0 & J2148+0657 & J2253+1608 & J2253+1608  \\[2pt]
2018-06-01 & 10:14:25.4  & 44.8 & 45 & 0.88-1.04  & 45.0-46.0 & J2148+0657 & J2253+1608 & J2253+1608  \\[2pt]
\enddata
\tablenotetext{a}{At exposure start.}
\end{deluxetable*}

%%% TABLE ACA

\begin{deluxetable*}{cc|cccc|ccc}[htbp] 
\tablecaption{Summary of our ALMA/ACA observations at 870 microns (Band 7). \label{tab:obslogACA}}
%\tablenum{2}
\tablewidth{0pt}
\tablehead{
\colhead{Date\tablenotemark{a}} &
\colhead{Time\tablenotemark{a}} &
\colhead{On source} & \colhead{$\mathrm{N_{Ant.}}$} & \colhead{PWV} & \colhead{Elevation} & \multicolumn{3}{c}{Calibrators} \\
\colhead{(YYYY-mm-dd)} & \colhead{(UTC)} &
\colhead{(min)} & \colhead{} & \colhead{(mm)} & \colhead{(deg)} & \colhead{Flux} & \colhead{Bandpass} & \colhead{Phase}
}
\startdata
2016-10-26 & 02:02:56.1 & 45.5 & 10 & 0.43-0.72 & 31.0-41.5 & Uranus & J2253+1608 & J2253+1608  \\[2pt]  %MS01
2016-11-08 & 00:06:03.1  & 45.5 & 12  & 0.51-0.78  & 40.5-45.5 & Neptune & J2253+1608 & J2253+1608  \\[2pt]  %MS02
2016-11-08 & 22:59:42.2  & 46.0 & 11 & 0.56-0.71  & 44.6-45.8 & Neptune & J2253+1608 & J2253+1608  \\[2pt]  %MS03
2016-11-22 & 23:56:27.5  & 40.5 & 11 & 0.63-0.83  & 34.0-43.0 & Uranus & J2253+1608 & J2253+1608  \\[2pt]  %MS04
2016-11-25 & 23:27:07.7 & 57.7 & 11 & 0.84-1.3 & 36.5-44.5 & Uranus & J2253+1608 & J2253+1608  \\[2pt]  %MS05
%2016-11-26 & 01:53:07.7 & 19.2 & 11 & 0.73-0.81 & --\tablenotemark{b} & Uranus & J0006-0623 & J2253+1608  \\[2pt]  %MS06, graph elevation not normal
2016-11-26 & 23:03:25.6 & 45.5 & 11 & 1.02-1.32 & 39.5-45.5 & Uranus & J0006-0623 & J2253+1608  \\[2pt]  %MS07
%2016-11-27 & 01:20:38.0 & 31.9 & 11 & 1.16-1.32 & --\tablenotemark{b} & Uranus & J0006-0623 & J2253+1608  \\[2pt] %MS08, graph elevation not normal
2016-12-02 & 21:59:36.8 & 43.8 & 11  & 1.45-1.65 & 43.8-46.0 & Neptune & J2253+1608 & J2253+1608  \\[2pt]
2016-12-03 & 22:53:19.4 & 45.5 & 10  & 0.39-1.13 & 36.5-44.5 & Uranus & J2253+1608 & J2253+1608  \\[2pt]
2016-12-17 & 23:00:46.8 & 46.0 & 11  & 1.01-1.36 & 34.5-38.5 & Mars & J2253+1608 & J2253+1608  \\[2pt]
2017-04-29 & 11:16:15.3 & 44.7 & 11  & 0.79-0.91 & 42.5-46.0 & Neptune & J2253+1608 & J2253+1608  \\[2pt]
2017-05-02 & 13:56:41.8 & 45.5 & 11 & 0.74-0.89 & 34.5-39.5 & Neptune & J2253+1608 & J2253+1608  \\[2pt]
2017-05-03 & 11:29:03.3 & 46.0 & 10 & 0.35-0.39 & 44.6-46.0 & Neptune & J2253+1608 & J2253+1608  \\[2pt]
2017-05-04 & 11:55:11.6 & 45.5 & 12 & 0.51-0.55 & 43.4-46.0 & Neptune & J2253+1608 & J2253+1608  \\[2pt]
2017-05-06 & 11:25:18.0 & 46.0 & 10  & 0.28-0.32 & 44.8-46.0 & Neptune & J2253+1608 & J2253+1608  \\[2pt]
2017-05-07 & 12:04:09.2 & 45.5 & 12  & 1.23-1.43 & 41.5-46.0 & Neptune & J2253+1608 & J2253+1608  \\[2pt]
2017-05-09 & 12:20:22.8 & 46.0 & 12 & 0.80-0.87 & 38.5-45.5 & Neptune & J2253+1608 & J2253+1608  \\[2pt]
2017-05-16 & 11:07:06.7 & 12.8 & 10 & 0.56-0.59 & 45.6-46.0 & Neptune & J2253+1608 & J2253+1608  \\[2pt]
2017-05-17 & 11:16:55.6 & 45.0 & 10 & 0.95-1.04 & 42.5-46.0 & Neptune & J2253+1608 & J2253+1608  \\[2pt]
2017-05-20 & 10:23:19.1 & 45.0 & 11  & 0.99-1.05 & 44.6-46.0 & Neptune & J2253+1608 & J2253+1608  \\[2pt]
2017-06-23 & 07:56:54.2 & 47.0 & 9  & 0.43-0.46 & 43.8-46.0 & Neptune & J2253+1608 & J2253+1608  \\[2pt]
2017-06-24 & 09:25:36.7 & 12.8 & 9 & 0.68-0.78 & 43.8-45.0 & Neptune & J2253+1608 & J2253+1608  \\[2pt]
2017-06-27 & 09:12:31.2 & 45.5 & 10 & 0.51-0.61 & 38.0-45.0 & Neptune & J2253+1608 & J2253+1608  \\[2pt]
2017-07-03 & 06:56:50.7 & 45.5 & 10 & 0.33-0.44 & 42.0-46.0 & Neptune & J2253+1608 & J2253+1608  \\[2pt]
2017-07-03 & 09:00:56.1 & 46.0 & 10  & 0.25-0.36 & 36.5-44.5 & Neptune & J2253+1608 & J2253+1608  \\[2pt]
2017-07-04 & 08:48:11.6 & 45.5 & 10 & 0.29-0.38 & 38.0-45.0 & Neptune & J2253+1608 & J2253+1608  \\[2pt]
2017-07-05 & 09:03:21.1 & 45.5 & 10 & 0.63-0.70 & 35.0-44.0 & Uranus & J2253+1608 & J2253+1608  \\[2pt]
2017-07-06 & 07:37:49.1 & 45.5 & 10 & 0.52-0.59 & 44.0-46.0 & Neptune & J2253+1608 & J2253+1608  \\[2pt]
2017-07-07 & 07:54:5097 & 46.0 & 11 & 0.50-0.61 & 42.5-46.0 & Neptune & J2253+1608 & J2253+1608  \\[2pt]
\enddata
\tablenotetext{a}{At exposure start.}
%\tablenotetext{b}{\virginie{[The plot returned does not display all the fields. I am unable to provide the elevation at which the source has been observed for those measurement sets. After inspecting the data more thoroughly, it appears that the reason why the elevation of the source does not appear in the plot is that all the data of field number 3 (the source), were flagged. Consequently, should I still include these measurement sets in the observations? Is it possible that they have been flagged by mistake? I would have expected that a MS where all the data from the source have been flagged would not pass the quality test and would be removed entirely from the delivery.]}}
\end{deluxetable*}

%%%%%%%%%%%%%%%%%%%%%%%%%%%%%%%%%%%%%%%%%%%%%

\subsection{Results}

\begin{figure*}
\makebox[\textwidth]{
\includegraphics[scale=0.24]{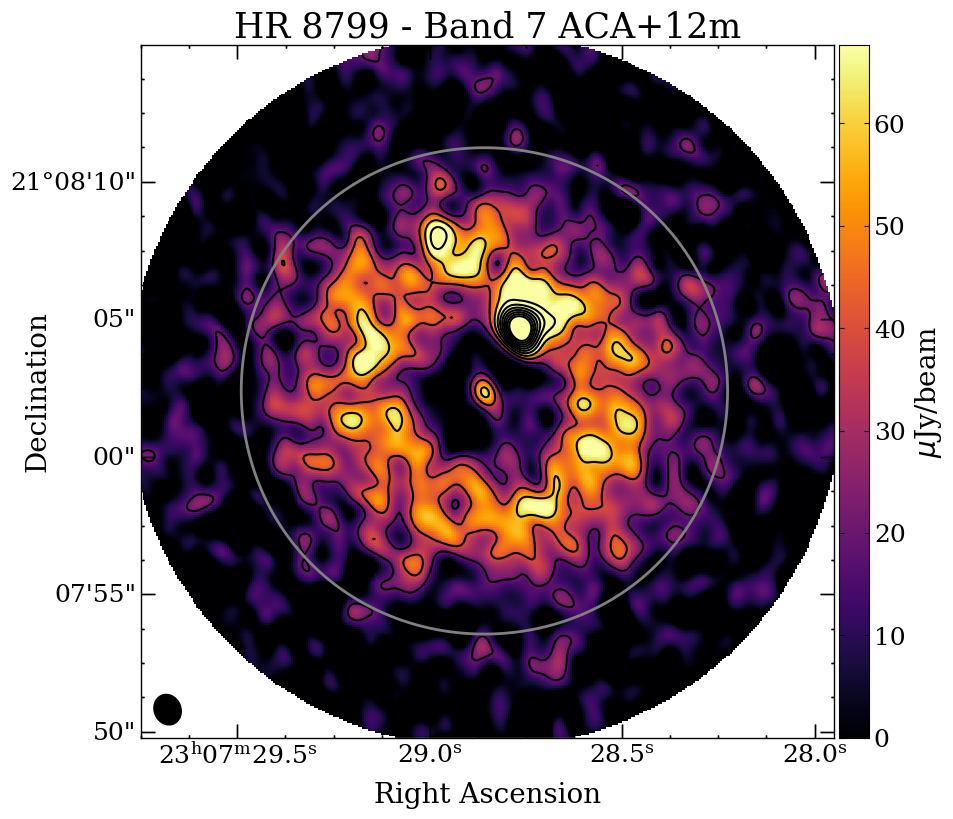}
\includegraphics[scale=0.24]{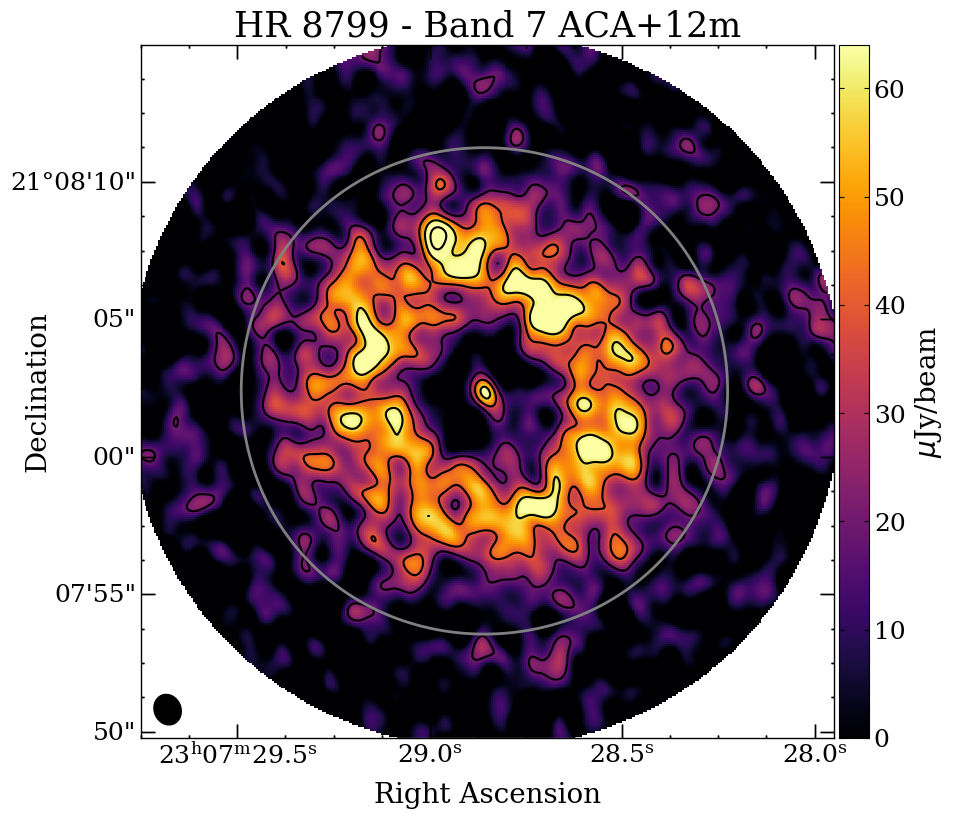}
\includegraphics[scale=0.48]{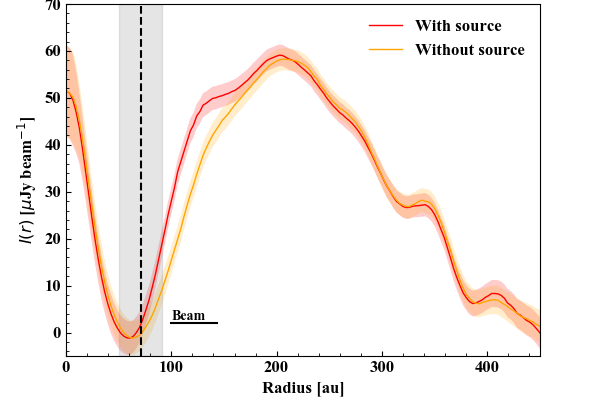}}
\caption{\emph{Left:} ALMA 880 microns continuum observations of HR 8799, combining the 12m and ACA data. North is up and East is left, with contours showing the $\pm$2, 4,... $\sigma$ significance levels. The synthesized beam is shown on the lower left side of images. The grey circle indicates the 50\% response of the 12m Array primary beam and the color bar shows the fluxes in $\mu\mathrm{Jy\,beam}^{-1}$. This image was obtained using a natural weighting scheme, resulting in a sensitivity of $\sigma=9.8\,\mu\mathrm{Jy.beam}^{-1}$ and a synthesized beam of dimensions $1\farcs13 \times 0\farcs97$, with position angle $22^{\circ}$. Central emission appears inner to the ring and is detected with SNR of 7, and a bright source (SNR $> 10$) appears within the ring, northwest of the star. \emph{Middle:} Same as left panel, with the best fit bright source (see Section\,\ref{sec:model}) having been subtracted from our observations. \emph{Right:} Corresponding intensity profile, corrected for the primary beam, obtained by integrating azimuthally over $360^\circ$ (in red) and over a range of angles that exclude the bright source seen at the inner edge of the ring (in orange), assuming the ring has an inclination of $26.8^\circ$ and PA of $68.3^\circ$, the same orientation as the orbit of the planets \citep[see Table 4 of][]{WangJason2018}. The vertical dashed line corresponds to the semi-major axis of the outermost planet HR 8799 b as derived by \citet{WangJason2018}, while the grey region corresponds to its chaotic zone given by $a_p \pm 1.7 a_p (M_p/M_\star)^{0.31}$, which holds for high mass planets such as HR 8799 b \citep[][with $a_p$ as the semi-major axis of the planet, $M_p$ as its mass, and $M_\star$ as the mass of the star; these parameters were given best fit values as found by \citet{WangJason2018}]{Morrison2015}.}
\label{fig:HR8799_continuum}
\end{figure*}

We recalibrated the raw data using the pipeline and calibration script provided by ALMA. The calibrators are listed in Tables~\ref{tab:obslog12m} and \ref{tab:obslogACA}, for the 12m Array and the ACA observations, respectively. We used the \emph{TCLEAN} algorithm and task in CASA version 5.1.1 \citep{McMullin2007} to perform the image reconstruction of the continuum emission, that is, to obtain the inverse Fourier transform of the observed visibilities. Note that we excluded the channels spanning the CO emission.
To recover the maximum SNR, we combined the four spectral windows. We used a natural weighting scheme, resulting in a beam size of $1\farcs13 \times 0\farcs97$ ($\sim 43\,$au at $d=41.3\,$pc), with position angle of $22^{\circ}$, while the image possesses an RMS of $\sigma\,=\,9.8\,\mu\mathrm{Jy\,beam^{-1}}$.\footnote{Since the disk occupies a significant portion of the field of view, it is difficult to find a large region away from the source that is solely noise and thus from which to measure the RMS. Therefore, as in \citet{Booth2016}, we measure the RMS in a dirty image pointing $90\arcsec$ away from the target in Declination.} 
%\textbf{This approach is validated as the RMS found in our residual maps after subtracting our best fit models (see Section~\ref{sec:model} and ~\ref{sec:app_residuals}) is close enough to the value found here.}}
We show the resulting image -- non corrected with the primary beam -- in the left panel of Figure \ref{fig:HR8799_continuum}.

\paragraph{The disk}

 The disk is well resolved; it has a noisy appearance and is seen with SNR per beam of up to 6 and a peak emission of $\sim\,70\,\mu\mathrm{Jy\,beam^{-1}}$; it also appears close to face-on (or only moderately inclined), and axisymmetric. It exhibits a ring-like architecture, extending from $\sim$ $2\arcsec \, (\sim$ $80\,\mathrm{au})$ to $\sim$ $7\arcsec \, (\sim$ $300\,\mathrm{au})$ in radius, with a clear central cavity.

\paragraph{The central emission}

In addition to the bright outer ring, a point-like source is detected at the center of the cavity.
A Gaussian ellipse fit to this central emission performed in CASA (using the \texttt{imfit} task over an aperture 2\farcs5 in diameter) is centered on $23^{\mathrm{h}}07^{\mathrm{m}}28^{\mathrm{s}}.852 \pm 0.004$ in Right Ascension, and $+21^{\circ}08\arcmin 02\farcs26 \pm 0.08$ in Declination, with peak emission and integrated flux measured as $65.8\pm 6.7$ $\mu\mathrm{Jy} \mathrm{beam}^{-1}$ and $73\pm 13$ $\mu\mathrm{Jy}$, respectively. This central emission is thus detected with a SNR $\simeq$ 7. Since the integrated flux and peak flux are consistent, we conclude that the central emission is unresolved.

According to \textit{Gaia}'s findings on the position and proper motion of HR 8799 \citep{GaiaCollaboration2018}, the star's position is expected to be $23^{\mathrm{h}}07^{\mathrm{m}}28^{\mathrm{s}}.86$ in Right Ascension, and $+21^{\circ}08\arcmin 02\farcs39$ in Declination (at the time of the ALMA 12m Array observations which resolved the central emission from that of the cold debris disk; see Table \ref{tab:obslog12m}). Therefore, we can conclude that the position of the central emission as returned by ALMA is consistent with the expected position of the star. Note that the astrometric accuracy returned by our Gaussian fitting procedure is consistent (within a factor of two) with ALMA's nominal astrometric accuracy.\footnote{ALMA's astrometric accuracy is given as 70\arcsec $\times(\nu \times B \times SNR)^{-1}$, where $\nu$ is the frequency of the observations (in GHz) and $B$ is the largest used baseline (in km) (Equation 10.7 of the Cycle 7 ALMA Technical Handbook). This source is seen at SNR 7, $\nu=345\,$GHz, and the longest baseline is $B= 0.3137$km (in the 12m array observations), which, according to the formula above, leads to an expected astrometric accuracy of 92.4 mas.}  The centroid of the ring emission is offset no more than 0.22" or 9 au from the central emission source, implying an upper limit to the ring eccentricity of 0.05.

\paragraph{A bright source}

A bright source (with peak flux $\sim 270 $ $\mu\mathrm{Jy beam}^{-1}$) is present within the ring and close to its inner edge in the northwest direction (position angle $334^\circ \pm 2$). Since its emission is mingled with that of the disk, it is difficult to characterize it directly from the image.
In the right panel of Figure \ref{fig:HR8799_continuum}, we present the intensity profile of the ring, isolating the effect of this bright source on it by integrating azimuthally over the whole ring and then by a range of angles excluding this bright source. Our goal here is to qualitatively assess the effect of this bright source on the disk profile, therefore, we assumed here that the ring has the same orientation as the most recent one found for the  orbit of the planets, with an inclination of $26.8^\circ$ and PA of $68.3^{\circ}$ \citep[see Table 4 of][]{WangJason2018}.
We note that the presence of this bright source pushes the inner edge of the ring further inwards as compared to when it is not included. This means that the presence of this bright source would bias any fit that does not include it in the model. In addition, including this source in our models will disentangle it from the disk and facilitate its characterization. We will further discuss its nature in Section~\ref{sec:bright_source_discussion}.

\medskip

In the next section we will fit the emission with a detailed model, to quantify the inner edge location and to characterize all the observed components (outer disk, central emission, bright source).

\section{Modelling} \label{sec:model}

\subsection{Model Description}

In order to characterize more precisely the disk's architecture we now fit, in visibility space, parametric debris disk models to the available ALMA observations, i.e., we simultaneously fit three datasets: our 12m and ACA Band 7 observations, along with the 12m Band 6 observations from Cycle 1 \citep[project 2012.1.00482.S, presented in][]{Booth2016}. The calibrated dataset for this was retrieved from the ALMA archive and treated similarly to our Band 7 data (removal of the CO line emission, and combination of the four spectral windows).

We follow the method used in \citet{Marino2018,Marino2019,Marino2020}, combining radiative transfer simulations \citep[RADMC-3D,][]{Dullemond2012} with a Markov chain Monte Carlo (MCMC) procedure \citep[\texttt{emcee};][]{Foreman-Mackey2013} to find the ranges of disk parameters that are consistent with the observations. More specifically, the radiative transfer code is used to produce model images which are multiplied by the primary beam, and then translated into model visibilities at the same $uv$ points as the 12m Band 7, ACA Band 7, and 12m Band 6 observations for comparison with those observations through a $\chi^2$. This is then fed into the MCMC procedure to sample the parameter space. The final $\chi^2$ is defined as the sum of the individual $\chi^2$ from the band 7 12m, ACA and band 6 12 m data \citep[as in][]{Marino2018,Marino2020}. The weights of each individual data set are renormalised to ensure the reduced $\chi^2$ of each data set is about unity \citep[for a detailed justification of this procedure see][and in particular, their Equation 4]{Booth2021}.

Guided by previous work and the radial profile extracted from the ALMA image (Figure~\ref{fig:HR8799_continuum}), we consider three models for the disk surface density distribution, as described below. 

We first adopt a disk model (hereafter referred to as Model 1) made of a single power-law. Our goal here is to use the same type of simple parametric model that both \citet{Booth2016} and \citet{Wilner2018} used to fit the ALMA Band 6 observations, and facilitate comparison with those previous works. The surface density $\Sigma(r)$ of the disk is parametrized as
\begin{equation}\label{eq:1plaw}
 \Sigma(r)=  \Sigma_0
\left\{
\begin{array}{lr}
0  & \qquad  r < r_{min}, \\
 & \\
\left(\frac{r}{r_{min}} \right)^{\gamma_1} & \qquad  r_{min} < r < r_0, \\
 & \\
0 & \qquad  r > r_{max},
\end{array}
\right.
\end{equation}
\noindent where $r_{min}$ and $r_{max}$ are the disk inner and outer edges, respectively, and $\gamma$ is the power-law index.

As pointed out in Section~\ref{sec:intro}, the inner edge of the disk is of particular interest, potentially exhibiting planet-driven structure. In particular, the most recent theoretical study of planet-disk interactions for the system of HR 8799 specifically predicts that the disk radial profile should be inconsistent with a single power-law \citep{Gozdziewski2020}. We thus seek to determine how well our Band 7 observations can constrain the inner edge radial profile. We also aim to determine whether more complex models are a better fit to the observations as compared to a simple single power-law profile. Therefore, we performed two additional fittings, using the following parametric models. 

For our second model (Model 2) we consider a more gradual falloff, as compared to the sharp edges in Model 1. The surface density for this model consists of three power laws:

\begin{equation}\label{eq:3plaw}
 \Sigma(r)=  \Sigma_0
\left\{
\begin{array}{lr}
\left(\frac{r}{r_{1}} \right)^{\gamma_1}  & \qquad  r < r_1, \\
 & \\
\left(\frac{r}{r_{1}} \right)^{\gamma_2} & \qquad  r_1 < r < r_2, \\
 & \\
\left(\frac{r_2}{r_{1}} \right)^{\gamma_2} \left(\frac{r}{r_{2}} \right)^{\gamma_3} & \qquad  r > r _2  \\
\end{array}
\right.
\end{equation}
\noindent where $r_1$ and $r_2$ mark the locations where the slope changes.

Our third model for the disk surface density (Model 3) considers yet another possible profile for the disk edges -- Gaussian declines, with a single power law in between.
\begin{equation}\label{eq:1plaw_gauss}
 \Sigma(r)=  \Sigma_0
\left\{
\begin{array}{lr}
e^{ -\frac{1}{2} \left( \frac{r-r_1}{\sigma_{1}} \right)^2 }  & \qquad  r < r_1, \\
 & \\
\left(\frac{r}{r_{1}} \right)^{\gamma} & \qquad  r_1 < r < r_2, \\
 & \\
\left(\frac{r_2}{r_{1}} \right)^{\gamma} e^{ -\frac{1}{2} \left( \frac{r-r_2}{\sigma_{2}} \right)^2 } & \qquad  r > r _2  \\
\end{array}
\right.
\end{equation}
Additionally, we set the surface density to zero interior to 70~au for both Model 2 and Model 3, since debris further in would have been cleared by the giant planets, and beyond 600~au because our observations are not sensitive to emission beyond that separation ($\sim14\arcsec$).
 
The surface densities defined above are then fed into radiative transfer calculations with RADMC3D which allows to compute images for a given random selection of surface density parameters, disk orientation (inclination and position angle), and total dust mass. For this we also need to define a certain dust opacity, which will determine the final dust mass that we infer from the disk emission. The dust is assumed to have a size distribution and composition as in \citet{Marino2018} -- the dust grains are a mix of 70\% astrosilicates \citep{Draine2003}, 15\% amorphous carbon, and 15\% water ice \citep{Li1998}, with a size distribution following a power law of index $-$3.5 over sizes from 1 micron to 1 centimeter -- resulting in a Band 7 dust opacity of 1.9 cm$^2$/g. To save computational cost the disk surface brightness is only computed at band 7, and it is scaled to band 6 via a spectral index, $(\alpha_\mathrm{mm})$, which is left as free parameter. This is justified since the disk emission is in the Rayleigh-Jeans regime at these wavelengths, and thus the spectral index is uniform across the disk if the dust opacity is uniform as well. In addition, we leave the flux of the central star as a free parameter as a way to account for extra emission due to the unresolved warm belt or stellar chromospheric emission. 
%Note that this flux is not related to the calculation of the dust emission. For this, we used a PHOENIX model from the library of \citet{2013A&A...553A...6H}, and choosing the one which temperature was the closest to the best fit of 7380 K obtained by \citet{2014ApJ...780...97M}, that is, $T_\star=7250\,$K.
Finally, in order for our models not to be biased towards small inner edges due to the bright source, we include in the images an extra source with surface brightness defined as an elliptical Gaussian. The shape (standard deviations along its major and minor axis, and position angle), fluxes (at band 7 and band 6) and its relative position to the star at the mean epoch of the band 7 observations are also left as free parameters\footnote{Its position relative to the star at the epoch of the band 6 observations is calculated based on HR~8799's proper motion.}. This adds a 7 free parameters to our global model.

These model images are then multiplied by their corresponding primary beams, and Fourier transformed to obtain model visibilities at the same \emph{uv} points as the observations. Then we allow for phase center offsets in Right Ascension and Declination, and separately for the three observation sets (Band 7 12m, Band 7 ACA, Band 6 12m), adding six free parameters. This gives us a total of 22 free parameters for Model 1, and 24 for Models 2 and 3. These model visibilities are then compared directly with the observed visibilities via a $\chi^2$, and with a MCMC procedure the parameter space is explored to obtain a Bayesian posterior distributions for all the free parameters.

We assume uniform priors except for the disk inclination and position angle, for which we make the (reasonable) assumption that it is co-planar with the orbits of the planets (see discussion in Section \ref{sec:discussion_disk_orientation}). These priors take the form of normal distributions, which, using Table 4 of \citet{WangJason2018}, leads to $i \sim \mathcal{N} (26.8^{\circ}, 2.3^{\circ})$ and $\Omega \sim \mathcal{N} (68.25^{\circ}, 5.55^{\circ})$. Assuming co-planarity is justified by the expected secular evolution of a planet and a less massive disk. Even if initially misaligned by $\lesssim30^{\circ}$, a less massive debris disk will realign in a few secular time-scales ($\sim2-3$ Myr at 200~au) and end up with a high scale height that is of the order of the initial misalignment \citep{Pearce2014}.

%%%%%%%%%%%%%%%%%%%%%%%%%%%%%%%%%%%%%%%%%%%%%%%%%%

\subsection{Best-Fit Model Results}

The best-fit values and uncertainties are given for each parameter in Tables~\ref{tab:ring_models} and \ref{tab:ring_mcmc_results}. The interested reader will find the cornerplots that show the correlations between the disk surface density parameters for our three models in Appendix \ref{sec:app_cornerplot}, as well as the residual maps obtained by subtracting our best fit models to the observations in Appendix \ref{sec:app_residuals}.

We list the best-fit $\chi^2$ value for each of our models in Table~\ref{tab:ring_models}. Note that these $\chi^2$ are very large due to the equally large number of degrees of freedom involved when fitting a set of visibilities in the $(u,v)$ space (of the order $10^5$ for the ACA data and $10^6$ for the 12m data). We list as well the reduced $\chi^2$, which are all extremely similar. It is thus difficult to assess the success of each model via examination of the $\chi^2$ or reduced $\chi^2$ values as they are.

A sense of which model is more adequate than another can be quantified using the individual likelihood function $\exp({-\chi^2/2})$ and considering $\Delta \chi^2$, that is, the difference between $\chi^2$ (also displayed in Table \ref{tab:ring_models}). From there, taking Model 3 -- which has the smallest $\chi^2$ -- as reference, it is easily shown that $\exp{(-\Delta \chi^2/2)}$ will tell us how many times less likely a model will be a good fit to our data than Model 3. We find that Model 3 is about 60 times more likely than Model 2, which is only a marginal preference as compared to the likelihood of Model 1, which is nearly 10 billion times less likely than Model 3. Using the same reasoning and taking Model 2 as reference, we find that Model 2 is about 200 million times more likely to fit our data than Model 1. Even after considering the lower number of free parameters of Model 1, e.g. through the Bayesian Information Criterion, it is still extremely disfavoured compared to Models 2 and 3.

%%%%%%%%%%%%%%%%%%%%%%%%%%%%%%%%%%%%%%%%%%%
%            INDIVIDUAL RESULTS
%%%%%%%%%%%%%%%%%%%%%%%%%%%%%%%%%%%%%%%%%%%

\paragraph{The disk}

We find consistent results in the left panel of Figure \ref{fig:surfaceDensities}, where we compare the radial profiles in Band 7 of the three models we tested. We find that our Model 1 clearly underestimates the disk flux between $\sim 100-130$ au, as well as the peak location, which it places at $135 \pm 4\,$au. This is a direct consequence of it being described with a sharp inner edge. From there, the single power-law portion of the model struggles to reproduce the bulk and tail of the disk. This overall yields significant residuals.  
It is thus more adequately represented by Models 2 and 3, which both describe a disk with smooth edges and a bulk peaking at $\sim 200$ au. 

These differences are barely visible in the Band 6 observations (see right panel of Figure \ref{fig:surfaceDensities} which displays the same comparison in Band 6). From our Band 7 higher resolution and more sensitive observations, it is clear that the structure is more complex than a single power law can describe, with an emission profile that rises to a peak before falling off. 
The three models, on the other hand, all agree well upon the orientation of the disk (inclination $\sim 30^\circ$ and position angle $\sim 60^\circ$)\footnote{Note that we imposed strong priors on these parameters based on the orbits of the planets. See Section~\ref{sec:discussion_disk_orientation} for a discussion.}, its mass (of the order of $10^{-1}\,\mathrm{M_\oplus}$), and its millimeter spectral index $\alpha_\mathrm{mm}$, which in our preferred model (Model 3) was found to be $2.5\pm0.2$ (see Table~\ref{tab:ring_mcmc_results}). 

Finally, we measure the total disk flux directly from the best fit model images produced for each of our three models used in the MCMC fit. We report these values in Table~\ref{tab:ring_mcmc_results}. 
Model 1 returns the smallest fluxes, and Model 2 the highest. Our preferred model (Model 3) finds the disk flux to be $3.1 \pm 0.5\,$mJy and $8.1\pm 1.0\,$mJy in Band 6 and Band 7, respectively (including a 10\% absolute flux calibration uncertainty). 

\paragraph{Additional sources}

Our three models also agree well on the characteristics of the central emission and those of the bright source located near the disk inner edge.
According to our preferred model (Model 3), the central emission was found to possess a flux of $71.3^{+10.2}_{-11.1}\,\mu\mathrm{Jy}$ in Band 7 and $37.9^{+16.0}_{-16.2}\,\mu\mathrm{Jy}$ in Band 6. The Band 7 flux agrees with what was directly retrieved from the observations via a Gaussian fit within CASA. 

The bright source is found offset northwest from the star, and more specifically by $-1\farcs28\,\pm\,0.05$ in Right Ascension and $2\farcs34^{+0.06}_{-0.05}$ in Declination  at the epoch of the Band 7 observations. Deconvolved from the beam, it is best described by an elliptical Gaussian of standard deviation $0\farcs24 \pm 0.04$ along its major axis, $0\farcs15 ^{+0.04}_{-0.05}$ along its minor axis, and position angle $18^{\circ\,+21}_{-20}$. 

Note that it is only marginally resolved. Its FWHM along its major axis is $0\farcs57$, which is 20\% smaller than the beam major axis using uniform weights ( $0\farcs69$). Its FWHM along its minor axis is $0\farcs35$, although still consistent with zero. This means that we marginally resolve its major axis. We also rule out that its major axis is constrained due to beam smearing or the combination of data from multiple dates. These two effects could only contribute to a 0\farcs01 smearing of a point source at a $\sim3\arcsec$ separation. Its total flux is found to be $316^{+20}_{-19}$ $\mu\mathrm{Jy}$ at 880 microns (Band 7) and $58 \pm 18$ $\mu\mathrm{Jy}$ at 1.3 mm (Band 6). This results in a spectral slope of $4.34^{+1.11}_{-0.85}$.

\begin{figure}
\movetabledown=50mm
\begin{rotatetable*}
\begin{deluxetable*}{ccccccccc}
\tablecaption{Comparison of our best fit models -- Disk profiles  \label{tab:ring_models}}
%\tablenum{5}
\tablewidth{0pt}
\tablehead{ 
\multicolumn{3}{c}{Model 1} & \multicolumn{3}{|c|}{Model 2} & \multicolumn{3}{c}{Model 3}
}
\startdata
\multicolumn{3}{c}{Single power law and edges as step functions}  & \multicolumn{3}{|c|}{One power law and edges as power laws} & \multicolumn{3}{c}{One power law and edges as Gaussians} \\[2pt]
\multicolumn{3}{c}{$ \Sigma(r)=  \Sigma_0
\left\{
\begin{array}{lr}
0  & \qquad  r < r_{min}, \\
 & \\
\left(\frac{r}{r_{min}} \right)^{\gamma_1} & \qquad  r_{min} < r < r_0, \\
 & \\
0 & \qquad  r > r_{max},
\end{array}
\right.$}  & \multicolumn{3}{|c|}{$ \Sigma(r)=  \Sigma_0
\left\{
\begin{array}{lr}
\left(\frac{r}{r_{1}} \right)^{\gamma_1}  & \qquad  r < r_1, \\
 & \\
\left(\frac{r}{r_{1}} \right)^{\gamma_2} & \qquad  r_1 < r < r_2, \\
 & \\
\left(\frac{r_2}{r_{1}} \right)^{\gamma_2} \left(\frac{r}{r_{2}} \right)^{\gamma_3} & \qquad  r > r _2  \\
\end{array}
\right.$} & \multicolumn{3}{c}{$ \Sigma(r)=  \Sigma_0
\left\{
\begin{array}{lr}
e^{ -\frac{1}{2} \left( \frac{r-r_1}{\sigma_{1}} \right)^2 }  & \qquad  r < r_1, \\
 & \\
\left(\frac{r}{r_{1}} \right)^{\gamma} & \qquad  r_1 < r < r_2, \\
 & \\
\left(\frac{r_2}{r_{1}} \right)^{\gamma} e^{ -\frac{1}{2} \left( \frac{r-r_2}{\sigma_{2}} \right)^2 } & \qquad  r > r _2  \\
\end{array}
\right.$} \\[2pt]
\hline
Parameter & Description & \multicolumn{1}{c|}{Value} & Parameter & Description & \multicolumn{1}{c|}{Value} & Parameter & Description & Value \\[2pt]
\hline
$r_{\min}$ & Inner Radius (au) & \multicolumn{1}{c|}{$135 \pm 4$} &  
$\gamma_1$ &  Power-law Index & \multicolumn{1}{c|}{$3.0_{-0.5}^{+0.9}$} & 
$r_{1}$ & Peak Radius Inner Edge (au)  & $172_{-40}^{+26}$  \\[2pt]
%%%%%%%%%%%%%%%%%%%%%%%%%%%%%%%%%%%%%%%%%%%%
$\gamma$ & Power-law Index & \multicolumn{1}{c|}{$-0.23 \pm 0.13$} & 
$r_{1}$ & Threshold Radius (au)  & \multicolumn{1}{c|}{$182_{-16}^{+11}$} & 
$\sigma_1$ & Standard Deviation Inner edge (au) & $37_{-21}^{+14}$  \\[2pt]
%%%%%%%%%%%%%%%%%%%%%%%%%%%%%%%%%%%%%%%%%%%%
$r_{\max}$ & Outer Radius  (au)& \multicolumn{1}{c|}{$360 \pm 6$} & 
$\gamma_2$ & Power-law Index &  \multicolumn{1}{c|}{$-0.6_{-0.3}^{+0.5}$} & 
$\gamma$ & Power-law Index  & $1.2_{-1.4}^{+0.9}$ \\[2pt]
%%%%%%%%%%%%%%%%%%%%%%%%%%%%%%%%%%%%%%%%%%%%
  &   & \multicolumn{1}{c|}{ }  & 
$r_{2}$ & Threshold Radius (au)  & \multicolumn{1}{c|}{$325_{-30}^{+18}$} & 
$r_{2}$ & Peak Radius Outer edge (au) &  $206_{-13}^{+19}$\\[2pt]
%%%%%%%%%%%%%%%%%%%%%%%%%%%%%%%%%%%%%%%%%%%%
 &  & \multicolumn{1}{c|}{ }  & 
$\gamma_3$ & Power-law Index  & \multicolumn{1}{c|}{$-5.6_{-2.1}^{+1.6}$} & 
$\sigma_2$ & Standard Deviation Outer Edge (au) & $118_{-15}^{+13}$\\[2pt]
\hline
\hline
\multicolumn{3}{c}{Goodness of fit} & \multicolumn{2}{c}{Model 1} & \multicolumn{2}{c}{Model 2} & \multicolumn{2}{c}{Model 3} \\[2pt]
\hline
\multicolumn{3}{c}{$\chi^2$} & \multicolumn{2}{c}{8818667.62} & \multicolumn{2}{c}{8818629.15} & \multicolumn{2}{c}{8818621.05} \\[2pt]
\multicolumn{3}{c}{Reduced $\chi^2$} & \multicolumn{2}{c}{1.12354} & \multicolumn{2}{c}{1.12354} & \multicolumn{2}{c}{1.12353} \\[2pt]
\multicolumn{3}{c}{$\Delta \chi^2$ (Model 3 taken as reference)} & \multicolumn{2}{c}{46.57} & \multicolumn{2}{c}{8.1} & \multicolumn{2}{c}{0} \\[2pt]
\multicolumn{3}{c}{$\exp{(\Delta \chi^2/2)}$} & \multicolumn{2}{c}{$1.3 \times 10^{10}$} & \multicolumn{2}{c}{57.4} & \multicolumn{2}{c}{1} \\[2pt]
\enddata
\end{deluxetable*}
\end{rotatetable*}
\end{figure}

\begin{deluxetable*}{ccc|cc|cc|cc}
\tablecaption{Comparison of our best fit models -- Other disk and background source  parameters. Note that aside from the flux of the disk, the uncertainties on the fluxes presented here stem solely from the MCMC process and do not take into account the 10\% flux calibration uncertainty of ALMA. Fluxes of our disk best fit models are as measured directly in the model images. Uncertainties are retrieved using the uncertainties on the disk mass -- the flux density is linearly proportional to the mass -- and a 10\% flux calibration uncertainty. Since the flux in Band 6 was scaled from the best fit in Band 7 using the spectral index, uncertainties on the flux in Band 6 includes as well uncertainties on the spectral index \label{tab:ring_mcmc_results}}
%\tablenum{5}
\tablewidth{0pt}
\tablehead{ 
%& & & \multicolumn{2}{c}{Model 1} & \multicolumn{2}{c}{Model 2} & \multicolumn{2}{c}{Model 3}
\multicolumn{9}{c}{Other Disk Characteristics}  
%\\[2pt]
}
\startdata
%\hline
Parameter & \multicolumn{2}{c}{Description} & \multicolumn{2}{c}{Value Model 1} & \multicolumn{2}{c}{Value Model 2} & \multicolumn{2}{c}{Value Model 3} \\[2pt]
\hline
$\mathrm{M_{dust}}$ & \multicolumn{2}{c}{Mass $\mathrm{(M_{\oplus})}$} &  
\multicolumn{2}{c}{$(9.0 \pm 0.5)\times10^{-2}$} & 
\multicolumn{2}{c}{$(1.3 \pm 0.1)\times10^{-1}$} & 
\multicolumn{2}{c}{$(1.3 \pm 0.1)\times10^{-1}$} \\[2pt]
%%%%%%%%%%%%%%%%%%%%%%%%%%%%%%%%%%%%%%%%%%%%
$i_d$ & \multicolumn{2}{c}{Inclination $(^\circ)$} & 
\multicolumn{2}{c}{$28.8_{-1.9}^{+1.8}$} & 
\multicolumn{2}{c}{$28.3_{-2.1}^{+1.9}$} &
 \multicolumn{2}{c}{$28.3_{-1.9}^{+1.8}$}   \\[2pt]
 %%%%%%%%%%%%%%%%%%%%%%%%%%%%%%%%%%%%%%%%%%%%
$\Omega_d$ & \multicolumn{2}{c}{Position Angle $(^\circ)$} & 
\multicolumn{2}{c}{$60.3_{-4.2}^{+4.0}$} & 
\multicolumn{2}{c}{$60.1_{-4.1}^{+4.5}$} & 
\multicolumn{2}{c}{$60.4 \pm 4.2$}   \\[2pt]
%%%%%%%%%%%%%%%%%%%%%%%%%%%%%%%%%%%%%%%%%%%%
$\alpha_\mathrm{mm}$ & \multicolumn{2}{c}{Spectral Index} & 
\multicolumn{2}{c}{$2.4 \pm 0.2$} & 
\multicolumn{2}{c}{$2.5 \pm 0.2$} & 
\multicolumn{2}{c}{$2.5 \pm 0.2$}  \\[2pt]
\hline
\hline
\multicolumn{9}{c}{Bright Source Characteristics} \\[2pt]
\hline
Parameter & \multicolumn{2}{c}{Description} & \multicolumn{2}{c}{Value Model 1} & \multicolumn{2}{c}{Value Model 2} & \multicolumn{2}{c}{Value Model 3} \\
\hline
$\mathrm{F}_{s,7}$ & \multicolumn{2}{c}{Flux in Band 7 ($\mu$Jy) } & 
\multicolumn{2}{c}{$320_{-18}^{+20}$}  &  
\multicolumn{2}{c}{$319_{-18}^{+20}$}  &  
\multicolumn{2}{c}{$316_{-19}^{+20}$}  \\[2pt]
%%%%%%%%%%%%%%%%%%%%%%%%%%%%%%%%%%%%%%%%%%%%
$\mathrm{F}_{s,6}$ & \multicolumn{2}{c}{Flux in Band 6 ($\mu$Jy) } & 
\multicolumn{2}{c}{$63_{-17}^{+18}$}  &  
\multicolumn{2}{c}{$60_{-19}^{+18}$}  & 
 \multicolumn{2}{c}{$58 \pm 18$}  \\[2pt]
 %%%%%%%%%%%%%%%%%%%%%%%%%%%%%%%%%%%%%%%%%%%%
$\mathrm{R}_{s,maj}$  &  \multicolumn{2}{c}{Gaussian Emission Major axis $(\arcsec)$} & 
\multicolumn{2}{c}{$0.23 \pm 0.04$}  & 
\multicolumn{2}{c}{$0.24 \pm 0.04$} & 
\multicolumn{2}{c}{$0.24 \pm 0.04$} \\[2pt]
%%%%%%%%%%%%%%%%%%%%%%%%%%%%%%%%%%%%%%%%%%%%
$\mathrm{R}_{s,min}$ &  \multicolumn{2}{c}{Gaussian Emission Minor axis $(\arcsec)$} & 
\multicolumn{2}{c}{$0.15_{-0.05}^{+0.04}$} & 
\multicolumn{2}{c}{$0.15_{-0.05}^{+0.04}$} & 
\multicolumn{2}{c}{$0.15_{-0.05}^{+0.04}$} \\[2pt]
%%%%%%%%%%%%%%%%%%%%%%%%%%%%%%%%%%%%%%%%%%%%
$\Omega_s$&  \multicolumn{2}{c}{Gaussian Emission Position Angle $(^\circ)$} & 
\multicolumn{2}{c}{$25_{-21}^{+19}$} & 
\multicolumn{2}{c}{$27_{-15}^{+24}$} & 
\multicolumn{2}{c}{$18_{-20}^{+21}$} \\[2pt]
%%%%%%%%%%%%%%%%%%%%%%%%%%%%%%%%%%%%%%%%%%%%
$\Delta\mathrm{RA}_{s}$  &  \multicolumn{2}{c}{Offset from the Star in Right Ascension $(\arcsec)$} & 
\multicolumn{2}{c}{$-1.26 \pm 0.05$} & 
\multicolumn{2}{c}{$-1.28 \pm 0.05$} & 
\multicolumn{2}{c}{$-1.28 \pm 0.05$} \\[2pt]
%%%%%%%%%%%%%%%%%%%%%%%%%%%%%%%%%%%%%%%%%%%%
$\Delta\mathrm{Dec}_{s}$ & \multicolumn{2}{c}{Offset from the Star in Declination $(\arcsec)$} & 
\multicolumn{2}{c}{$2.30  \pm 0.06$} & 
\multicolumn{2}{c}{$2.33  \pm 0.06$} & 
\multicolumn{2}{c}{$2.34_{-0.05}^{+0.06}$} \\[2pt]
\hline
\hline
\multicolumn{9}{c}{Central Emission Characteristics} \\[2pt]
\hline
Parameter & \multicolumn{2}{c}{Description} & \multicolumn{2}{c}{Value Model 1} & \multicolumn{2}{c}{Value Model 2} & \multicolumn{2}{c}{Value Model 3} \\
\hline
$\mathrm{F}_{c,7}$ & \multicolumn{2}{c}{Flux in Band 7 ($\mu$Jy) } & 
\multicolumn{2}{c}{$60.7_{-10.4}^{+10.3}$}  &  
\multicolumn{2}{c}{$71.6_{-10.7}^{+9.8}$}  &  
\multicolumn{2}{c}{$71.3_{-11.1}^{+10.2}$}  \\[2pt]
%%%%%%%%%%%%%%%%%%%%%%%%%%%%%%%%%%%%%%%%%%%%
$\mathrm{F}_{c,6}$ & \multicolumn{2}{c}{Flux in Band 6 ($\mu$Jy) } & 
\multicolumn{2}{c}{$31.1_{-14.3}^{+15.7}$}  &  
\multicolumn{2}{c}{$37.0_{-14.1}^{+14.6}$}  & 
 \multicolumn{2}{c}{$37.9_{-16.1}^{+16.0}$}  \\[2pt]
\hline
\hline
\multicolumn{9}{c}{Model Disk Fluxes (mJy)} \\[2pt]
%Need to be retrieved from images non multiplied by PB
%Numbers: for model 1, 5.6% uncertainty comes from that on dust mass. Adding 10% flux calibration gives a total of 11.5% in Band 7. Adding 8.3% of uncertainty stemming from uncertainty on alpha to Band 6, there's a total of 14.2% uncertainty on Band 6.
%for model 2, 7.7% uncertainty comes from that on dust mass. Adding 10% flux calibration gives a total of 12.6% in Band 7. Adding 8% of uncertainty stemming from uncertainty on alpha to Band 6, there's a total of 14.9% uncertainty on Band 6.
%for model 3, 7.7% uncertainty comes from that on dust mass. Adding 10% flux calibration gives a total of 12.6% in Band 7. Adding 8% of uncertainty stemming from uncertainty on alpha to Band 6, there's a total of 14.9% uncertainty on Band 6.
\hline
Parameter & \multicolumn{2}{c}{Description} & \multicolumn{2}{c}{Value Model 1} & \multicolumn{2}{c}{Value Model 2} & \multicolumn{2}{c}{Value Model 3} \\[2pt]
\hline 
%%%%%%%%%%%%%%%%%%%%%%%%%%%%%%%%%%%%%%%%%%%%
$\mathrm{F_{7}}$ & \multicolumn{2}{c}{Disk Flux in Band 7} &  
\multicolumn{2}{c}{$6.3 \pm 0.7$} & 
\multicolumn{2}{c}{$9.0 \pm 1.1$} & 
\multicolumn{2}{c}{$8.1 \pm 1.0$} \\[2pt]
 %%%%%%%%%%%%%%%%%%%%%%%%%%%%%%%%%%%%%%%%%%%%
$\mathrm{F_{6}}$ & \multicolumn{2}{c}{Disk Flux in. Band 6} & 
\multicolumn{2}{c}{$2.3 \pm 0.3$} & 
\multicolumn{2}{c}{$3.4 \pm 0.5$} & 
\multicolumn{2}{c}{$3.1 \pm 0.5$} \\[2pt]
\enddata
\end{deluxetable*}

%%%%%%%%%%%%%%%%%%%%%%%%%%%%%%%%%%%%%%%%%%%
%           RADIAL PROFILES
%%%%%%%%%%%%%%%%%%%%%%%%%%%%%%%%%%%%%%%%%%%

% \begin{figure}
% \makebox[\columnwidth]{
% \includegraphics[scale=0.6]{B7_Radial_profile_residuals}
% }
% \caption{Radial profile of the residuals found after subtracting each of our three models from the Band 7 observations (in the visibility plane).}
% \label{fig:residuals}
% \end{figure}

\begin{figure*}
\makebox[\textwidth]{
\includegraphics[scale=0.6]{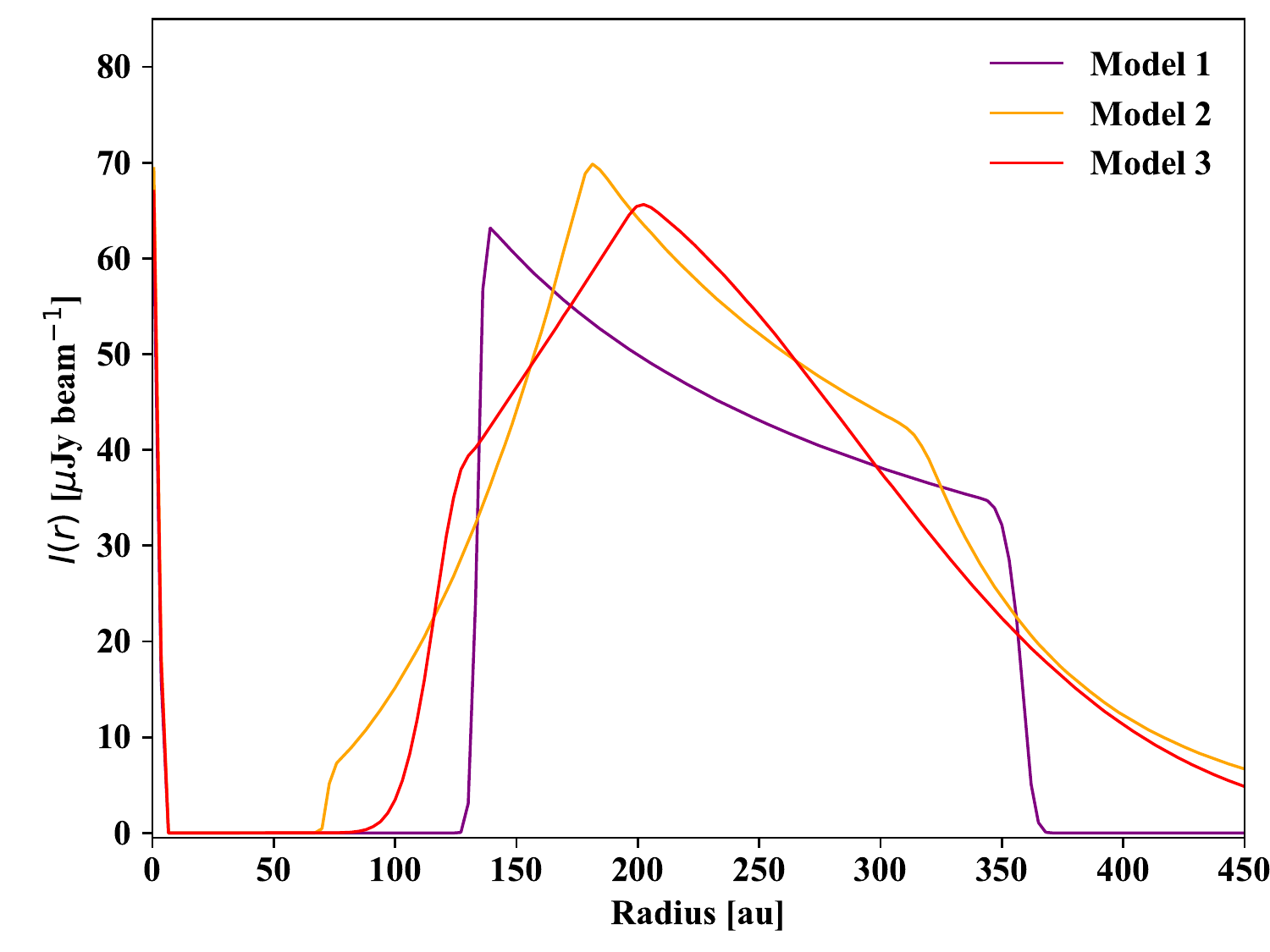}
\includegraphics[scale=0.6]{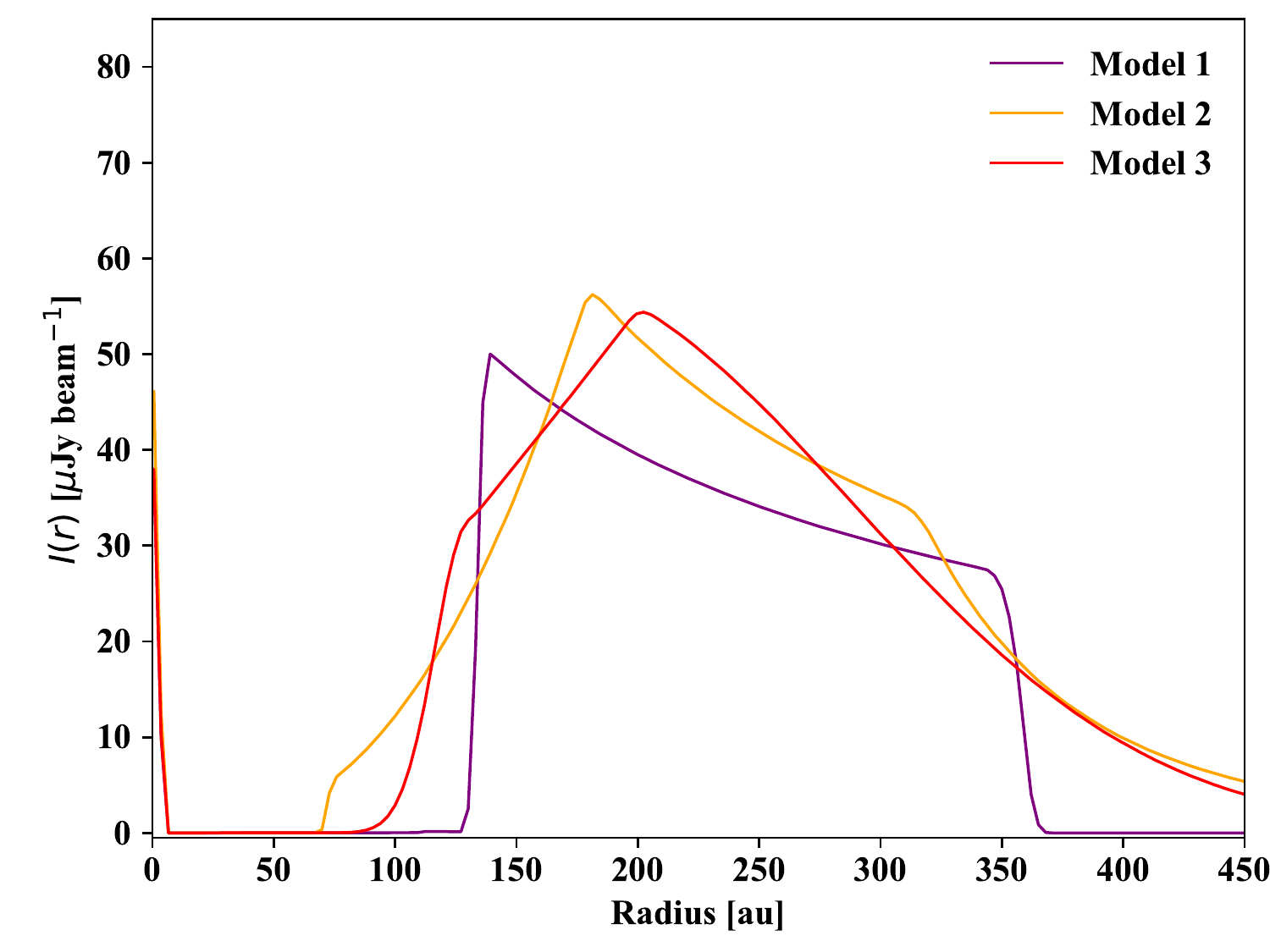}
}

\makebox[\textwidth]{
\includegraphics[scale=0.6]{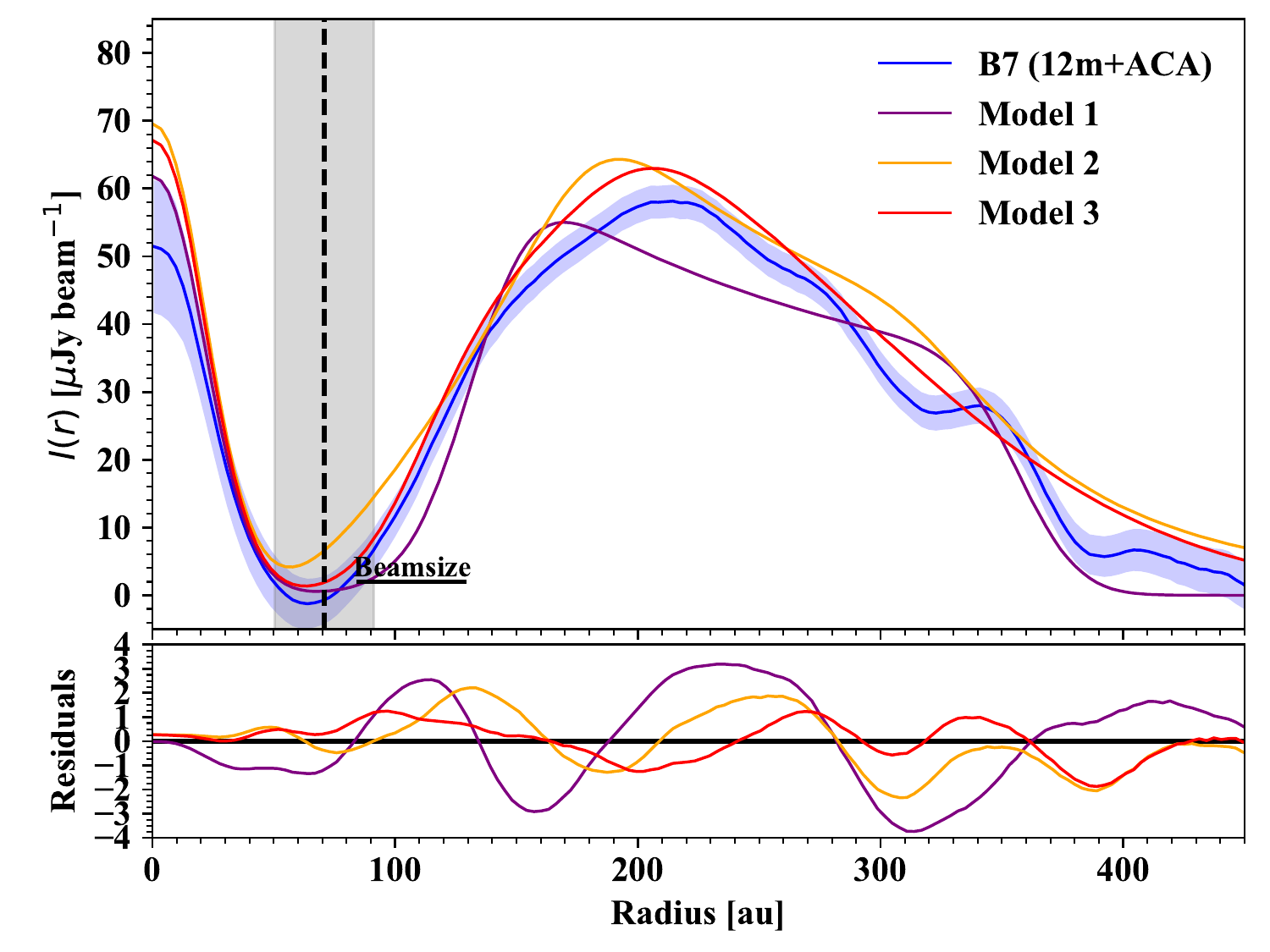}
\includegraphics[scale=0.6]{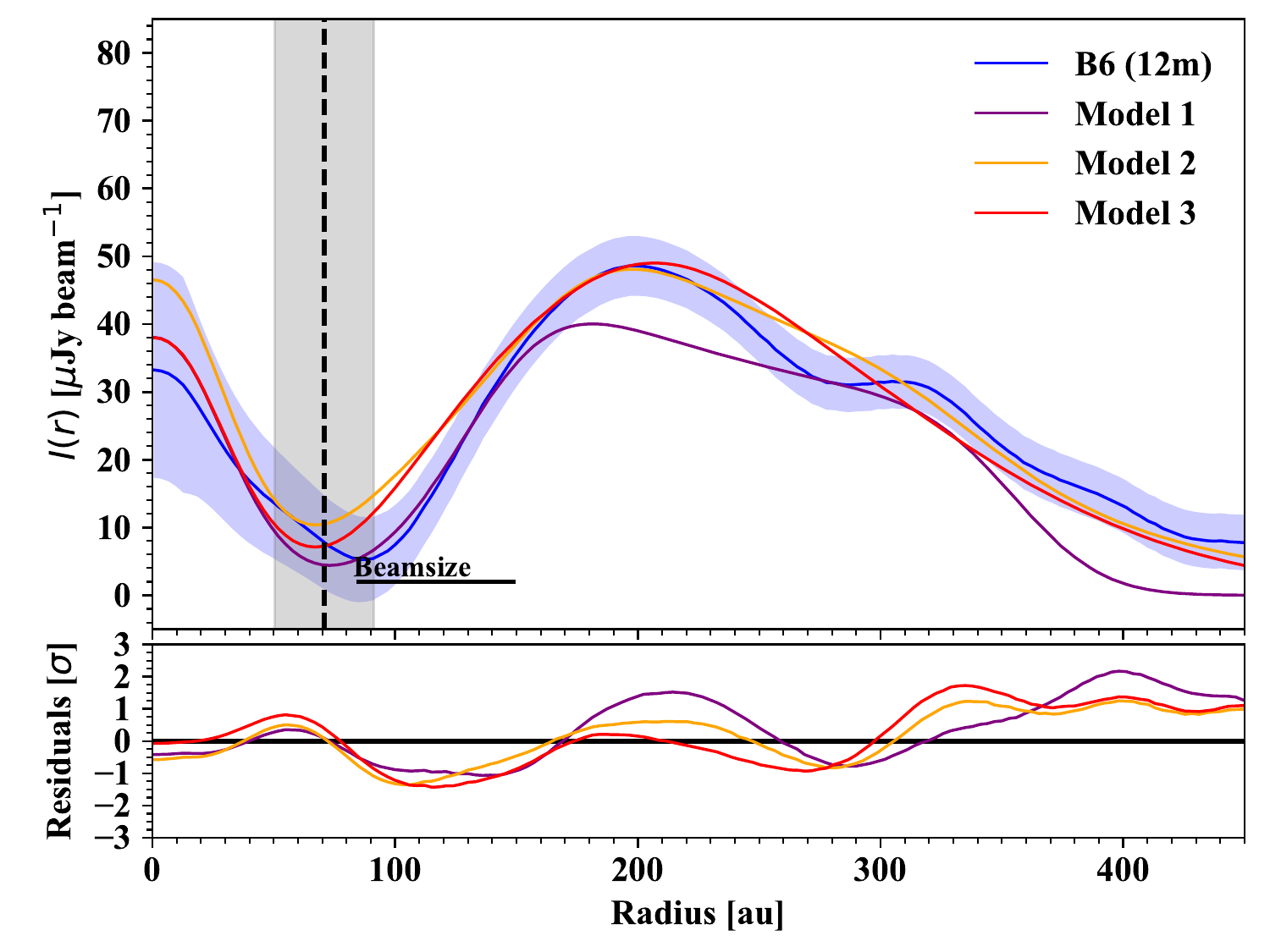}
}
\caption{\emph{Top:} Unconvolved surface brightness distributions of our three best-fit models, in Band 7 (\emph{left}), and Band 6 (\emph{right}). The central emission is included, while the bright source is excluded.
\emph{Middle:} The surface brightness distributions convolved with the beam for the three best-fit models along with the surface brightness distribution for the CLEAN image (shown in blue), for both Band 7 (\emph{left}) and Band 6 (\emph{right}). The background source has been subtracted from the images. The vertical dashed line corresponds to the semi-major axis of the outermost planet HR 8799 b as derived by \citet{WangJason2018}, while the grey region corresponds to its chaotic zone (as defined in Figure~\ref{fig:HR8799_continuum}). The lower panels display the radial profile of the residuals (in units of $\mu$Jy/beam) found after subtracting each of our three models from the observations (in the visibility plane). Note that the residuals are a direct comparison with the observations and not a comparison between the models and the CLEAN image, which is dependent on the CLEAN model. Therefore, the middle panels allow for a qualitative comparison of our best fit models with our observations, while the bottom panels - residuals - provide an accurate quantitative comparison.
}
\label{fig:surfaceDensities}
\end{figure*}
  
\section{Discussion} \label{sec:discussion}

\subsection{The disk geometry}

By observing the debris disk at higher resolution and sensitivity in Band 7 than in Band 6, our primary goal was to settle the dispute over the inner edge location stemming from the analysis of the Band 6 data. The fact that Band 7 observations are at smaller wavelengths than in Band 6 is not expected in itself to affect the location of this inner edge. Band 6 and Band 7 both trace large millimeter-sized dust grains which are not affected by the effect of stellar radiation; we thus can expect that the location of the inner edge is the same across both bands \citep[see, e.g., the debris disks of HD 107146 and HD 206893][]{Marino2018,Marino2020}.

As both the models of \citet{Booth2016} and \citet{Wilner2018} consisted of a single power law with step function edges, we can compare their results to those yielded by our Model 1, which uses the same parameters.  
Our Model 1 returns an inner edge location of $135 \pm 4\,$au. In Figure \ref{fig:surfaceDensities}, we display the location of planet HR 8799 b as well as the extent of its chaotic zone -- using the upper limit of $93_{-2}^{+3}\,$au derived in \citet{WangJason2018} -- hence showing the minimum radius at which the disk inner edge should be truncated if sculpted by planet b. The inner edge of Model 1 is located beyond this distance, which would tend to support the conclusions of \citet{Booth2016}. 

However, we also importantly find that this debris disk is not adequately represented by such a simple model, and in particular, that the inner edge is not sharp. As can be seen in the left panel of Figure \ref{fig:surfaceDensities}, the radial profile of the disk is much better fit by our Models 2 and 3, with an inner edge that is not abrupt as in Model 1, extending farther inwards towards planet b's chaotic zone. Therefore, we conclude that the concept of inner edge as used in both \citet{Booth2016} and \citet{Wilner2018}, i.e., a step-function, simply does not apply to this debris disk. In addition, Models 2 and 3 both show a peak emission, followed by a smooth outer tail with a change of slope that is required by model 2.

Our results support the theoretical findings of \citet{Gozdziewski2020}, who recently suggested that this debris disk could not be fit by a single power law, and should instead exhibit a radial peak emission. This would be expected in the presence of planetesimals on high eccentricity orbits, and is in accordance with the results of \citet{Geiler2019}, who, using a fully collisional model of the debris disk, find that a disk that can fit both \textit{Herschel} and ALMA Band 6 data needs to contain a such a high eccentricity component. Such a high eccentricity component, as pointed out by \cite{Marino2021}, is directly reflected in the smooth outer edge of this disk and would suggest typical eccentricities between $0.2-0.6$ near its outer edge.

This high eccentricity population could originate from scattering. As pointed out in \citet{Geiler2019}, this is unlikely to have occurred with planet b as it is now, because it is too massive and would lead to unbound orbits. The scattering of planetesimals should originate from a lower mass planet, either a younger planet b, or a low mass additional planet beyond planet b. On the other hand, the high eccentricity planetesimals seen in \citet{Gozdziewski2020} are rather of resonant origin. However, the dynamical modelling of \citet{Gozdziewski2020} consists in determining where and how bodies could orbit in a stable manner given the presence of the planets as they are today. This means that the maps of the disk they present correspond to where material would be in the ideal situation where the planets were entirely immersed in a disk of planetesimals at the end of the protoplanetary phase, i.e., it does not take into account what could have happened to this solid material during the protoplanetary phase, e.g., during gap opening or migration. Consequently, whether this accurately represents reality is debatable. 
Their work is nevertheless an excellent starting point to understand how a debris disk profile might look like when it possesses a high eccentricity component, which helps to interpret the profile we observe with our ALMA Band 7 data. 

In their Figure C2 (upper left panel), they show a snapshot of the disk at a given time and color code its eccentricity. One can distinguish an annulus of high eccentricity particles at 140-160 au. Since this is a snapshot, this annulus should correspond to where these eccentric particles have the most chances to be caught. In other words, this annulus logically represents the apastron of these eccentric particles, since this is the portion of their orbits they will spend the most time at. 

Since these particles at apastron are colocated with non-excited particles on circular orbits, then there should be a high velocity dispersion, with eccentric particles at apastron exhibiting a smaller velocity than circular particles. This is what happens at 140-160 au, and why this annulus of high eccentricity particles correspond to a high velocity dispersion annulus in the lower left panel. Higher density and higher velocity dispersion are both conducive to enhanced dust production. Hence \citet{Gozdziewski2020} conclude that the disk density profile should peak at 140-160 au.

Consequently, if the model of \citet{Gozdziewski2020} were entirely correct about the dynamical composition of the disk, the disk should be described by (from the inner most to the outermost parts): highly asymmetric features at the inner edge due to large clumps of particles in 1:1 and 3:2 mean-motion resonances with planet b, a smooth inner edge due to a low density associated with unstable particles in 2:1 mean-motion resonance with planet b, a radial peak emission at 140-160 au, and finally, a smooth tail.

We find no evidence in our observations for the large resonant clumps and highly asymmetric inner edge, however, we indeed find an inner edge smoother than a step function, a primary peak emission (though at $\sim 200\,$au), and a smooth tail. This does support a scenario in which there is a high eccentricity population on top of a low eccentricity one, but does not necessarily support a resonant origin for them. The smooth inner edge could also be due to the presence of an additional low mass planet beyond planet b, and again, not necessarily related to resonances.

%%%%%%%%%%%%%%%%%%%%%%%%%%%%%%%%%%

\subsection{The disk orientation}\label{sec:discussion_disk_orientation}

In preliminary MCMC fits (not presented here), we noticed that the position angle and inclination of the disk were correlated with the disk outer edge. Therefore, instead of arbitrarily truncating the disk outer edge, we chose to impose strong priors on the disk inclination and position angle, based on the best fit of these quantities for the planets orbits, assuming that the disk and the planets were coplanar.

Therefore, we further investigate the disk orientation by running an additional MCMC fit using our preferred model (Model 3), this time leaving the disk inclination and position angle unconstrained and with uniform priors. Aside from these parameters, there was no noticeable change in any of the other model parameters.

\begin{figure}
\makebox[\columnwidth]{
\includegraphics[scale=0.7]{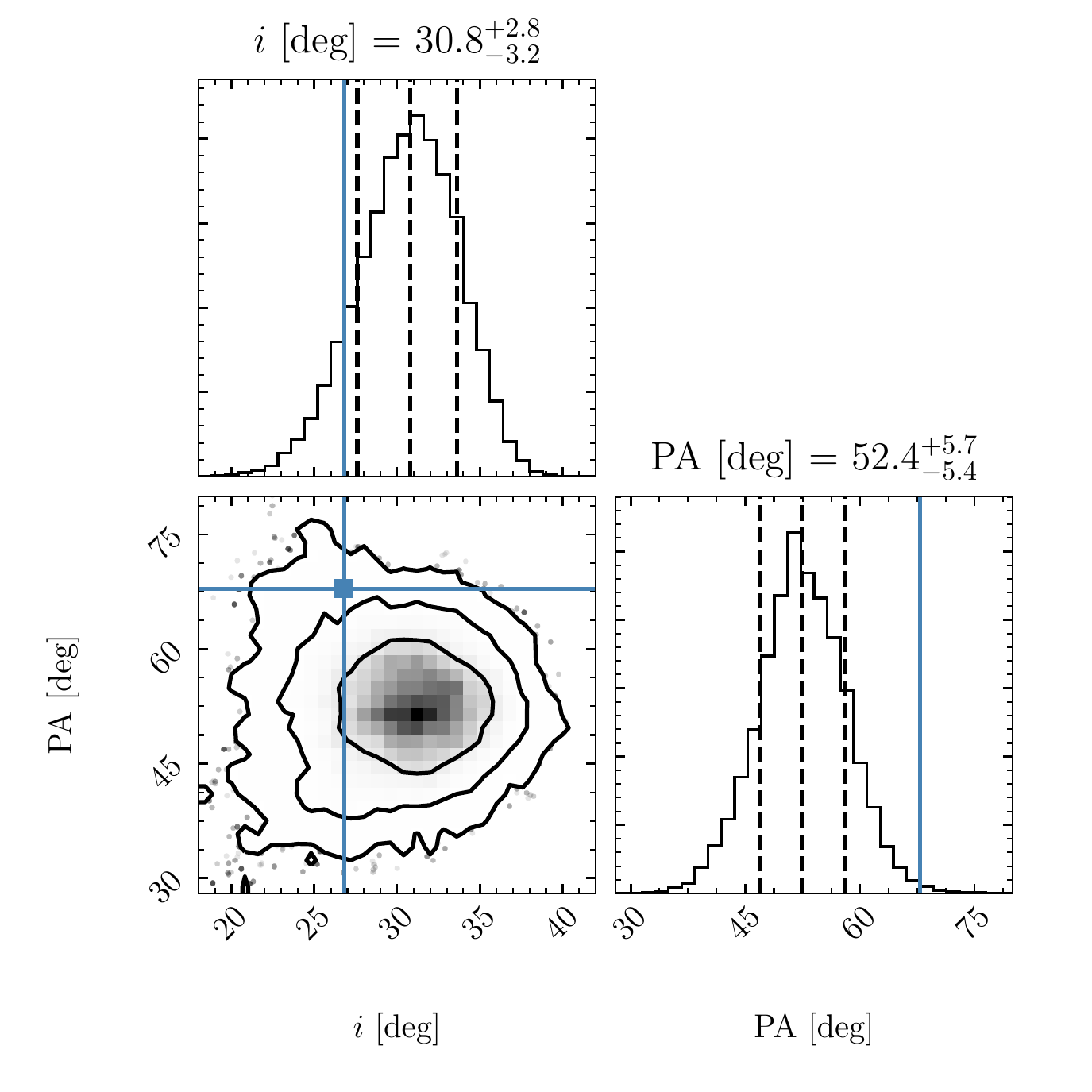}
}
\caption{Posterior distribution for the position angle and inclination of the disk when left as free parameters in our preferred model (Model 3; one power law with Gaussian edges). We display the marginalized probability distributions (top histograms) and the correlations between them (central panel), with 68, 95 and 99.7\% confidence levels represented as contours. The blue lines show the position angle and inclination of the planets found by \citet{WangJason2018} (stable coplanar solution; $i=26.8^{\circ} \pm 2.3; \Omega=67.9^{\circ\,+5.9}_{-5.2}$).
}
\label{fig:posterior_PA_Inc}
\end{figure}

When removing the prior on the disk inclination and position angle ($i=26.8^{\circ} \pm 2.3; \Omega=67.9^{\circ\,+5.9}_{-5.2}$), we recover a disk inclination of $30.8^{\circ\,+2.8}_{-3.2}$ and position angle of $52.4^{\circ\,+5.7}_{-5.4}$. These are comparable to the previously derived values (inclination $28.3^{\circ\,+1.8}_{-1.9}$ and position angle $60.4^{\circ}\pm 4.2$), and still marginally consistent with the best fit orientation of the planets found by \citet{WangJason2018} considering only stable coplanar solutions, which we used as a prior (see Figure \ref{fig:posterior_PA_Inc}).

%In contrast with our previous findings (inclination $28.3^{\circ\,+1.8}_{-1.9}$ and position angle $60.4^{\circ}\pm 4.2$), we now %find a best fit inclination and position angle of $30.8^{\circ\,+2.8}_{-3.2}$ and $52.4^{\circ\,+5.7}_{-5.4}$, respectively (see %Figure \ref{fig:posterior_PA_Inc})}. There is thus a $\sim\,2-3\sigma$ discrepancy between those values and the orientation of the %planets found by \citet{WangJason2018} (stable coplanar solution; $i=26.8^{\circ} \pm 2.3; \Omega=67.9^{\circ\,+5.9}_{-5.2}$).

We further translate these results into a mutual inclination with the orbital plane of the planets\footnote{Note that the mutual inclination is not simply the difference between the disk and planets' respective inclinations, $i_d$ and $i_p$. This statement holds only if they both have the same position angle, that is, if $\Omega_d=\Omega_p$. If the disk and planets have a different position angle, then this must be taken into account when determining the mutual inclination between their orbital planes. This inclination, $i_m$, is defined as: $\cos(i_m)=\cos(i_d)\cos(i_p)+\sin(i_d)\sin(i_p)\cos(\Omega_d-\Omega_p)$.}. Ideally, we would use pairs of disk inclination and position angle taken from our MCMC posterior distribution and compute the mutual inclination with pairs taken from the posterior distribution for the orbital plane of the planets, which are, unfortunately, not readily available. Therefore we use the values quoted in the literature for the best fit planet inclination and position angle, and approximate their posterior distribution by a Gaussian distributions.

\begin{figure}
\makebox[\columnwidth]{
\includegraphics[scale=0.7]{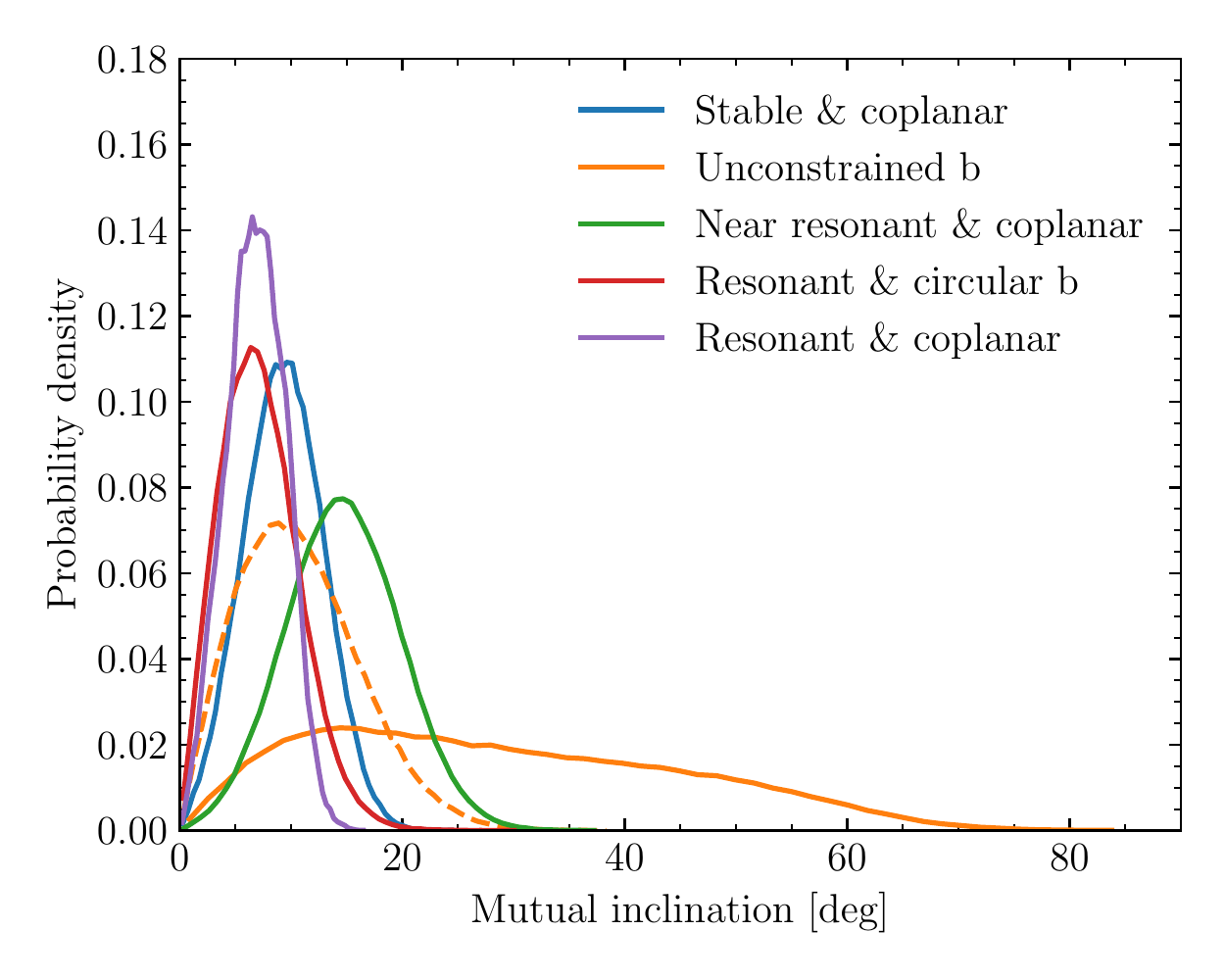}
}
\caption{Best fit mutual inclination between the debris disk and the orbital plane of the planets as found by previous best-fit to their astrometric positions. The mutual inclination is shown in the case the four planets were assumed to be coplanar, either i) on stable orbits \citep[in blue; Table 4 of][]{WangJason2018}, ii) in near resonant orbits \citep[in green; Table 3 of][]{WangJason2018}, and iii) in a Double Laplace resonant configuration \citep[in purple; Table 1 of][]{Gozdziewski2020}. We also display the mutual inclination with the orbital plane of planet b in the case the planets were not assumed to be coplanar; we show the cases where no constraint has been applied (in yellow, solid line for the findings of  \citet{WangJason2018} and dashed line for those of \citet{Wertz2017}), and finally, if planet b was assumed to be in resonance with planet c, and on a circular orbit \citep[in red; Table 7 of][]{Zurlo2016}. See Table~\ref{tab:mutual_inc} for a summary of the numerical values.
}
\label{fig:mutual_inclination}
\end{figure}

We display the resulting posterior distribution of the mutual inclination in Figure \ref{fig:mutual_inclination} (blue curve). We find a median mutual inclination of $9^{\circ}$, with a $3\sigma$ lower and upper limit of $1^{\circ}$ and $19^{\circ}$, respectively. This means that both orientations are roughly consistent with each other and there is no evidence of a misalignment. Given the large uncertainties of the mutual inclination, we can only exclude a misalignment larger than $19^{\circ}$.  Note that this result can heavily depend on  the priors imposed to constrain the orbital plane of the planets, e.g.\ assuming co-planarity, circular orbits or near-resonant configurations. As a matter of fact, and due to these short arc observations, it is nearly always necessary to impose constraints in the orbital fitting procedures. In particular, there is a degeneracy between coplanar and eccentric orbits, and mutually inclined and circular ones. Therefore, we carry out the same comparison with inclinations and position angles (longitude of ascending nodes) as found by other orbital fits.

%there is a 99.7\% chance of a misalignment greater than $1^{\circ}$.} \virginie{[Would appreciate anothr set of eyes on that sentence. Seba?]}

%Note that it does not necessarily mean that the planets' orbits and the disk are actually misaligned since strong priors are typically imposed to constrain the orbital plane of the planets, e.g.\ assuming co-planarity, circular orbits or near-resonant configurations. 

In the case of planets assumed to be coplanar, \citet{WangJason2018} also carried out an orbital fitting for planets in near resonant configuration, while \citet{Gozdziewski2020} investigated an exact Double Laplace resonance configuration. With the former, we find a median mutual inclination of $15^{\circ}$ with $3\sigma$ lower and upper limits $2^{\circ}$ and $29^{\circ}$, respectively (green curve in Figure \ref{fig:mutual_inclination}), while the latter is found mutually inclined with the disk by $7^{\circ}$, again with $3\sigma$ lower and upper limits $1^{\circ}$ and $14^{\circ}$, respectively (purple curve in Figure \ref{fig:mutual_inclination}). Similar to our previous conclusion, we find no strong evidence of a misalignment when considering these orbital solutions, although the data is still consistent with a misalignment smaller than $29^{\circ}$ and $14^{\circ}$, respectively.

%\textbf{With the former, we find a median mutual inclination of $15^{\circ}$ with $3\sigma$ lower and upper limits $2^{\circ}$ and $29^{\circ}$, respectively (green curve in Figure \ref{fig:mutual_inclination}), while the latter is found mutually inclined with the disk by $7^{\circ}$, again with $3\sigma$ lower and upper limits $1^{\circ}$ and $14^{\circ}$, respectively (purple curve in Figure \ref{fig:mutual_inclination}). }

In the case the planets are not assumed to be coplanar, we investigated mutual inclination with the best fit orbital plane of planet b, i.e., the planet closest to the debris disk.
We find a mutual inclination of $7^{\circ}$ -- with $3\sigma$ lower and upper limits $0.5^{\circ}$ and $19^{\circ}$, respectively -- in the case planet b is in Laplace resonance with planets c and d, as reported by \citet{Zurlo2016} (red curve in Figure \ref{fig:mutual_inclination}).

Finally, in the case the orbital fit were carried out without any underlying assumption, we compared again with the best fit for the orbit of planet b. We find a mutual inclination of $27^{\circ}$ with the fit of \citet{WangJason2018} (solid yellow curve in Figure \ref{fig:mutual_inclination}; $3\sigma$ lower and upper limits $1^{\circ}$ and $71^{\circ}$, respectively), and of $10^{\circ}$ with the fit of \citet{Wertz2017} (dashed yellow curve in Figure \ref{fig:mutual_inclination}; $3\sigma$ lower and upper limits $1^{\circ}$ and $28^{\circ}$, respectively). Note however, the following caveat: the posterior distribution of inclination and position angle for HR 8799 b in \citet{Wertz2017} is not well represented by a Gaussian. It is asymmetric and has wide wings. This means that using their 16th and 84th percentiles is not enough to represent their posterior, and thus we are likely missing the wide wings in their distributions. This explains why, though b's orbit is not well constrained, the distribution of mutual inclinations appears well peaked. In reality, it should be spread wider, similar to what we found with the unconstrained case of \citet{WangJason2018}.

We summarize these values in Table~\ref{tab:mutual_inc}, and overall find no strong evidence for misalignment, although a small misalignment of $\sim10^{\circ}$ cannot be excluded yet. In mature systems such as HR 8799, which age is much larger than the secular timescales involved, it is theoretically expected that disks and planets should be aligned as a result of secular interactions \citep[e.g.][]{Pearce2014}. However, if HR~8799's planets are mutually inclined, the disk could be forced to a misaligned configuration and even be warped \citep{Wyatt1999}. Note that there is a growing body of evidence of mutual misalignment ($\gtrsim4^\circ$) between debris disks and planets, both in the Solar System, and in extrasolar systems -- Beta Pic \citep{Mouillet1997}, HD113337 and HD38529 \citep{Xuan2020}, and HD106906 \citep{Nguyen2021}. Further observations that constrained better both the HR~8799 planets orbits and disk orientation could shed light on the degree of orbital alignment in this system. 

%\virginie{[Is this still something we can claim?]}}

%We only marginal evidence for misalignment. In mature systems such as HR 8799, which age is much larger than the secular timescales involved, it is theoretically expected that disks and planets should be aligned as a result of secular interactions. However, our result is in line with a growing body of evidence of mutual misalignment ($\gtrsim4^\circ$) between debris disks and planets, both in the Solar System, and in extrasolar systems -- Beta Pic \citep{Mouillet1997}, HD113337 and HD38529 \citep{Xuan2020}, and HD106906 \citep{Nguyen2021}. \virginie{[Is this still something we can claim?]}

\begin{deluxetable*}{c|ccc|c}[htbp] 
\tablecaption{Best fit mutual inclination between the debris disk and the orbital plane of the planets as found by previous fits to their astrometric positions. In the case the planets were not assumed to be coplanar, we used the orbit of the planet HR8799 b -- the closest to the cold debris disk -- as a reference. See Figure~\ref{fig:mutual_inclination} for the corresponding plots. \label{tab:mutual_inc}}
%\tablenum{2}
\tablewidth{0pt}
\tablehead{ \colhead{Case} &
\colhead{Median } &
\colhead{$-3\sigma$} &
\colhead{$+3\sigma$} & \colhead{Reference}\\
\colhead{} & \colhead{$(^\circ)$} & \colhead{$(^\circ)$} &
\colhead{$(^\circ)$} & \colhead{}
}
\startdata
Stable \& Coplanar & 9 & 1 & 19 & \citet{WangJason2018}  \\[2pt]
Near resonant \& Coplanar & 15 & 2 & 29 & \citet{WangJason2018}  \\[2pt]
Resonant \& Coplanar & 7  & 1 & 14 &  \citet{Gozdziewski2020}  \\[2pt]
Circular \& Resonant & 7 & 0.5 & 19 &  \citet{Zurlo2016}  \\[2pt]
Unconstrained & 27 & 1 & 71 & \citet{WangJason2018}  \\[2pt]
Unconstrained & 10  & 1 & 28 & \citet{Wertz2017}  \\[2pt]
\enddata
\end{deluxetable*}

% the mutual inclination 99.7% confidence lower limits, medians and 99.7% confidence upper limits are
% # Wang+2-18 table 4 [ 0.93876394 9.32501645 18.94044047]
% # Wang+2018 table 3 unconstrained b [ 1.44389553 26.73295899 71.35815997]
% # Wetz+2017 unconstrained b [ 0.68533266 10.21817471 28.01154489]
% # Wang+2018 table 3 near resonance and co-planar [ 2.00421006 14.62946222 28.92850298]
% # Zurlo b circular, resonant and non-coplanar [ 0.49689594 6.94883805 18.73509779]
% # Gozdziewski coplanar and resonant[ 0.73921585 6.89442862 13.69305833]

%%%%%%%%%%%%%%%%%%%%%%%%%%%%%%%%%%%%%%%%%%%%%%%%%%%%%%%%%%%%%%%

\subsection{The disk flux}

Measurements with JCMT/SCUBA-2 reported a disk total flux of $17.4 \pm 1.5$ mJy at 850 \micron\ \citep{Holland2017}. If we account for the 30 \micron\ of difference between JCMT/SCUBA (850 \micron) and ALMA Band 7 (880 \micron), we expect the disk in Band 7 to possess a flux of 16.0 mJy (using a $\lambda^{-2.0}$ flux dependency, as found in our MCMC fit). 

This is almost a factor of 2 higher than what we found (8.1 mJy in our Model 3). One factor that could cause the SCUBA-2 flux to be too high is the contribution of others sources of emission. \citet{Holland2017} measure the flux within a 60\arcsec aperture whereas we are considering the best fit model flux. The aperture will include flux from more than just the disk. It will also include flux from the central emission, the bright source and any other background emission. Our model fits the first two of these as free parameters and combined they contribute 0.35$\pm$0.03 mJy. This is clearly not enough to account for the difference. An additional source of background emission comes from the background molecular cloud that we detect in CO emission (see Appendix \ref{sec:app_gas}). This is clearly seen in continuum emission between 160 and 500 microns \citep{Matthews2014}, but its emission at longer wavelengths is harder to determine. There is no obvious sign of the cloud in the SCUBA-2 images \citep{Holland2017} but it may still contribute at a low level, whilst any contribution to the ALMA flux should be even lower since the larger scale of the emission should be filtered out due to the interferometric nature of the observations leading to a disparity between the reported total fluxes.

On the other hand, one factor that could cause the ALMA flux to be too low is that it is model dependent. This is necessary because one cannot directly measure the total flux of such an extended disk from interferometric observations. Nonetheless, it does mean that if our model is missing a crucial component, then we can underestimate the total flux. In particular, if there is disk emission on a scale larger than the maximum angular scale of the observations (12\arcsec{} for the 12m-array and 20\arcsec{} for the ACA) then this would not be detected in the ALMA observations but would be detected in the SCUBA-2 observations.

Further analysis that is beyond the scope of this paper is required to determine the relative contributions of disk and background flux. Here we simply conclude that the total disk flux density at 880 microns is between 8.1 and 15.6 mJy.

%SCUBA-2 peak flux is $10.9 \pm 1.0$ mJy

%%%%%%%%%%%%%%%%%%%%%%%%%%%%%%%%%%%%%%%%%%%%%%%%%%%%%%%%%%%%%%%

\subsection{The bright source}\label{sec:bright_source_discussion}
We detect a bright source to the northwest of the star, coincident with the disk. Whilst it is possible that this is emission coming from a clump of material in the disk, such bright sources are often seen at these long wavelengths and typically attributed to background galaxies \citep[e.g.][]{Su2017,Bayo2019,Faramaz2019}.
If it were a background sub-millimeter galaxy, then the easiest way to determine this conclusively is to see how the position relative to the star changes over time.
Given that HR~8799's proper motion is towards the southeast, then if the source is a background galaxy, the star is moving away from it and the galaxy will appear further and further out in the disk over time.

Whilst a faint source in the same position was noted as a significant residual by \citet{Booth2016}, the proper motion of HR 8799 is also quite small. Between the Band 6 epoch (2015) and Band 7 epoch (2018), HR 8799 has only moved about $0\farcs3$ in RA and $-0\farcs015$ in Declination. Since the the faint source was detected with a low S/N in 2015 its position was not well constrained. Our new observations constrained its relative position with respect to the star with a precision of 50~mas. This means that by 2021, this bright source would have moved by 0.36\arcsec, and thus similar band 7 observations could rule-out that it is co-moving with HR~8799 with a 5$\sigma$ significance.

Since our MCMC procedure returned the source's best fit flux in both Band 6 and Band 7, we were able to derive its millimeter spectral slope -- $3.98^{+0.96}_{-0.80}$. This is too steep to be consistent with the typical millimeter spectral index expected from debris disks \citep[see Table 4 of][]{MacGregor2016}, and is more consistent with typical values expected for extragalactic dust emission \citep[for example][find a typical spectral emissivity index of 1.6$\pm$0.4 for their sample of galaxies, which equates to a millimeter spectral index of 3.6$\pm$0.4]{Casey2012}
This bright source is thus likely to be a background submillimeter galaxy.

Finally, and according to millimeter-counts in Band 6 and using the Schechter function \citep{Carniani2015}, we would expect $\sim\,2$ background sources as bright or brighter than this source within the 12m Array Half Power Beam Width (HPBW) of diameter $25\farcs3$ in Band 6. Therefore, though it has not been reported before, its presence is however not surprising. 

\subsection{The central emission}
From our CLEAN band 7 image (Section \ref{sec:obs}), we found an integrated flux density at the location of the star of $73\,\pm\,13\,\mu\mathrm{Jy}$. In our parametric models, we assumed a central point source that was left as a free parameter. These resulted in flux densities of $61\,\pm\,10\,\mu\mathrm{Jy}$, $72\,\pm\,10\,\mu\mathrm{Jy}$ and $71^{+10}_{-11}\,\mu\mathrm{Jy}$ for models 1, 2 and 3 respectively (Table \ref{tab:ring_mcmc_results}). Despite the differences between these, all three models can be seen to be consistent with the observed central flux from the lower left plot of Figure \ref{fig:surfaceDensities}. This means that what we infer for the flux density at the location of the star is dependent on the model assumed for the disc (whether that is a parametric model or a CLEAN model). However, the differences are within the uncertainties and so for the following discussion we shall make use of the flux density inferred from the CLEAN image.

Fitting optical and near-IR photometry with a PHOENIX stellar photosphere model, G. Kennedy (private communication) finds an expected emission from the stellar photosphere of $36.0\,\pm\,3.6\,\mu$Jy at 880 microns \citep[see also Section 3 of ][for details on the method used]{Yelverton2019}, which is about half of our observed central flux. Subtracting the stellar photosphere from our observed flux, we measure an excess emission of $37\,\pm\,13\,\mu\mathrm{Jy}$ flux at 880 microns. 
%Best fit central flux in Band 7 for our Model 3: $71.3^{+10.2}_{-11.1}\,\mu\mathrm{Jy}$
%Observed 73 pm 13

One possibility is that this excess emission is due to the warm dust previously identified based on mid-IR spectroscopy \citep{Chen2006}. Indeed, the warm belt's emission is expected to be mingled with that of the stellar photosphere, as its dimensions prevent it from being resolved in our images; it is located interior to HR 8799 e, which orbits at 15 au = $0\farcs36$ from the star, inferior to the size of the beam in our observations. 
The expected flux for this warm dust can be extrapolated from the mid-IR detection to longer wavelengths. The slope for this extrapolation is quite uncertain, however. If the warm dust is fit with a 150 K blackbody \citep[e.g.\ Fig.~3 in][]{Su2009}, the expected flux at 880 microns is $\sim$200\,$\mu\mathrm{Jy}$, however this extrapolation does not account for the disk geometry, nor for how dust opacities decrease with wavelength for the expected grain size distributions, and thus their emission will fall off faster than a blackbody.  An additional $\lambda^{-1}$ drop-off (i.e.\ $\beta\,=\,1$) results in an expected flux an order of magnitude smaller, consistent with our observed value.

However, at these long wavelengths, the emission from the star may not just come from its photosphere but can also include significant contributions from the chromosphere and corona. So far, only a few stars have been studied in detail at these wavelengths and so it is not yet clear exactly how the relative contributions depend on the stellar properties. None the less, it is clear that there is a strong dependence on spectral type \citep[see e.g.][]{Liseau2015,White2019,White2020}. Of the few stars studied in detail at these wavelengths, $\gamma$ Vir A and $\gamma$ Vir B are the closest in spectral type to HR 8799 (both are F0IV) so provide the best example for what we might expect from HR 8799's emission. They have been studied by \citet{White2020} who find that they both have a band 7 ALMA flux density equivalent to about twice that predicted from photospheric models. If this is also true for HR 8799, then the central emission can be explained entirely by stellar emission and we do not have any significant detection of dust close to the star.

\section{Conclusion} \label{sec:conclusion}

With ALMA Band 7 (880 \micron), we have imaged the extended low-surface-brightness ring of debris orbiting HR 8799 at $\sim$100--400 au, -- still the largest and widest debris disk observed to date -- at an unprecedented combination of resolution and sensitivity.

Unlike previous sub-mm and far-IR images of the HR 8799 system, we have successfully detected the stellar emission in the center of the debris ring. This central source includes flux from both the star itself, as well as an excess that could be either due to the inner belt of warm dust at $\sim$10 au, already known from the mid-IR spectrum, or the star's chromosphere.

A bright background galaxy is detected and resolved at a separation of 2.6\arcsec\ from the star, within the broad disk and close to its inner edge. Based on its spectral slope, we conclude that the source is an extragalactic background submillimeter galaxy.

We find that the disk radial profile is more complex than a single power-law model, which was enough to describe the previous Band 6 observations that had a lower resolution and sensitivity.
In particular, we find that the inner edge is smoother than a step function, and is best reproduced by a Gaussian peaking at $\sim 170\,$au and with standard deviation $\sim 40\,$au.

As argued by \citet{Gozdziewski2020}, a low density at $\sim 100\,$au could be expected from material trapped in 2:1 mean-motion resonance with planet b. Nevertheless, a low mass planet orbiting outer to planet b could also produce a shallow gap and a smooth edge. The ideal way to test these hypotheses would be to use a combination of N-body simulations and radiative transfer as in \citet{Read2018} for comparison and best fit with our Band 7 radial profile.

The peak profile is also in accordance with the predictions of \citet{Gozdziewski2020}, i.e., the presence of a population on high eccentricity orbits should make the disk profile deviate from a simple single power law and have it exhibit a peak instead. This supports the results of \citet{Geiler2019}, who first suggested the cold debris disk of HR 8799 should include such an excited population in order to explain multi-wavelength data. This excited population, as shown by \cite{Marino2021}, could explain the smooth disc outer edge. The origin of such a population cannot however be pinned down. It could be due to scattering by an extra planet beyond planet b, or due to resonance trapping with planet b, which does not require the presence of an additional planet.

The question of whether there is an extra planet beyond planet b remains thus entirely open. There is the need for physically motivated models to compare with our observations, rather than just an ad hoc power-law models, and we advocate the use of our Band 7 observations for studies similar to that of \citet{Read2018}.

%%%%% Acknowledgements %%%%%%

\acknowledgments

We thank the anonymous referee for their helpful comments. 
This paper makes use of the following ALMA data: ADS/JAO.ALMA\#2016.1.00907.S and ADS/JAO.ALMA\#2012.1.00482.S.
ALMA is a partnership of ESO (representing its member states), NSF (USA) and NINS (Japan), together with NRC (Canada), MOST and ASIAA (Taiwan), and KASI (Republic of Korea), in cooperation with the Republic of Chile. 
The Joint ALMA Observatory is operated by ESO, AUI/NRAO and NAOJ.
The National Radio Astronomy Observatory is a facility of the National Science Foundation operated under cooperative agreement by Associated Universities, Inc.
VF's postdoctoral fellowship is supported by the Exoplanet Science Initiative at the Jet Propulsion Laboratory, California Institute of Technology, under a contract with the National Aeronautics and Space Administration (80NM0018D0004).
MB acknowledges support from the Deutsche Forschungsgemeinschaft (DFG) through project Kr 2164/13-2. %
JC acknowledges support by ANID, -- Millennium Science Initiative Program --
NCN19\_171. AZ acknowledges support from the FONDECYT Iniciaci\'on en investigaci\'on project number 11190837
We thank G. Kennedy for providing a stellar photosphere model of the star.
This work has made use of data from the European Space Agency (ESA) mission {\it Gaia} (\url{https://www.cosmos.esa.int/gaia}), processed by the {\it Gaia} Data Processing and Analysis Consortium (DPAC, \url{https://www.cosmos.esa.int/web/gaia/dpac/consortium}).
Funding for the DPAC has been provided by national institutions, in particular the institutions participating in the {\it Gaia} Multilateral Agreement.
EM acknowledges support from NASA award 17-K2GO6-0030.

\vspace{1cm}

Copyright 2021. All rights reserved.

%%%%% Appendix %%%%%%
\appendix

%%%%%%%%%%%%%%%%%%%%%%%%%%%%%%%%%%%%%%%%%%%%%%%%%%%%%%%%%%%%%%%%%%%%%%%%%%%%%%%%%%

%%%%%%%%%%%%%%%%%%%%%%%%%%%%%%%%%%%%%%%%%%%%%%%%%%%%%

\section{Stellar parameters} \label{sec:stellar}

Here we take advantage of the Gaia EDR3 astrometry -- released around the time this study was completed -- in order to update HR 8799's stellar parameters. 
We also use this new astrometry to investigate its group membership and search for possible stellar companions, which could help constrain the age of the system.
Our adopted stellar parameters for HR 8799 are summarized in Table \ref{tab:star}. 
We adopt several values from the detailed interferometric study of HR 8799 of \citet{Baines2012} and update them using the improved Gaia EDR3 astrometry \citep{GaiaEDR3}. 

\begin{deluxetable}{lll}[tbp]
\tablecaption{Adopted Stellar Parameters for HR\,8799\label{tab:star}}
%\tablenum{5}
\tablewidth{0pt}
\tablehead{
\colhead{Parameter} & \colhead{Value} & \colhead{Reference} 
}
\startdata
Spec. Type & F0VkA5mA5 $\lambda$\,Boo & \citet{Gray2014}\\
$\varpi$ (mas) & 24.4620\,$\pm$\,0.0455 & \citet{GaiaEDR3}\\
$D$ (pc) & 40.851\,$\pm$\,0.076 & 1\\
\ebv & 0.002 & \citet{Lallement2019}\\
$T_{\rm eff}$ (K) & 7193\,$\pm$\,87 & \citet{Baines2012}\\
$f_{\rm bol}$ (W\,m$^{-2}$) & 1.043\,$\pm$\,0.012 ($\times$10$^{-10}$) & \citet{Baines2012}\\
$\theta_{LD}$ ($\mu as$) & 342\,$\pm$\,8 & \citet{Baines2012}\\
$R$ ($R_{\odot}$) & 1.502\,$\pm$\,0.035 & 2\\
$m_{\rm bol}$ (mag) & 5.957\,$\pm$\,0.012 & 3\\
$M_{\rm bol}$ (mag) & 2.901\,$\pm$\,0.013 & 4\\
log(L/L$_{\odot}$) & 0.7356\,$\pm$\,0.0053 & 4\\
L (L$_{\odot}$) & 5.441\,$\pm$\,0.066 & 4\\
Mass ($M_{\odot}$) & 1.516$^{+0.038}_{-0.024}$ & \citet{Baines2012}\\
$\alpha_{ICRS}$ (deg) & 346.8696488373135 & \citet{GaiaEDR3}, 5\\
$\delta_{ICRS}$ (deg) & +21.1342529909730 & \citet{GaiaEDR3}, 5\\
$v_R$ (\kms)        & $-10.5^{+0.5}_{-0.6}$ & \citet{Ruffio2019}\\
$v_R^{LSRK}$ (\kms) & $-2.96^{+0.5}_{-0.6}$ & \citet{Ruffio2019}, 6\\
$\mu_{\alpha}$ (mas\,yr$^{-1}$) &  108.284\,$\pm$\,0.056 & \citet{GaiaEDR3}\\
$\mu_{\delta}$ (mas\,yr$^{-1}$) &  $-$50.040\,$\pm$\,0.059 & \citet{GaiaEDR3}\\
$U$ (\kms) & $-$12.81\,$\pm$\,0.15 & 7\\
$V$ (\kms) & $-$19.98\,$\pm$\,0.29 & 7\\
$W$ (\kms) &  $-$8.99\,$\pm$\,0.44 & 7\\
Age (Myr)  & 33$^{+7}_{-13}$ & \citet{Baines2012}\\
$X,Y,Z$ (pc) & $-$1.6, 33.2, $-$23.8 & 8\\
\hline
\enddata
\tablecomments{
1) calculated as $D$ = 1/($\varpi$ $-$ $-$0.017 mas) using Gaia EDR3 parallax $\varpi$ \citep{GaiaEDR3} with median quasar parallax $-$0.017\,mas from \citet{Lindegren2020}. Note that the Gaia EDR3 was not available at the time the modelling work of this paper was carried out, and that it hence made use of the distance derived from Gaia DR2. With the difference between the distances being 1.0\%, it has negligible impact on the results of our modelling.
2) Calculated using $\theta_{LD}$ from \citet{Baines2012} and corrected \citet{GaiaEDR3} distance (see $D$).
3) Using $f_{\rm bol}$ from \citet{Baines2012} on IAU 2015 $m_{bol}$ scale.
4) Combining $m_{\rm bol}$ and corrected Gaia EDR3 parallax (see $D$), on IAU 2015 systems with nominal solar values $M_{Bol,\odot}$ = 4.74 and 
$L^{N}_{\odot}$ = 3.828$\times$10$^{26}$\,W.
5) Gaia EDR3 ICRS position \citep{GaiaEDR3} adjusted to epoch J2000.0 by Vizier. 
6) $v_R^{LSRK}$ is the star's radial velocity in LSRK frame (Local Standard of Rest - Kinematic), for which ALMA adopts solar apex of B1900 18h +30$^{\circ}$.
7) Calculated using Gaia EDR3 astrometry \citep{GaiaEDR3} and $v_R$ from \citet{Ruffio2019} following \citet{ESA1997}. 
6) Barycentric Galactic Cartesian coordinates (pc) calculated using Gaia EDR3 astrometry \citep{GaiaEDR3} following \citet{ESA1997}.
}
\end{deluxetable}

\subsection{Kinematics and Group Membership} %%%%

In the original discovery paper for the first three HR 8799 planets, \citet{Marois2008} had noted the similarity in velocity between HR 8799 and the $\sim$30 Myr-old Columba and Carina associations.
\citet{Doyon2010} and \citet{Zuckerman2011} separately proposed that HR 8799 was a kinematic member of the Columba association.
\citet{Bell2015} estimated an updated isochronal age for Columba of 42$^{+6}_{-4}$ Myr, which was adopted as the HR 8799 system age by \citet{WangJason2018}.
Kinematic analysis by \citet{Hinz2010} calculated that HR 8799 did {\it not} appear to get much closer than about $\sim$40 pc from the centroids for either Columba or Carina over the past several tens of Myr, placing some doubt on their association with either group.
Recently, \citet{Lee2019} have claimed that HR 8799 is actually a high probability (87\%) member of the $\beta$ Pic Moving Group (BPMG).
Looking at the membership lists for nearby associations from \citet{Gagne2018}, HR 8799 appears to be remarkably isolated with respect to previously listed bona fide members of nearby young groups.
The nearest young association star in \citet{Gagne2018} to HR 8799 is 15$^{\circ}$ away (Wolf 1225 in AB Dor Group), and the nearest member of the $\beta$ Pic Group is 29$^{\circ}$.9 away (GJ 3076).
The nearest member of Columba -- to which HR 8799 has been purported to be a member -- is 32$^{\circ}$.7 away (HD 984).
So the current state of HR 8799's credentials for membership to any nearby young stellar groups which might provide useful age information is somewhat confusing.

We reexamine the kinematics of HR 8799 compared to the nearby young associations in light of the updated astrometry from Gaia.  
We use the BANYAN $\Sigma$ (Bayesian Analysis for Nearby Young AssociatioNs $\Sigma$) tool from \citet{Gagne2018} for estimating Bayesian kinematic membership probabilities to 27 nearby young associations within 150\,pc based on positions and velocities.
In Table \ref{tab:pmem} we list BANYAN $\Sigma$ membership probabilities to some nearby young stellar groups and the field, using astrometry from Gaia EDR3 \citep{GaiaEDR3}, Gaia DR2 \citep{GaiaCollaboration2018} and the revised Hipparcos reduction \citep{vanLeeuwen2007}, and radial velocity estimates from \citet{Ruffio2019}, \citet{WangJason2018}, and \citet{Gontcharov2006}. 
\citet{Gontcharov2006} and \citet{WangJason2018} provide two independent observed radial velocities for HR 8799 (\rv\, = $-$12.6\,$\pm$\,1.4 \kms\, and \rv\, = $-$10.9\,$\pm$\,0.5 \kms, respectively), and the most recent estimate provided by \citet{Ruffio2019} (\rv\,= $-$10.5$^{+0.5}_{-0.6}$ \kms) is a posterior accounting for these two previous estimates and accounting for the measured radial velocities of planets $b$ and $c$. 
Remarkably, when combining the exquisite astrometry provided by Gaia DR2 or EDR3 with the two most recent published radial velocities \citep[][as well as the top line of our Table \ref{tab:pmem}]{WangJason2018,Ruffio2019}, there does not seem to be a strong indication of membership to HR 8799 in {\it any} of the nearby young groups.
%\markb{[My understanding is that the BANYAN Sigma association models are based on DR1 astrometry. Could this be an issue that we should acknowledge in this paragraph?]}
%\eric{unlikely, and probably not worth mentioning. any offsets at ~ten of microarcsec level and given the proximity of these groups - the members mostly have parallaxes gt 10 mas or, zero pt differences = negligible}.

\begin{deluxetable}{llrrrr}[tbp]
\tablecaption{BANYAN $\Sigma$ Membership Probabilities for HR\,8799\label{tab:pmem}}
%\tablenum{5}
\tablewidth{0pt}
\tablehead{
\colhead{Ref(Astrom)} &
\colhead{Ref(RV)} & 
\colhead{P(Col)\%} & 
\colhead{P(BPMG)\%} & 
\colhead{P(Car)\%} & 
\colhead{P(Field)\%}
}
\startdata
{\bf G3} & {\bf R19} & {\bf 0.0} & {\bf 0.1} & {\bf 0.0} & {\bf 99.9}\\
\hline
G3 & W18 & 0.0 & 0.0 & 0.0 & 99.9\\
G3 & G06 & 62.2 & 0.0 & 0.0 & 37.7\\
G2 & R19 & 0.0 & 2.2 & 0.0 & 97.8\\
G2 & W18 & 0.0 & 0.3 & 0.0 & 99.7\\
G2 & G06 & 53.5 & 2.1 & 1.0 & 43.4\\
H2 & R19 & 0.0 & 1.2 & 0.0 & 98.8\\
H2 & W18 & 0.0 & 0.1 & 0.0 & 99.9\\
H2 & G06 & 62.9 & 0.5 & 0.0 & 36.6\\
\hline
\enddata
\tablecomments{
Membership probabilities for Columba, $\beta$ Pic Moving Group (BPMG), Carina, and "field" using BANYAN $\Sigma$ tool \citep{Gagne2018}. 
Probabilities for all other 26 groups within 150\,pc in BANYAN $\Sigma$ database are $<$0.1\%. 
References for astrometry and radial velocities (RV):
G3 = Gaia EDR3 \citep{GaiaEDR3},
G2 = Gaia DR2 \citep{GaiaCollaboration2018},
H2 = revised {\it Hipparcos} \citep{vanLeeuwen2007},
R19 = \citet{Ruffio2019} (\rv\, =\, -10.5$^{+0.5}_{-0.6}$\, \kms),
W18 = \citet{WangJason2018} (\rv\, =\, -10.9$\pm$0.5\, \kms),
G06 = \citet{Gontcharov2006} (\rv\, =\, -12.6$\pm$1.4\, \kms).}
\end{deluxetable}

We look closer at the relative velocities and positions of HR 8799 to the young associations compiled in \citet{Gagne2018} to see if there are any further clues to HR 8799's origin. 
Following \citet{Hinz2010}, we also calculate Galactic orbits for HR 8799 and some of associations in \citet[][using their velocities and centroids]{Gagne2018} using an epicycle approximation, adopting Oort constants from \citet{Bovy2016}, LSR velocity from \citet{Bland-Hawthorn2016}, and distance of the Sun above the Galactic plane from \citet{Karim2017}. 
We calculate 3D velocities and positions for HR 8799 using the \citet{GaiaEDR3} astrometry and the radial velocity from \citet{Ruffio2019} and list them in Table \ref{tab:star}. 
HR 8799's velocity with respect to Columba, Carina, and $\beta$ Pic groups are 3.7 \kms, 4.5 \kms, and 4.4 \kms, respectively.

\begin{figure}
\makebox[\textwidth]{
\includegraphics[scale=0.5]{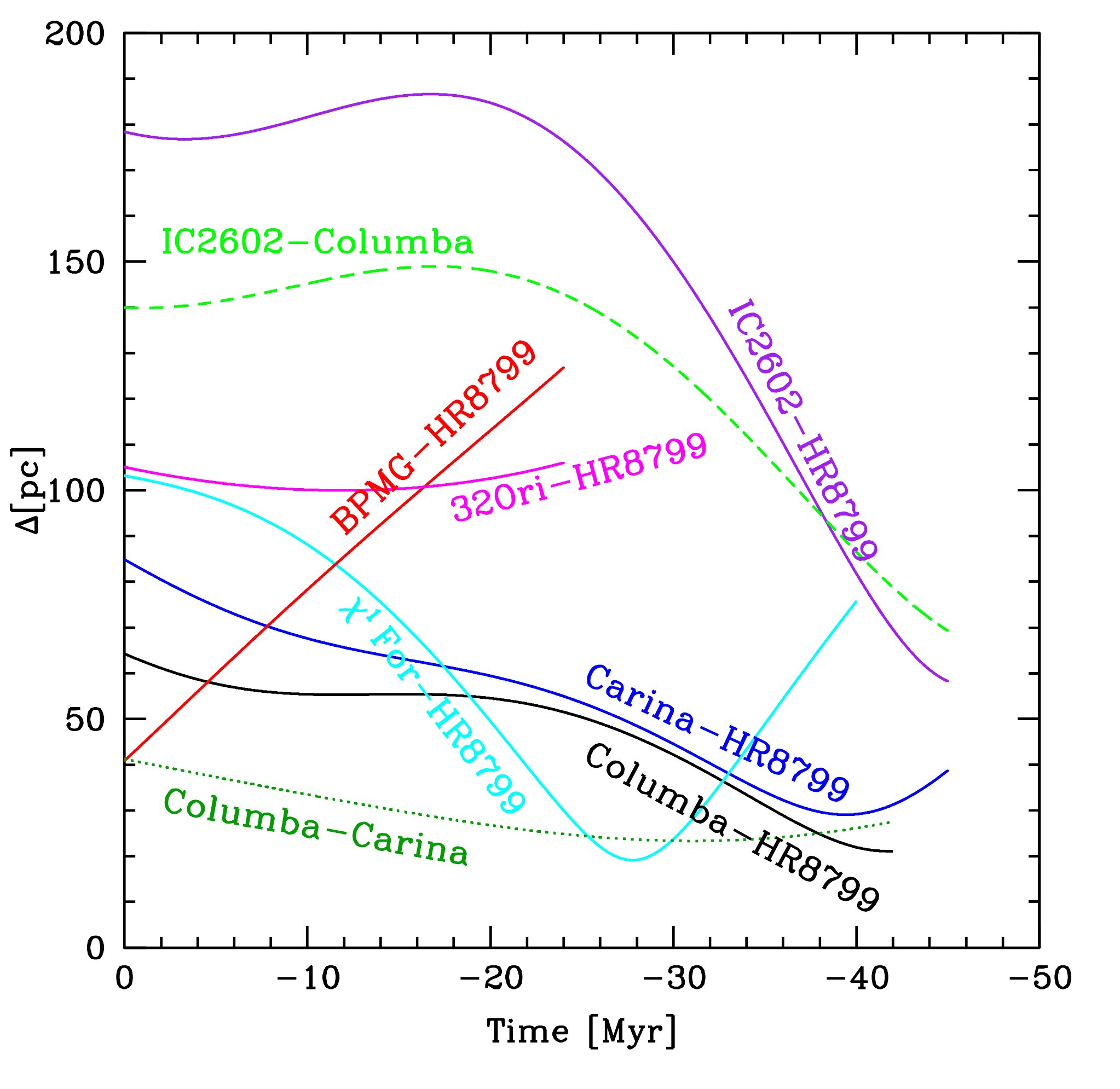}
}
\caption{Past separation between HR 8799 and several nearby young $\sim$20-45 Myr-old groups (solid lines) calculated using an epicyclic orbit approximation: the $\sim$40 Myr-old groups Columba (black), Carina (dark blue), $\chi^1$ Fornacis (cyan), and IC 2602 (purple), and the younger $\sim$20-25 Myr-old groups $\beta$ Pic Moving Group (BPMG; red) and 32 Ori Group (magenta).
Also plotted are the relative separations between Columba and Carina (dark green dotted) and Columba and IC 2602 (light green dashed). 
HR 8799 was within $\sim$30 pc of the centroids of the $\sim$40 Myr-old Carina and Columba associations $\sim$40 Myr ago, albeit with relative velocity $\sim$4-5\,\kms, and all of them may have been formed in the same molecular cloud complex (over $\sim$50-100 pc) that spawned the IC 2602 cluster.
}
\label{fig:close_young_groups}
\end{figure}

{\it Was HR 8799 much closer to these groups in the past?}
 In Figure~\ref{fig:close_young_groups} we plot the predicted separation between HR 8799 and several nearby young groups, as well as a mutual separations between some of the groups.
Running the orbits of HR 8799 and the Columba centroid back in time, we find that the star was closest to Columba 41.7 Myr ago at separation $\Delta$ = 21\,pc to the Columba centroid, with $\delta v$ = 3.9 \kms. 
The situation is similar, but a slightly worse match,
with the similarly aged Carina association \citep[45 $^{+11}_{-7}$ Myr;][]{Bell2015}\footnote{Although recent analyses have suggested a younger age for Carina \citep{Schneider2019,Booth2021b}. If correct, this younger age would rule out HR 8799 as a member.}.
HR 8799 is currently 85\,pc from the Carina centroid ($\Delta v$ = 4.5 \kms) 
and its closest passage to Carina was 39.4 Myr ago at separation $\Delta$ = 29\,pc with similar relative velocity ($\Delta v$ = 4.5 \kms). 
The story is quite different for HR 8799 with respect to the BPMG.
HR 8799 is currently $\Delta$ = 41\,pc from the BPMG centroid at present ($\Delta v$ = 4.4\,\kms), but was no closer to the BPMG centroid in the past (see Fig. 7). 
If one runs the clock back 20 Myr, representative of recent age estimates \footnote{Recent age estimates for BPMG bracket 18-24 Myr \citep{Mamajek2014, Bell2015, Binks2016, Shkolnik2017, Crundall2019, Miret-Roig2020}.}, one finds that HR 8799) was $\sim$113\,pc from the BPMG centroid at relative velocity $\Delta v$ = 5.9 \kms\, (i.e. not so different from its current situation, and certainly no closer to the core $\beta$ Pic membership). 
Given that the past trajectories of HR 8799 and BPMG are so clearly divergent, we find no support for the \citet{Lee2019} that HR 8799 could be a BPMG member.

Among the $<$150\,pc young groups in the \citet{Gagne2018} BANYAN $\Sigma$ database, HR 8799's velocity is within 5 \kms\, of several others groups as well (with their abbreviations and current velocity differences): 
118 Tau Group (``118TAU"; 0.9 \kms), 
32 Ori Cluster (``THOR"; 1.2 \kms), 
$\epsilon$ Cha Association (``EPSC"; 3.1 \kms), 
$\chi^{1}$ For Cluster (``XFOR"; 3.6 \kms),
TW Hya Association (``TWA"; 4.2 \kms), and
$\eta$ Cha Cluster (``ETAC"; 4.5 \kms). 
It is worth checking whether there is any 
past kinematic convergence with these groups as well, despite their low membership probabilities as assessed by BANYAN $\Sigma$. 
The 118 Tau, $\epsilon$ Cha, TW Hya, and $\eta$ Cha groups are all extremely young \citep[spanning $\sim$few-10 Myr;][]{Gagne2018}. 
HR 8799 has never been any closer than 100 pc to
either the $\epsilon$ Cha ($\sim$5 Myr) or $\eta$ Cha ($\sim$10 Myr) groups over the past 10 Myr, and
was $\sim$100 pc away from the TW Hya group (age $\sim$10 Myr) 10 Myr ago, so there is no evidence that the star was anywhere near these well-studied young groups on the periphery of the Sco-Cen complex. 
Similarly, HR 8799 has not been within 100 pc of the 118 Tau Group over the past 10 Myr. 
HR 8799's velocity agrees with that of the 32 Ori group ("THOR") within 1.2\,\kms, but remarkably it never appears to have been any closer than $\sim$100 pc (12 Myr ago) to the group's centroid since the group's birth $\sim$24 Myr ago \citep{Bell2017,Murphy2020} (Fig. 7). 
Although HR 8799 is currently $\sim$103 pc from the $\sim$40 Myr-old \citep{Zuckerman2011} $\chi^1$ For cluster ("XFOR" or Alessi 13), with current velocity difference 3.6\,\kms, it appears that the star was much closer to the cluster but with a larger velocity difference in the past. 
The closest pass we can find of HR 8799 to the $\chi^2$ For cluster was $\Delta$ = 19\,pc, 27.8\,Myr ago with velocity difference 6.0\,\kms.
We found also that while the nearby young cluster IC\,2602 has a velocity differing from that of HR\,8799 today by $\delta v$ = 9.6\,\kms\, (current separation $\Delta$ = 178 pc).
HR\,8799 was much closer to IC\,2602 at the time of the cluster's birth \citep[44\,$\pm$\,4 Myr;][]{Randich2018} - $\Delta$ = 61\,pc at $\delta v$ $\simeq$ 9.0\,\kms.
We conclude that a genetic tie between HR 8799 and most of these other nearby young stars can generally be ruled out, but we save $\chi^1$ For and IC 2602 clusters for further discussion.

We are left with a short list of nearby young groups (Columba, Carina, and perhaps $\chi^1$ For and IC 2602) which have very similar ages \citep[42$^{+6}_{-4}$ Myr,
45$^{+11}_{-7}$ Myr, $\sim$40 Myr, 44\,$\pm$\,4 Myr;][]{Bell2015,Zuckerman2011,Randich2018} and velocities, and which appear to have been within tens of pc of HR 8799 between 28 and 44 Myr ago, but with relative velocities at the 4-9\,\kms\, level (see Fig. ~\ref{fig:close_young_groups}).
{\it It is possible that HR 8799 and these young groups - Columba, Carina - and perhaps $\chi^1$ For and IC 2602 - all formed approximately contemporaneously in separate star-formation episodes within the same large-scale molecular cloud complex ($\sim$tens pc)?} 
Turbulence induces velocity dispersion in giant molecular cloud complexes at the $\sim$few km\,s$^{-1}$ level over length scales $\ell$ 
-- scaling roughly as $\sigma_v$ $\simeq$ 1 km\,s$^{-1}$($\ell$/pc)$^{0.4}$ \citep{Larson1981}.

%%%%%%%%%%%%%%%%%%%%%%%%%%%%%%%%%%%%%%%%%%%%%%%%%%%%%

Recent astrometric surveys of stellar associations with Gaia have confirmed that such velocity patterns are imprinted on subgroups within OB associations like Sco-Cen \citep[e.g.][]{Wright2020} and T associations like Taurus-Auriga \citep[e.g.][]{Galli2019}.
The example of Taurus-Auriga provided by \citet{Galli2019} is particularly striking, as they reveal approximately a dozen subgroups, each consisting of $\sim$4-24 stars each \citep{Joncour2018}, containing very few massive stars, in $\sim$pc-scale regions with intrinsic velocity dispersion of order $\sim$1\,\kms.
The subgroups have typical separations of $\sim$25\,pc, with typical velocity offsets of $\sim$1-5\,\kms.
The stars that formed in small subgroups with relative motions at the $<$few \kms\, level might be classified by future observers as members of the same stellar associations, but the ones that formed in groups with relative motions at the $\sim$few-10 \kms\, level might be classified as members of a separate groups (e.g. as we have with Columba, Carina, $\chi^1$ For, and IC 2602). 
The differences in 3D velocities between these young groups is similar in scale to the differences between the subgroups of the
$\sim$10$^2$\,pc-scale Sco-Cen complex ($\sim$3-5\,\kms) although the velocity dispersions within the individual subgroups is at the $\sim$1-2\,\kms\, level \citep{deBruijne1999,Wright2020}. 
We posit that HR 8799 formed either alone, or in a since-dispersed small cohort, within tens of pc of other stars forming in the same giant molecular cloud complex $\sim$40 Myr ago that spawned the Columba and Carina groups (and possibly including the $\chi^1$ For and IC 2602 clusters) but inheriting a velocity offset of $\sim$5\,\kms\, which has left it now somewhat isolated and $\sim$60-100\,pc away from the stars that formed elsewhere in the same cloud complex around the same time.

\subsection{Search for New Stellar Companions Using Gaia}

We also conducted a new search for co-moving companions to HR 8799 using astrometry from the Gaia EDR3 \citep{GaiaEDR3}, just in case any low-mass wide companions with useful age diagnostics could be found which could provide useful age constraints. 
Stellar companions may be loosely bound to a star with separations up to approximately the tidal (Jacobi) radius, which is defined by the mass of the star and local Oort parameters.
\citet{Jiang2010} estimate the tidal radius as: 

\begin{equation}
r_{\rm t}\,= \left\{ \frac{G(M_1 + M_2)}{4\Omega A} \right\}^{1/3}
\end{equation}

\noindent where G is the Newtonian constant, M$_1$ and
M$_2$ are stellar masses for the pair, 
$\Omega$ is the Galactic angular circular speed,
and $A$ is the Oort A parameter.
Adopting modern estimates for $\Omega$ and $A$ from
\citet{Li2019} that take advantage of the Gaia DR2 astrometry, we reparameterize the tidal radius equation as:

\begin{equation}
r_{\rm t}\,=\,1.357\,{\rm pc}\,\left\{ \frac{M_{\rm total}}{M_{\odot}} \right\}^{1/3}
\end{equation}

For prospective companion stars ranging in mass from negligible to 1 M$_{\odot}$, and adopting a mass of HR 8799 of 1.52 M$_{\odot}$
we estimate the relevant tidal radius to be 1.56 pc (for companion of negligible mass) to 1.85 pc (for 1\,M${\odot}$ companion). 
We start with a search for objects of common proper motion and parallax within approximately 1 tidal radius, for which we adopt an upper bound of 1.85 pc or 1$^{\circ}$.32 at the Gaia EDR3 distance for HR 8799 (40.85 pc). 
Over such a small region of the sky we ignore (for now) the convergent nature of the proper motions predicted for prospective objects that would share the 3D space velocity of HR 8799 and conduct a simple search for objects that would have tangential motions in $\alpha$ and $\delta$ within $\pm$2 \kms\, ($\pm$10.3 \masyr\, at $d$ = 40.85\,pc) of HR 8799 and parallaxes that would place the object within the tidal radius (for $\pm$1.85 pc we adopt generous parallax search limits of $\varpi$ = 24.4620 $\pm$ 1.16 mas).
% proper motion and parallax search:
% pmRA: 97.984 .. 118.584 (<pmRA> = 108.284)
% pmDec: -60.34 .. -39.74  (<pmdec> = -50.04)
% parallax: 23.302 .. 25.622 (<plx> = 24.462)
%
Using these criteria, there are {\it zero} Gaia EDR3 entries sharing proper motion and parallax with HR 8799.
We doubled the limits for selecting by search radius, proper motion, and distance limits, to see if any further interesting candidates could be found. 
% proper motion and parallax search:
% pmRA: 87.684 .. 128.884 (<pmRA> = 108.284)
% pmDec: -29.44 .. -70.64  (<pmdec> = -50.04)
% parallax: 22.039 .. 26.919 (<plx> = 24.462)
%
This yielded two candidates: StKM 1-2077 (Gaia EDR3 2835796794780262912) and 2MASS J23030459+2100553 (Gaia EDR3 2832642880035528192).   
StKM 1-2077 appears to be a K7 \citep{Gaidos2014}
dwarf ($M_G$ = 7.29) with faint companion ($\Delta G$ = 3.58; Gaia EDR3 2835796794780715392), likely a mid-M dwarf ($M_G$ = 10.87) resolved in Gaia DR2 and EDR3.
The calculated 3D velocity of the primary based on Gaia EDR3 astrometry ($U, V, W$ = -14.0, -24.7, -8.7 \kms) differ from HR 8799 by 4.9 \kms, and the star's rotation period 
\citep[$P_{rot}$ = 11.8554 day;][]{Oelkers2018}\footnote{Compare to Pleiades and Praesepe clusters in Fig. 12 of \citet{Curtis2020} for color ($G_{BP}$ - $G_{RP}$) = 1.617.} and modest
activity (log($L_{X}/L_{bol}$) = -4.02), suggest an age intermediate between the $\sim$120 Myr-old Pleiades and $\sim$670 Myr-old Praesepe clusters.
Regardless, the sizeable velocity offset (4.9 \kms)
and separation (3.17 pc from HR 8799) argue against a physical connection. 
2MASS J23030459+2100553 appears to be a late M dwarf 1$^{\circ}$ from HR 8799 whose tangential velocity is within 2.9\,$\pm$\,0.1\,\kms\, of HR 8799, and whose Gaia EDR3 position and parallax places it within 2.6\,pc separation.
Using its 2MASS and Gaia EDR3 photometry ($G$ = 18.26, $K_s$ = 12.81) and Gaia EDR3 parallax ($\varpi$ =  26.0828\, $\pm$\, 0.1971 mas), we estimate absolute magnitudes of $M_G$ = 15.34 and $M_{Ks}$ = 9.90, which are very similar to the main sequence primary M8V standard VB\,10 \citep[calculated as $M_G$ = 15.46 and $M_{Ks}$ = 9.90; using data from ][]{Cutri03,GaiaCollaboration2018}. 
Using the absolute $K_s$ magnitude-mass calibration of \citet{Mann2019}, this translates to a mass of only 0.086 $M_{\odot}$, and the theoretical isochrones of \citet{Baraffe2015} predict that such low-mass stars should take $\sim$2 Gyr to contract to the main sequence . 
This exceeds any previously quoted isochronal age estimates for HR 8799, as well as the main sequence lifetime for stars of its mass. 
Its relative velocity with respect to HR\,8799 (2.9\,\kms) is about 40$\times$ higher than the estimated escape velocity (0.073\,\kms), arguing against boundedness. 
Based on this dynamical argument, and considering that its older evolutionary state ($>$2 Gyr) conflicts with the MS nature of HR 8799, we exclude the late-M dwarf 2MASS J23030459+2100553 as a potential companion to HR 8799. 
Our search for wide companions of HR 8799 with Gaia EDR3 should have detected any objects with $G$ $<$ 20.0, or brighter than absolute magnitude $M_G$ $\simeq$ 16.9 at the distance of HR 8799 - comparable to that of the L1 standard 2MASS J14392836+1929149 \citep[$M_G$ = 16.79, $M_Ks$ = 10.79; mass = 0.076\,\Msun\, using calibration of ][]{Mann2019}, and near the hydrogen-burning limit \citep{Dieterich2014}. 
{\it We conclude that an exhaustive search of the Gaia EDR3 astrometric catalog finds no evidence for any co-distant, common proper motion companions down to about the hydrogen-burning limit to HR 8799 within at least two tidal radii}.
Besides its planetary system, HR\,8799 appears to remain a stellar single.

%%%%%%%%%%%%%%%%%%%%%%%%%%%%%%%%%%%%%%%%%%%%%%%%%%

\section{The gas content}\label{sec:app_gas}

CO emission consistent with the stellar location has been detected in previous JCMT \citep{Williams2006,Su2009}, SMA \citep{Hughes2011,Wilner2018} and ALMA \citep{Booth2016} observations. This CO extends well beyond the star and originates from the background cloud HLCG 92-35 \citep[which is directly behind HR8799,][]{Yamamoto2003}. Nonetheless, the high sensitivity of our observations relative to the previous observations may allow us to detect CO from the disk itself. In addition, the similarity between the radial velocity of the star and that of the CO led \citet{Su2009} to suggest that there may be some connection between HR 8799 and the background cloud.

\begin{figure*}
\makebox[\textwidth]{
\includegraphics[scale=0.35]{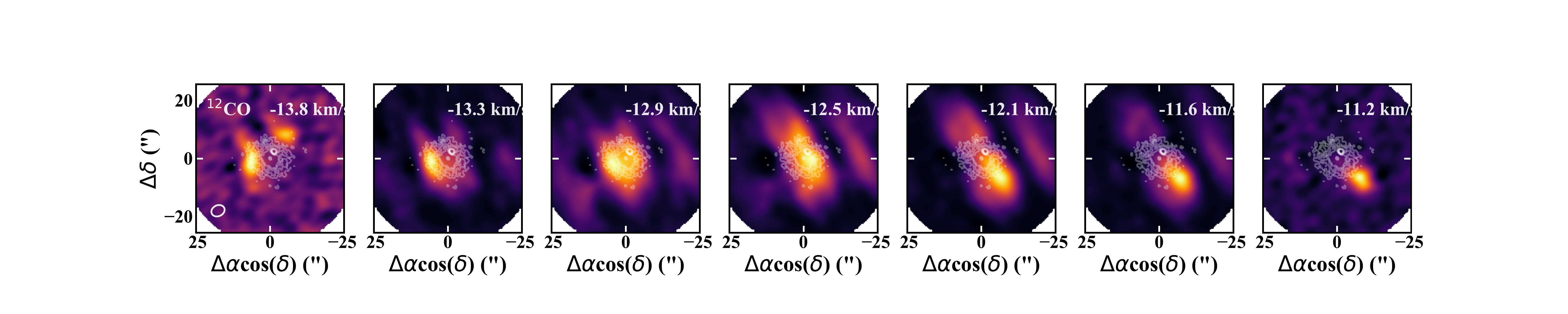}}
\makebox[\textwidth]{
\includegraphics[scale=0.35]{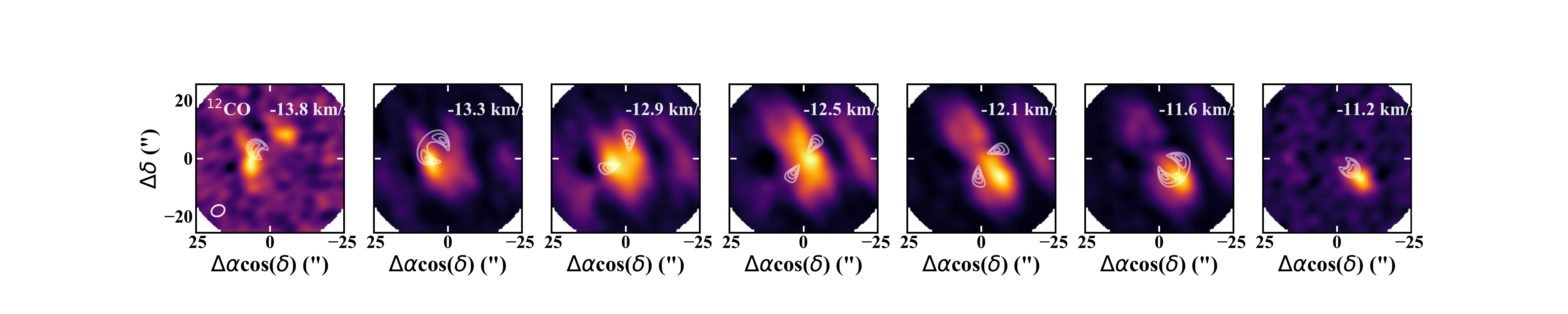}}
\caption{\emph{Top:} Channel maps with continuum overlaid show significant emission (perhaps an enhancement) along the line of sight to the continuum disk, as well as significantly extended emission (well beyond the continuum disk) out to the edges of the primary beam. These channel maps are not primary beam corrected, so note the enhancement at the disk location is not as pronounced as it appears in these maps. \emph{Bottom:} Channel maps with a toy model of Keplerian rotation (full resolution, not smoothed) overlaid show that the velocity pattern is clearly not consistent with Keplerian rotation expected from a disk with a CO radial surface density distribution similar to that of the HR 8799 disk. For the toy model, we assume a stellar velocity of $-$12.5 \kms, close to the centroid of the observed emission, but significantly (3.3$\sigma$) different from the stellar RV determined by \citet{Ruffio2019}.}
\label{fig:CO_maps}
\end{figure*}

Using the ACA data, we map the CO $J$=3-2 line emission, using the TCLEAN algorithm to create a full-spectral-resolution cube centered spatially on HR8799 and covering $-$40 to $+$15 \kms\, velocities in the barycentric frame. We used a natural weighting scheme and a pixel size of $0\farcs4$. The sensitivity of the resulting images is 3.8 mJy/beam for a 488.281 kHz (0.423 \kms) channel (native), and with a beam of size $4\farcs74 \times 3\farcs76$. We display these images in Figure~\ref{fig:CO_maps}, where we compare them to the debris disk continuum emission. 
Channel maps with continuum overlaid show significant emission (perhaps an enhancement) along the line of sight to the continuum disk, as well as significantly extended emission (well beyond the continuum disk) out to the edges of the primary beam. These channel maps are not primary beam corrected, so the enhancement at the disk location is not as pronounced as it appears in these maps. We can also see from these that the bright source does not appear in the CO maps. This is unsurprising as its point-like nature in the continuum emission means that it is most likely to be a background galaxy (as discussed in section \ref{sec:bright_source_discussion}) and so its CO emission will be red-shifted beyond the spectral range of our observations, whereas the diffuse CO comes from molecular clouds within our own galaxy.

\begin{figure}
\makebox[\columnwidth]{
\includegraphics[scale=0.35]{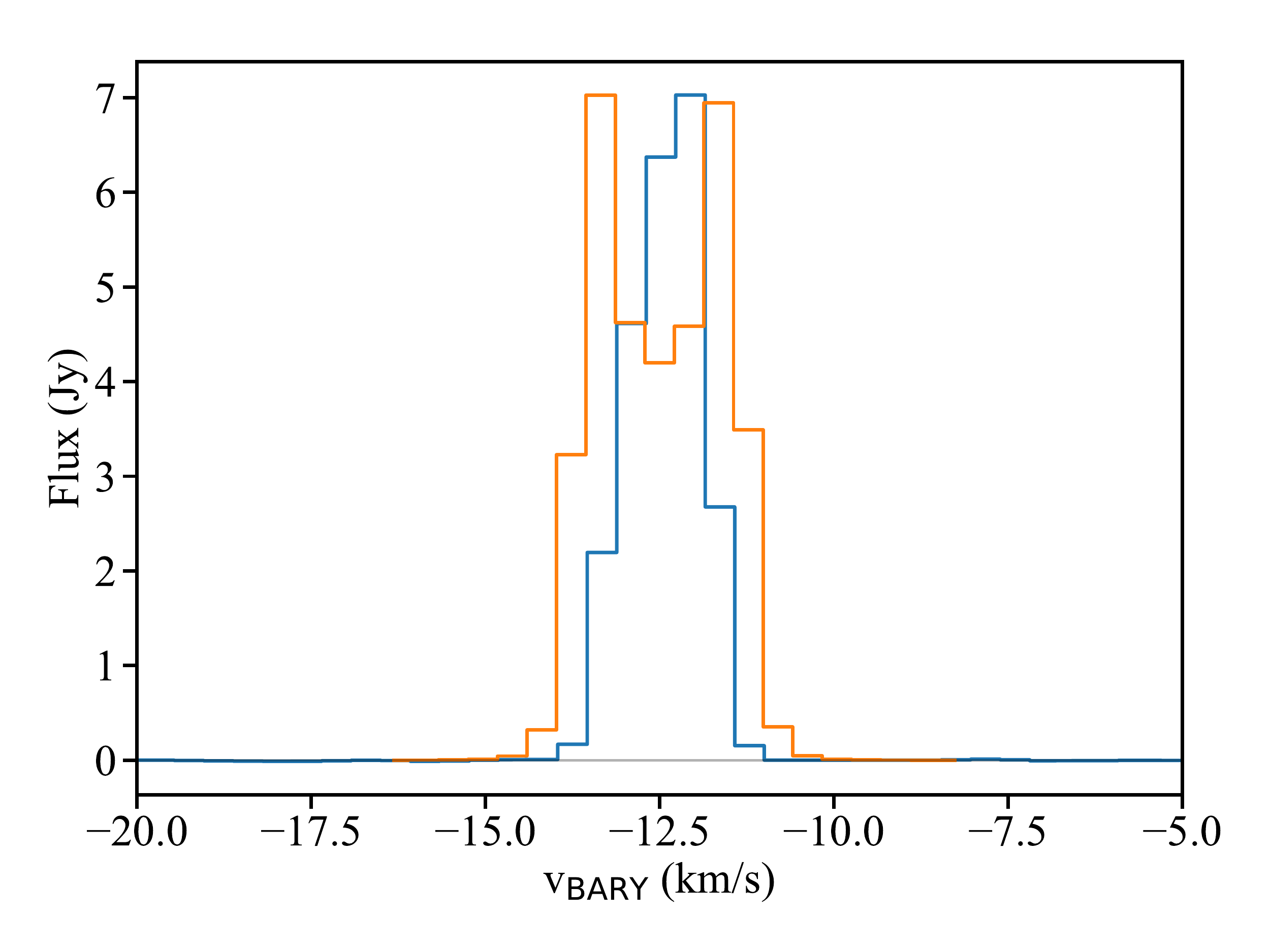}
}
\caption{\emph{Blue:} Observed CO emission line with a radial velocity centred on $\sim -12.5\,$\kms\, in the barycentric frame. \emph{Orange:} for comparison, the line profile expected from a toy model of a disk in Keplerian rotation and centred on the same radial velocity than the observed line. Note however that the radial velocity of the star is $-10.5^{+0.5}_{-0.6}\,$\kms \citep{Ruffio2019}.}
\label{fig:CO_line}
\end{figure}

In the lower row of Figure~\ref{fig:CO_maps}, we compare the channel maps with the emission expected from a toy model of a disk in Keplerian rotation (assumed inclination\,=\,33$^{\circ}$, stellar mass\,=\,1.56\,\msun, and mid-radius of 292\,au with Gaussian FWHM of 253\,au). For the toy model, we assumed a stellar velocity of $-$12.5\,\kms, close to the centroid of the observed emission. By spatially integrating within a 10\arcsec\ radius circular region (encompassing the full continuum disk), we extract the line profile shown in Figure~\ref{fig:CO_line}. The line profile of the emission shows only a single peak, clearly inconsistent with the distinct double peak seen in the toy model. In addition, although the radial velocity of the star has previously been reported as $-12.6\pm1.4\,$\kms \citep{Gontcharov2006}, matching that of the CO, \citet{Ruffio2019} has recently revised this, finding it to be $-10.5^{+0.5}_{-0.6}\,$\kms, clearly significantly different from the radial velocity of the gas.

We conclude that the morphology of the $^{12}$CO emission, line width, velocity shift with respect to the stellar velocity and brightness of the emission all confirm that the $^{12}$CO emission originates from the background cloud HLCG 92-35. Furthermore, the \emph{Gaia} survey has made it possible to refine the measurement of distances to molecular clouds by estimating interstellar extinction. Both \citet{Yan2019} and \citet{Zucker2019} have estimated the distance to the cloud MBM 54, which is in the same cloud complex and just a few degrees South of cloud HLCG 92-35, finding it to be $257^{+8}_{-7}\,$pc and $245\pm13\,$pc respectively, placing it far beyond HR 8799 and conclusively showing that the star is not associated with the background cloud.

%%%%%%%%%%%%%%%%%%%%%%%%%%%%%%%%%%%%%%%%%%%%%%%%%

\section{Cornerplots}\label{sec:app_cornerplot}

\begin{figure*}
\makebox[\textwidth]{
\includegraphics[scale=0.6]{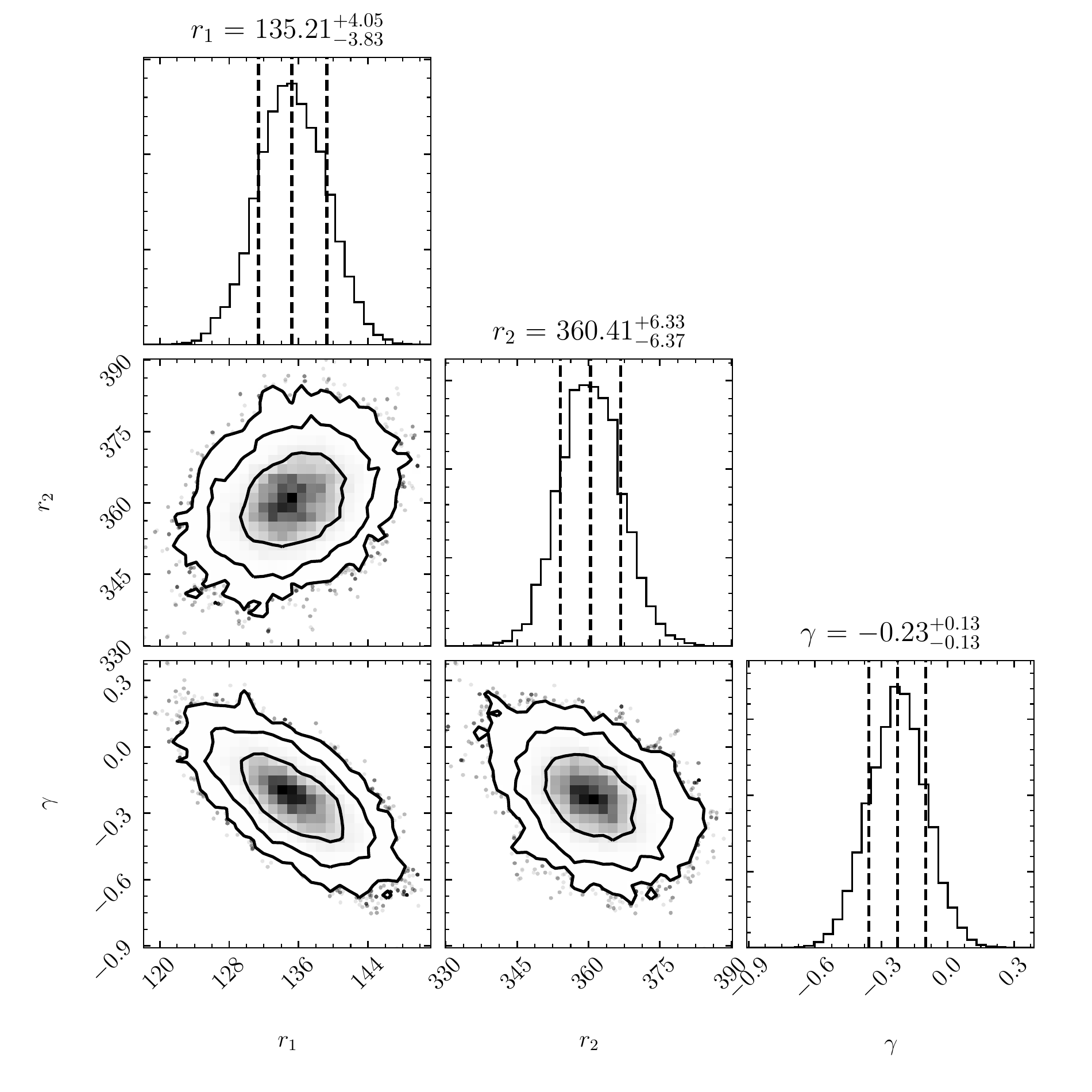}
}
\caption{Best-fit parameters for Model 1 (single power law with step-function edges) showing the marginalized probability distributions  for the three model parameters (top histograms) and  the correlations between pairs of parameters (central panels), with 68, 95 and 99.7\% confidence levels represented as contours.}
\label{fig:cornerplot_1plaw}
\end{figure*}

\begin{figure*}
\makebox[\textwidth]{
\includegraphics[scale=0.6]{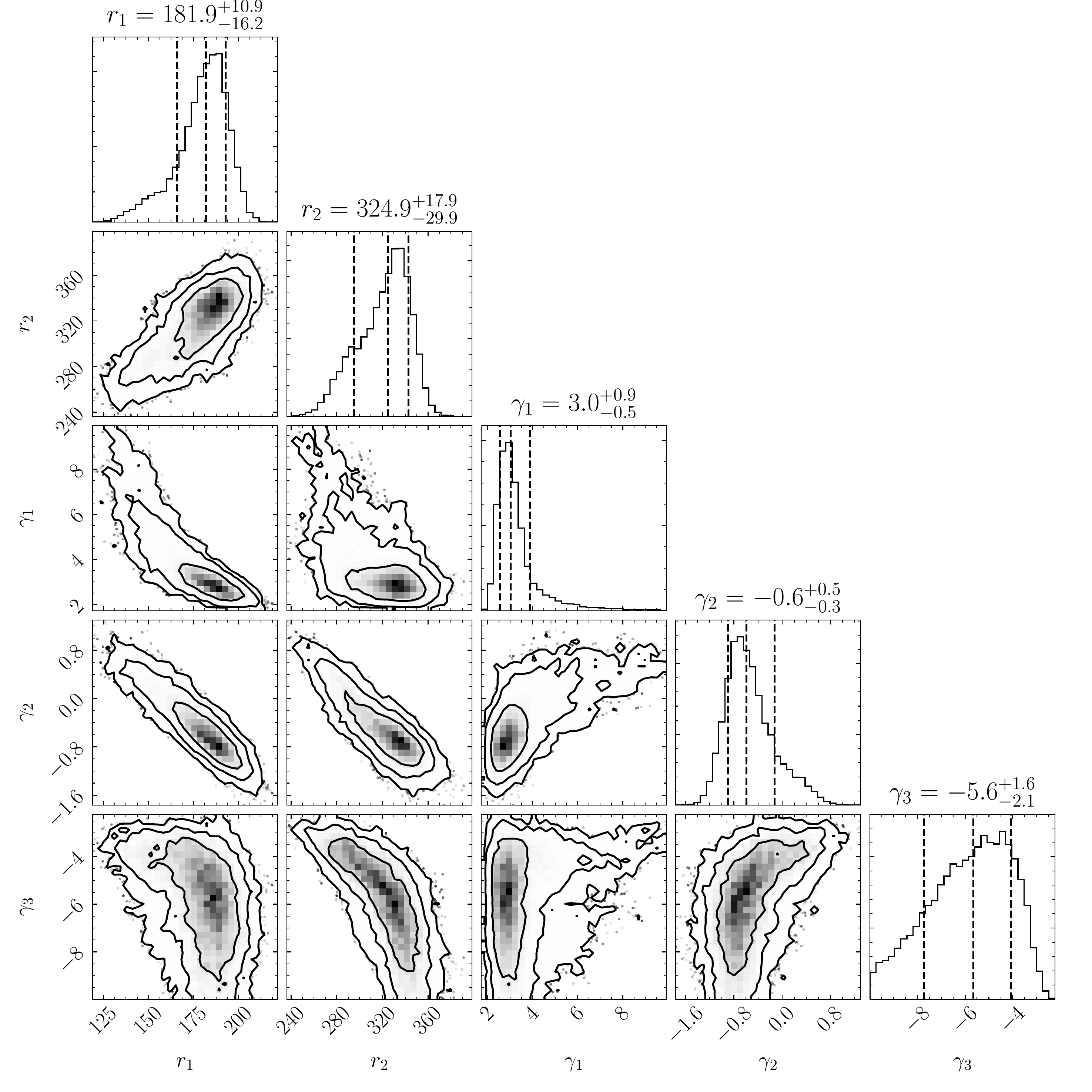}
}
\caption{Best-fit parameters for Model 2 (three power laws) showing the marginalized probability distributions for the five model parameters (top histograms) and the correlations between pairs of parameters (central panels), with 68, 95 and 99.7\% confidence levels represented as contours.}
\label{fig:cornerplot_3plaw}
\end{figure*}

\begin{figure*}
\makebox[\textwidth]{
\includegraphics[scale=0.6]{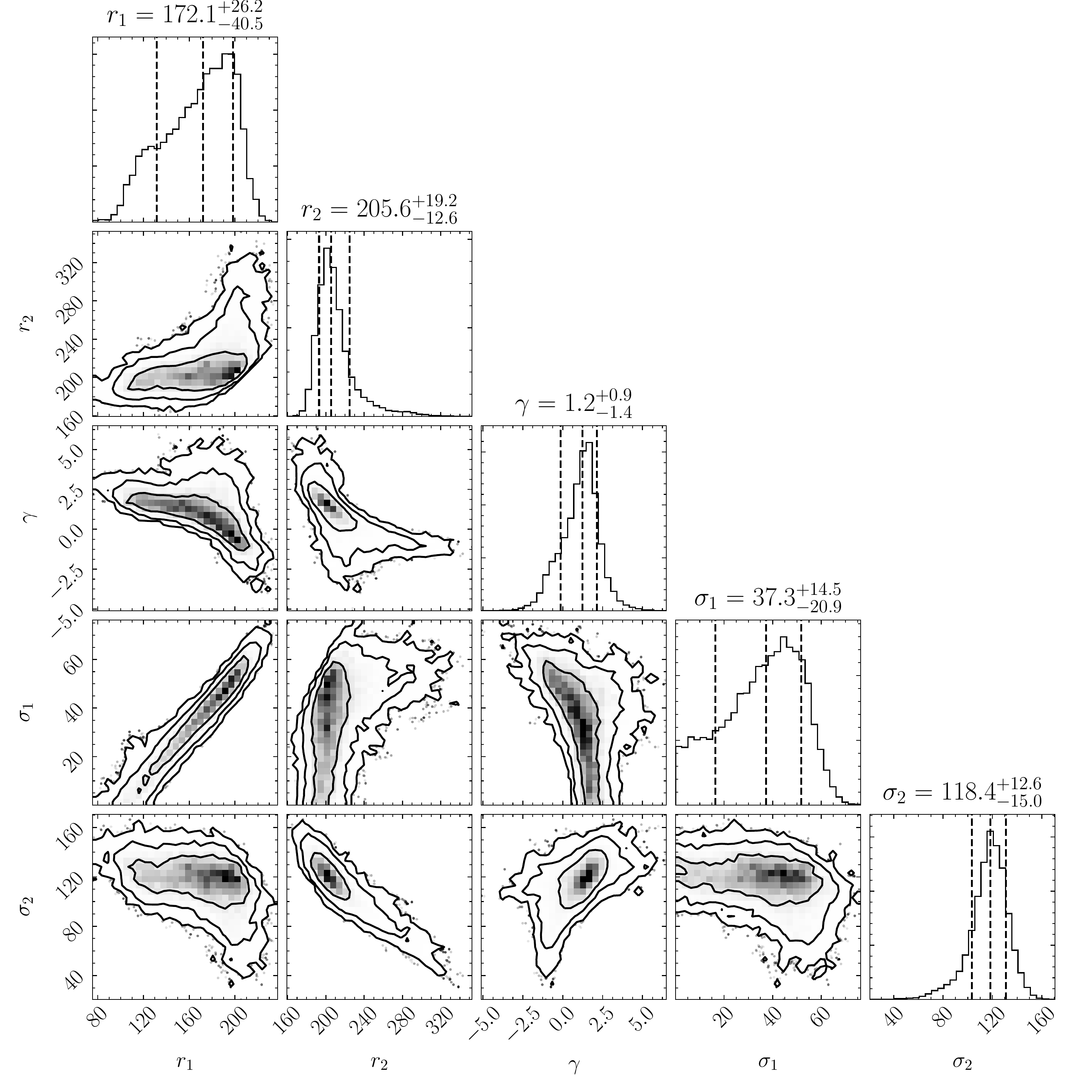}
}
\caption{Best-fit parameters for Model 3 (one power law with Gaussian edges) showing the marginalized probability distributions for the five model parameters (top histograms) and the correlations between pairs of parameters (central panels), with 68, 95 and 99.7\% confidence levels represented as contours.}
\label{fig:cornerplot_1plaw_gaussian}
\end{figure*}

Figures \ref{fig:cornerplot_1plaw}, \ref{fig:cornerplot_3plaw}, and \ref{fig:cornerplot_1plaw_gaussian} show the correlations between the disk surface density parameters for the single power-law, triple power-law, and Gaussian-edged models, respectively.
The full posterior distributions (22x22 or 24x24 cornerplots) are available upon request to the corresponding author.

%%%%%%%%%%%%%%%%%%%%%%%%%%%%%%%%%%%%%%%%%%%
%               CORNER PLOTS
%%%%%%%%%%%%%%%%%%%%%%%%%%%%%%%%%%%%%%%%%%%

%%%%%%%%%%%%%%%%%%%%%%%%%%%%%%%%%%%%%%%%%%%%%%%%%%%%%

\section{Residual Maps}\label{sec:app_residuals}

\begin{figure*}
\makebox[\textwidth]{
\includegraphics[scale=0.25]{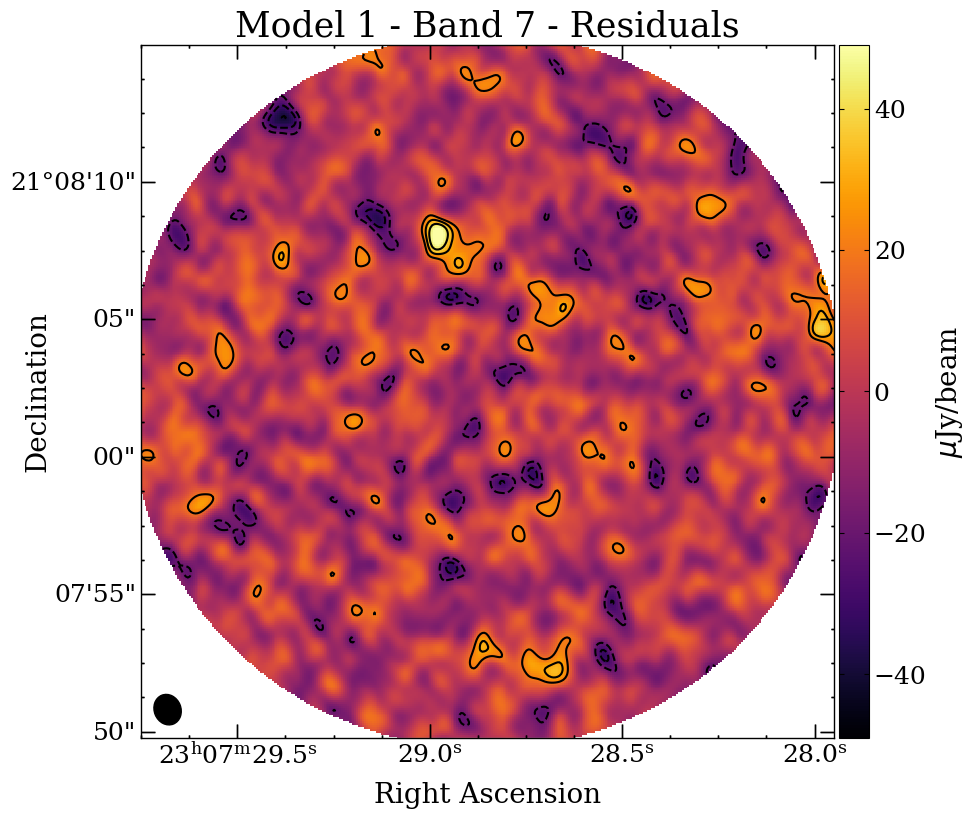}
\includegraphics[scale=0.25]{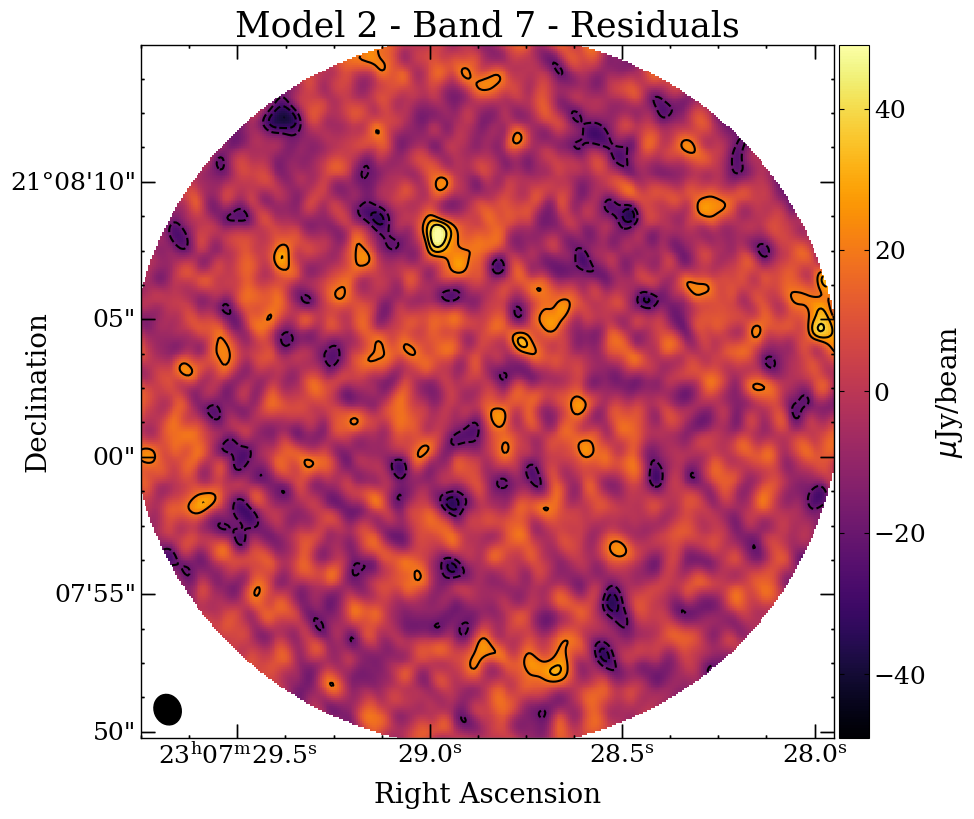}
\includegraphics[scale=0.25]{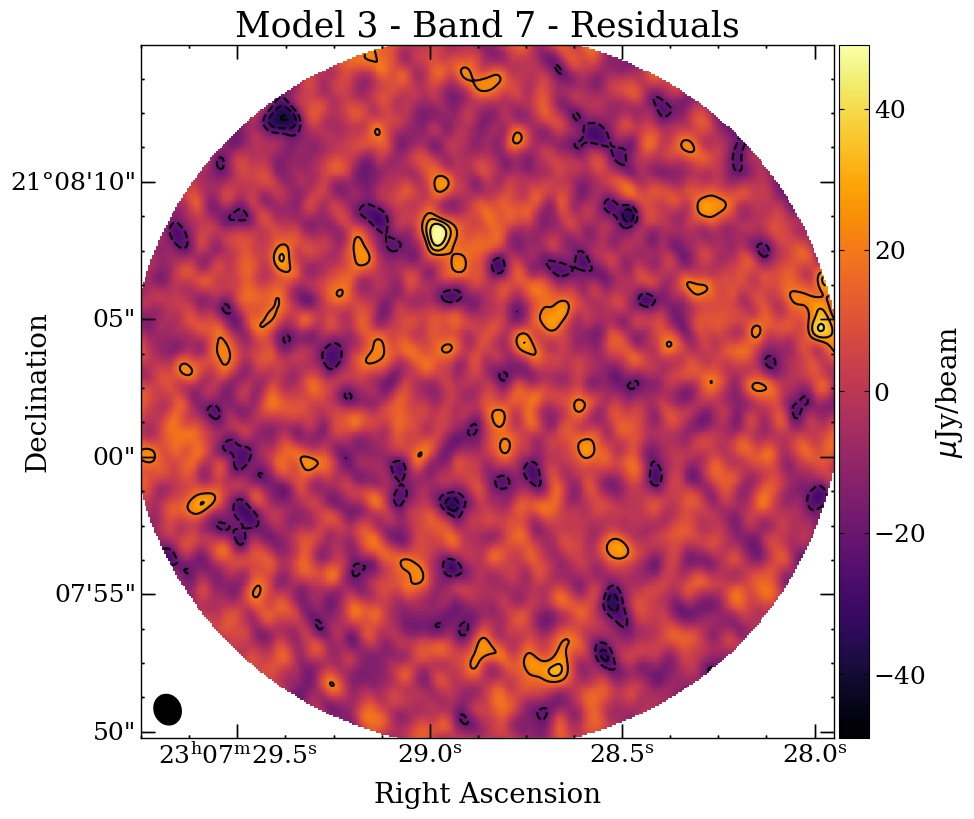}
}
\caption{ALMA 880 microns continuum residual maps, combining the 12m and ACA data. These were obtained by subtracting (in visibility space), our best fit models to our observations, and with residuals stemming from Model 1, Model 2, and Model 3, being shown from left to right, respectively. North is up and East is left, with contours showing the $\pm$2, 3, 4,... $\sigma$ significance levels. The synthesized beam is shown on the lower left side of images and the color bar shows the fluxes in $\mu\mathrm{Jy\,beam}^{-1}$. This image was obtained using a natural weighting scheme, resulting in a sensitivity of $\sigma=9.8\,\mu\mathrm{Jy.beam}^{-1}$ and a synthesized beam of dimensions $1\farcs13 \times 0\farcs97$, with position angle $22^{\circ}$. In all three cases, a $4\sigma$ residual is apparent North of the star. Note that this residual is not the bright source discussed earlier throughout the paper, and is instead located North East of it, at the disk visible outer edge.}
\label{fig:residualsB7}
\end{figure*}

\begin{figure*}
\makebox[\textwidth]{
\includegraphics[scale=0.25]{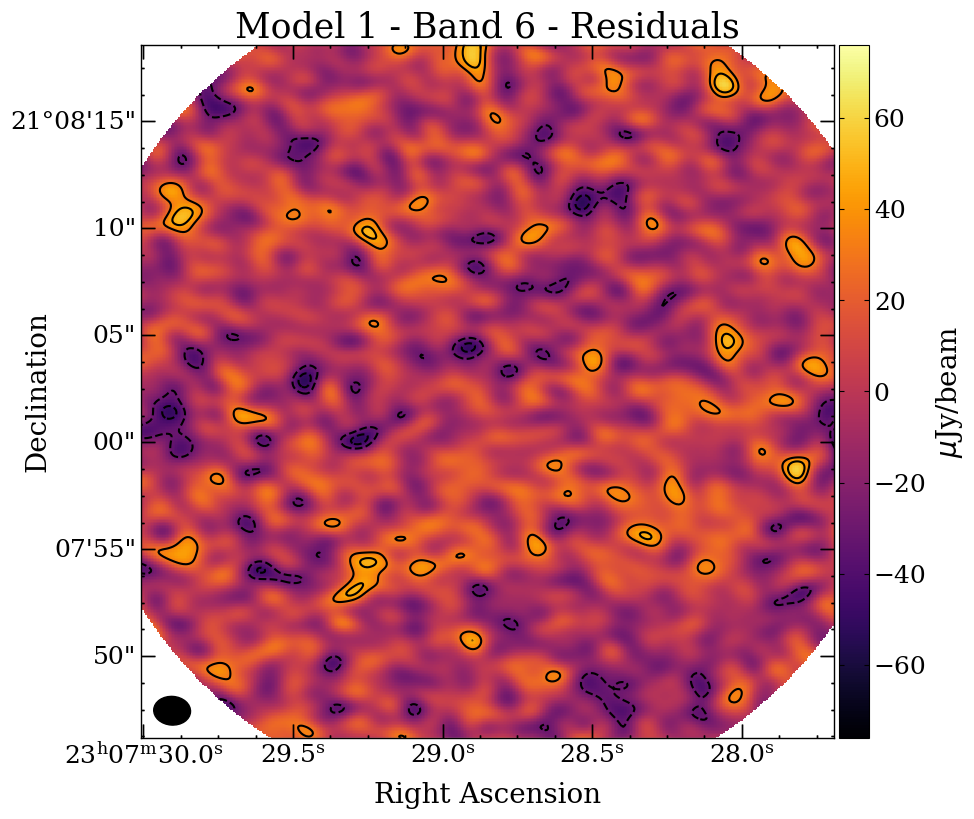}
\includegraphics[scale=0.25]{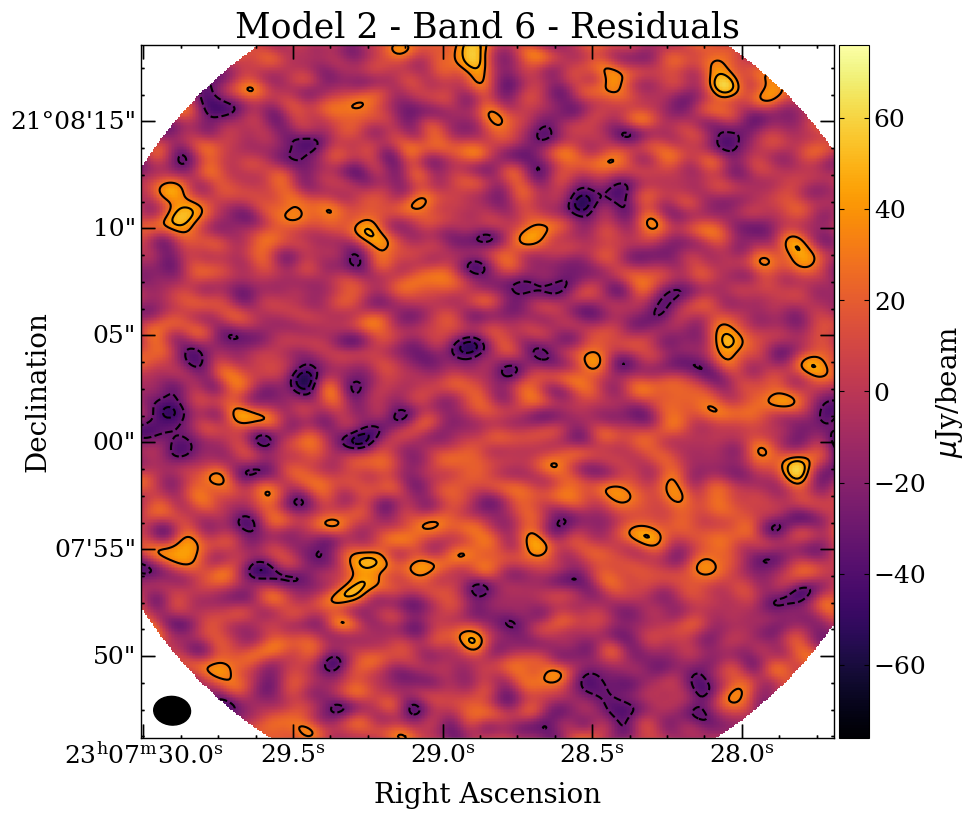}
\includegraphics[scale=0.25]{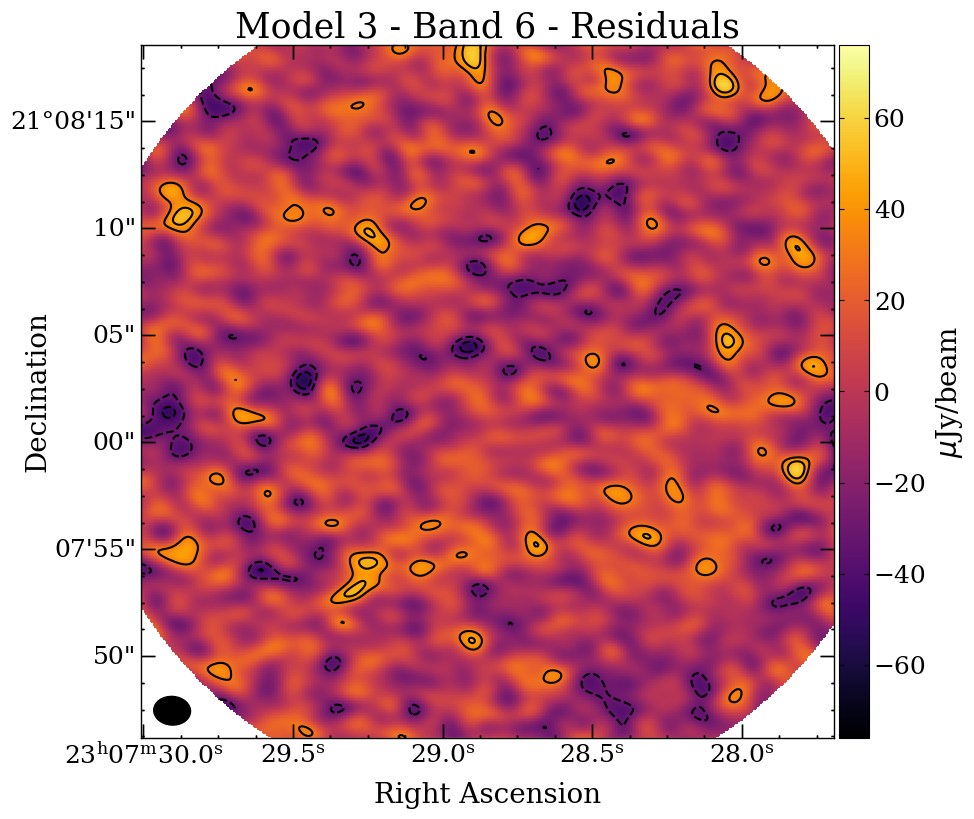}
}
\caption{ALMA 1.3 millimeter continuum residual maps. These were obtained by subtracting (in visibility space), our best fit models the ALMA Cycle 1 Band 6 observations, and with residuals stemming from Model 1, Model 2, and Model 3, being shown from left to right, respectively. North is up and East is left, with contours showing the $\pm$2, 3, 4,... $\sigma$ significance levels. The synthesized beam is shown on the lower left side of images and the color bar shows the fluxes in $\mu\mathrm{Jy\,beam}^{-1}$. This image was obtained using a natural weighting scheme, resulting in a sensitivity of $\sigma=15.9\,\mu\mathrm{Jy.beam}^{-1}$ and a synthesized beam of dimensions $1\farcs72 \times 1\farcs34$, with position angle $86^{\circ}$.}
\label{fig:residualsB6}
\end{figure*}

Figures \ref{fig:residualsB7} and \ref{fig:residualsB6} show the residual maps once our best fit models have been subtracted from the Band 7 and Band 6 observations, respectively.

%%%%%%%%%%%%%%%%%%%%%%%%%%%%%%%%%%%%%%%%%%%
%               RESIDUAL MAPS
%%%%%%%%%%%%%%%%%%%%%%%%%%%%%%%%%%%%%%%%%%%

\bibliography{HR8799}

\begin{thebibliography}{}
\expandafter\ifx\csname natexlab\endcsname\relax\def\natexlab#1{#1}\fi
\providecommand{\url}[1]{\href{#1}{#1}}
\providecommand{\dodoi}[1]{doi:~\href{http://doi.org/#1}{\nolinkurl{#1}}}
\providecommand{\doeprint}[1]{\href{http://ascl.net/#1}{\nolinkurl{http://ascl.net/#1}}}
\providecommand{\doarXiv}[1]{\href{https://arxiv.org/abs/#1}{\nolinkurl{https://arxiv.org/abs/#1}}}

\bibitem[{Aumann {et~al.}(1984)Aumann, Gillett, Beichman, de~Jong, Houck, Low,
  Neugebauer, Walker, \& Wesselius}]{Aumann1984}
Aumann, H.~H., Gillett, F.~C., Beichman, C.~A., {et~al.} 1984, \apjl, 278, L23,
  \dodoi{10.1086/184214}

\bibitem[{Bailer-Jones {et~al.}(2018)Bailer-Jones, Rybizki, Fouesneau,
  Mantelet, \& Andrae}]{Bailer-Jones2018}
Bailer-Jones, C. A.~L., Rybizki, J., Fouesneau, M., Mantelet, G., \& Andrae, R.
  2018, \aj, 156, 58, \dodoi{10.3847/1538-3881/aacb21}

\bibitem[{{Baines} {et~al.}(2012){Baines}, {White}, {Huber}, {Jones},
  {Boyajian}, {McAlister}, {ten Brummelaar}, {Turner}, {Sturmann}, {Sturmann},
  {Goldfinger}, {Farrington}, {Riedel}, {Ireland }, {von Braun}, \&
  {Ridgway}}]{Baines2012}
{Baines}, E.~K., {White}, R.~J., {Huber}, D., {et~al.} 2012, \apj, 761, 57,
  \dodoi{10.1088/0004-637X/761/1/57}

\bibitem[{{Baraffe} {et~al.}(2015){Baraffe}, {Homeier}, {Allard}, \&
  {Chabrier}}]{Baraffe2015}
{Baraffe}, I., {Homeier}, D., {Allard}, F., \& {Chabrier}, G. 2015, \aap, 577,
  A42, \dodoi{10.1051/0004-6361/201425481}

\bibitem[{{Bayo} {et~al.}(2019){Bayo}, {Olofsson}, {Matr{\`a}}, {Beam{\'\i}n},
  {Gallardo}, {de Gregorio-Monsalvo}, {Booth}, {Zamora}, {Iglesias}, {Henning},
  {Schreiber}, \& {C{\'a}ceres}}]{Bayo2019}
{Bayo}, A., {Olofsson}, J., {Matr{\`a}}, L., {et~al.} 2019, \mnras, 486, 5552,
  \dodoi{10.1093/mnras/stz1133}

\bibitem[{{Bell} {et~al.}(2015){Bell}, {Mamajek}, \& {Naylor}}]{Bell2015}
{Bell}, C. P.~M., {Mamajek}, E.~E., \& {Naylor}, T. 2015, \mnras, 454, 593,
  \dodoi{10.1093/mnras/stv1981}

\bibitem[{{Bell} {et~al.}(2017){Bell}, {Murphy}, \& {Mamajek}}]{Bell2017}
{Bell}, C. P.~M., {Murphy}, S.~J., \& {Mamajek}, E.~E. 2017, \mnras, 468, 1198,
  \dodoi{10.1093/mnras/stx535}

\bibitem[{{Binks} \& {Jeffries}(2016)}]{Binks2016}
{Binks}, A.~S., \& {Jeffries}, R.~D. 2016, \mnras, 455, 3345,
  \dodoi{10.1093/mnras/stv2431}

\bibitem[{{Bland-Hawthorn} \& {Gerhard}(2016)}]{Bland-Hawthorn2016}
{Bland-Hawthorn}, J., \& {Gerhard}, O. 2016, \araa, 54, 529,
  \dodoi{10.1146/annurev-astro-081915-023441}

\bibitem[{{Bohn} {et~al.}(2020){Bohn}, {Kenworthy}, {Ginski}, {Rieder},
  {Mamajek}, {Meshkat}, {Pecaut}, {Reggiani}, {de Boer}, {Keller}, {Snik}, \&
  {Southworth}}]{Bohn2020}
{Bohn}, A.~J., {Kenworthy}, M.~A., {Ginski}, C., {et~al.} 2020, \apjl, 898,
  L16, \dodoi{10.3847/2041-8213/aba27e}

\bibitem[{{Booth} {et~al.}(2021{\natexlab{a}}){Booth}, {del Burgo}, \&
  {Hambaryan}}]{Booth2021b}
{Booth}, M., {del Burgo}, C., \& {Hambaryan}, V.~V. 2021{\natexlab{a}}, \mnras,
  500, 5552, \dodoi{10.1093/mnras/staa3631}

\bibitem[{{Booth} {et~al.}(2021{\natexlab{b}}){Booth}, {Schulz}, {Krivov},
  {Marino}, {Pearce}, \& {Launhardt}}]{Booth2021}
{Booth}, M., {Schulz}, M., {Krivov}, A.~V., {et~al.} 2021{\natexlab{b}},
  \mnras, 500, 1604, \dodoi{10.1093/mnras/staa3362}

\bibitem[{{Booth} {et~al.}(2016){Booth}, {Jord{\'a}n}, {Casassus}, {Hales},
  {Dent}, {Faramaz}, {Matr{\`a}}, {Barkats}, {Brahm}, \& {Cuadra}}]{Booth2016}
{Booth}, M., {Jord{\'a}n}, A., {Casassus}, S., {et~al.} 2016, \mnras, 460, L10,
  \dodoi{10.1093/mnrasl/slw040}

\bibitem[{{Booth} {et~al.}(2017){Booth}, {Dent}, {Jord{\'a}n}, {Lestrade},
  {Hales}, {Wyatt}, {Casassus}, {Ertel}, {Greaves}, {Kennedy}, {Matr{\`a}},
  {Augereau}, \& {Villard}}]{Booth2017}
{Booth}, M., {Dent}, W. R.~F., {Jord{\'a}n}, A., {et~al.} 2017, \mnras, 469,
  3200, \dodoi{10.1093/mnras/stx1072}

\bibitem[{{Bovy} {et~al.}(2016){Bovy}, {Rix}, {Schlafly}, {Nidever},
  {Holtzman}, {Shetrone}, \& {Beers}}]{Bovy2016}
{Bovy}, J., {Rix}, H.-W., {Schlafly}, E.~F., {et~al.} 2016, \apj, 823, 30,
  \dodoi{10.3847/0004-637X/823/1/30}

\bibitem[{Carniani {et~al.}(2015)Carniani, Maiolino, De~Zotti, Negrello,
  Marconi, Bothwell, Capak, Carilli, Castellano, Cristiani, Ferrara, Fontana,
  Gallerani, Jones, Ohta, Ota, Pentericci, Santini, Sheth, Vallini, Vanzella,
  Wagg, \& Williams}]{Carniani2015}
Carniani, S., Maiolino, R., De~Zotti, G., {et~al.} 2015, \aap, 584, A78,
  \dodoi{10.1051/0004-6361/201525780}

\bibitem[{{Casey}(2012)}]{Casey2012}
{Casey}, C.~M. 2012, \mnras, 425, 3094,
  \dodoi{10.1111/j.1365-2966.2012.21455.x}

\bibitem[{{Chen} {et~al.}(2006){Chen}, {Sargent}, {Bohac}, {Kim},
  {Leibensperger}, {Jura}, {Najita}, {Forrest}, {Watson}, {Sloan}, \&
  {Keller}}]{Chen2006}
{Chen}, C.~H., {Sargent}, B.~A., {Bohac}, C., {et~al.} 2006, \apjs, 166, 351,
  \dodoi{10.1086/505751}

\bibitem[{{Crundall} {et~al.}(2019){Crundall}, {Ireland}, {Krumholz},
  {Federrath}, {{\v{Z}}erjal}, \& {Hansen}}]{Crundall2019}
{Crundall}, T.~D., {Ireland}, M.~J., {Krumholz}, M.~R., {et~al.} 2019, \mnras,
  489, 3625, \dodoi{10.1093/mnras/stz2376}

\bibitem[{{Curtis} {et~al.}(2020){Curtis}, {Ag{\"u}eros}, {Matt}, {Covey},
  {Douglas}, {Angus}, {Saar}, {Cody}, {Vanderburg}, {Law}, {Kraus}, {Latham},
  {Baranec}, {Riddle}, {Ziegler}, {Lund}, {Torres}, {Meibom}, {Aguirre}, \&
  {Wright}}]{Curtis2020}
{Curtis}, J.~L., {Ag{\"u}eros}, M.~A., {Matt}, S.~P., {et~al.} 2020, \apj, 904,
  140, \dodoi{10.3847/1538-4357/abbf58}

\bibitem[{{Cutri} {et~al.}(2003){Cutri}, {Skrutskie}, {van Dyk}, {Beichman},
  {Carpenter}, {Chester}, {Cambresy}, {Evans}, {Fowler}, {Gizis}, {Howard},
  {Huchra}, {Jarrett}, {Kopan}, {Kirkpatrick}, {Light}, {Marsh}, {McCallon},
  {Schneider}, {Stiening}, {Sykes}, {Weinberg}, {Wheaton}, {Wheelock}, \&
  {Zacarias}}]{Cutri03}
{Cutri}, R.~M., {Skrutskie}, M.~F., {van Dyk}, S., {et~al.} 2003, {2MASS All
  Sky Catalog of point sources.}

\bibitem[{{de Bruijne}(1999)}]{deBruijne1999}
{de Bruijne}, J. H.~J. 1999, \mnras, 310, 585,
  \dodoi{10.1046/j.1365-8711.1999.02953.x}

\bibitem[{{Dieterich} {et~al.}(2014){Dieterich}, {Henry}, {Jao}, {Winters},
  {Hosey}, {Riedel}, \& {Subasavage}}]{Dieterich2014}
{Dieterich}, S.~B., {Henry}, T.~J., {Jao}, W.-C., {et~al.} 2014, \aj, 147, 94,
  \dodoi{10.1088/0004-6256/147/5/94}

\bibitem[{{Doyon} {et~al.}(2010){Doyon}, {Lafreni{\`e}re}, {Artigau}, {Malo},
  \& {Marois}}]{Doyon2010}
{Doyon}, R., {Lafreni{\`e}re}, D., {Artigau}, E., {Malo}, L., \& {Marois}, C.
  2010, in In the Spirit of Lyot 2010, ed. A.~{Boccaletti}, E42

\bibitem[{{Draine}(2003)}]{Draine2003}
{Draine}, B.~T. 2003, \apj, 598, 1017, \dodoi{10.1086/379118}

\bibitem[{{Dullemond} {et~al.}(2012){Dullemond}, {Juhasz}, {Pohl}, {Sereshti},
  {Shetty}, {Peters}, {Commercon}, \& {Flock}}]{Dullemond2012}
{Dullemond}, C.~P., {Juhasz}, A., {Pohl}, A., {et~al.} 2012, {RADMC-3D: A
  multi-purpose radiative transfer tool}.
\newblock \doeprint{1202.015}

\bibitem[{{ESA}(1997)}]{ESA1997}
{ESA}. 1997, in ESA Special Publication, Vol. 1200, ESA Special Publication

\bibitem[{{Faramaz} {et~al.}(2019){Faramaz}, {Krist}, {Stapelfeldt}, {Bryden},
  {Mamajek}, {Matr{\`a}}, {Booth}, {Flaherty}, {Hales}, {Hughes}, {Bayo},
  {Casassus}, {Cuadra}, {Olofsson}, {Su}, \& {Wilner}}]{Faramaz2019}
{Faramaz}, V., {Krist}, J., {Stapelfeldt}, K.~R., {et~al.} 2019, \aj, 158, 162,
  \dodoi{10.3847/1538-3881/ab3ec1}

\bibitem[{Foreman-Mackey {et~al.}(2013)Foreman-Mackey, Hogg, Lang, \&
  Goodman}]{Foreman-Mackey2013}
Foreman-Mackey, D., Hogg, D.~W., Lang, D., \& Goodman, J. 2013, \pasp, 125,
  306, \dodoi{10.1086/670067}

\bibitem[{{Gagn{\'e}} {et~al.}(2018){Gagn{\'e}}, {Mamajek}, {Malo}, {Riedel},
  {Rodriguez}, {Lafreni{\`e}re}, {Faherty}, {Roy-Loubier}, {Pueyo}, {Robin}, \&
  {Doyon}}]{Gagne2018}
{Gagn{\'e}}, J., {Mamajek}, E.~E., {Malo}, L., {et~al.} 2018, \apj, 856, 23,
  \dodoi{10.3847/1538-4357/aaae09}

\bibitem[{{Gaia Collaboration} {et~al.}(2020){Gaia Collaboration}, {Brown},
  {Vallenari}, {Prusti}, {de Bruijne}, {Babusiaux}, \& {Biermann}}]{GaiaEDR3}
{Gaia Collaboration}, {Brown}, A.~G.~A., {Vallenari}, A., {et~al.} 2020, arXiv
  e-prints, arXiv:2012.01533.
\newblock \doarXiv{2012.01533}

\bibitem[{{Gaia Collaboration} {et~al.}(2016){Gaia Collaboration}, Prusti,
  de~Bruijne, Brown, Vallenari, Babusiaux, Bailer-Jones, Bastian, Biermann,
  Evans, Eyer, Jansen, Jordi, Klioner, Lammers, Lindegren, Luri, Mignard,
  Milligan, Panem, Poinsignon, Pourbaix, Randich, Sarri, Sartoretti, Siddiqui,
  Soubiran, Valette, van Leeuwen, Walton, Aerts, Arenou, Cropper, Drimmel,
  H{\o}g, Katz, Lattanzi, O'Mullane, Grebel, Holland, Huc, Passot, Bramante,
  Cacciari, Casta{\~n}eda, Chaoul, Cheek, De~Angeli, Fabricius, Guerra,
  Hern{\'a}ndez, Jean-Antoine-Piccolo, Masana, Messineo, Mowlavi, Nienartowicz,
  Ord{\'o}{\~n}ez-Blanco, Panuzzo, Portell, Richards, Riello, Seabroke, Tanga,
  Th{\'e}venin, Torra, Els, Gracia-Abril, Comoretto, Garcia-Reinaldos, Lock,
  Mercier, Altmann, Andrae, Astraatmadja, Bellas-Velidis, Benson, Berthier,
  Blomme, Busso, Carry, Cellino, Clementini, Cowell, Creevey, Cuypers,
  Davidson, De~Ridder, de~Torres, Delchambre, Dell'Oro, Ducourant, Fr{\'e}mat,
  Garc{\'{\i}}a-Torres, Gosset, Halbwachs, Hambly, Harrison, Hauser,
  Hestroffer, Hodgkin, Huckle, Hutton, Jasniewicz, Jordan, Kontizas, Korn,
  Lanzafame, Manteiga, Moitinho, Muinonen, Osinde, Pancino, Pauwels, Petit,
  Recio-Blanco, Robin, Sarro, Siopis, Smith, Smith, Sozzetti, Thuillot, van
  Reeven, Viala, Abbas, Abreu~Aramburu, Accart, Aguado, Allan, Allasia,
  Altavilla, {\'A}lvarez, Alves, Anderson, Andrei, Anglada~Varela, Antiche,
  Antoja, Ant{\'o}n, Arcay, Atzei, Ayache, Bach, Baker, Balaguer-N{\'u}{\~n}ez,
  Barache, Barata, Barbier, Barblan, Baroni, Barrado~y Navascu{\'e}s, Barros,
  Barstow, Becciani, Bellazzini, Bellei, Bello~Garc{\'{\i}}a, Belokurov,
  Bendjoya, Berihuete, Bianchi, Bienaym{\'e}, Billebaud, Blagorodnova,
  Blanco-Cuaresma, Boch, Bombrun, Borrachero, Bouquillon, Bourda, Bouy,
  Bragaglia, Breddels, Brouillet, Br{\"u}semeister, Bucciarelli, Budnik,
  Burgess, Burgon, Burlacu, Busonero, Buzzi, Caffau, Cambras, Campbell,
  Cancelliere, Cantat-Gaudin, Carlucci, Carrasco, Castellani, Charlot, Charnas,
  Charvet, Chassat, Chiavassa, Clotet, Cocozza, Collins, Collins, Costigan,
  Crifo, Cross, Crosta, Crowley, Dafonte, Damerdji, Dapergolas, David, David,
  De~Cat, de~Felice, de~Laverny, De~Luise, De~March, de~Martino, de~Souza,
  Debosscher, del Pozo, Delbo, Delgado, Delgado, di~Marco, Di~Matteo, Diakite,
  Distefano, Dolding, Dos~Anjos, Drazinos, Dur{\'a}n, Dzigan, Ecale,
  Edvardsson, Enke, Erdmann, Escolar, Espina, Evans, Eynard~Bontemps, Fabre,
  Fabrizio, Faigler, Falc{\~a}o, Farr{\`a}s~Casas, Faye, Federici, Fedorets,
  Fern{\'a}ndez-Hern{\'a}ndez, Fernique, Fienga, Figueras, Filippi, Findeisen,
  Fonti, Fouesneau, Fraile, Fraser, Fuchs, Furnell, Gai, Galleti, Galluccio,
  Garabato, Garc{\'{\i}}a-Sedano, Gar{\'e}, Garofalo, Garralda, Gavras,
  Gerssen, Geyer, Gilmore, Girona, Giuffrida, Gomes, Gonz{\'a}lez-Marcos,
  Gonz{\'a}lez-N{\'u}{\~n}ez, Gonz{\'a}lez-Vidal, Granvik, Guerrier, Guillout,
  Guiraud, G{\'u}rpide, Guti{\'e}rrez-S{\'a}nchez, Guy, Haigron,
  Hatzidimitriou, Haywood, Heiter, Helmi, Hobbs, Hofmann, Holl, Holland, Hunt,
  Hypki, Icardi, Irwin, Jevardat~de Fombelle, Jofr{\'e}, Jonker, Jorissen,
  Julbe, Karampelas, Kochoska, Kohley, Kolenberg, Kontizas, Koposov,
  Kordopatis, Koubsky, Kowalczyk, Krone-Martins, Kudryashova, Kull, Bachchan,
  Lacoste-Seris, Lanza, Lavigne, Le~Poncin-Lafitte, Lebreton, Lebzelter,
  Leccia, Leclerc, Lecoeur-Taibi, Lemaitre, Lenhardt, Leroux, Liao, Licata,
  Lindstr{\o}m, Lister, Livanou, Lobel, L{\"o}ffler, L{\'o}pez, Lopez-Lozano,
  Lorenz, Loureiro, MacDonald, Magalh{\~a}es~Fernandes, Managau, Mann,
  Mantelet, Marchal, Marchant, Marconi, Marie, Marinoni, Marrese,
  Marschalk{\'o}, Marshall, Mart{\'{\i}}n-Fleitas, Martino, Mary, Matijevi{\v
  c}, Mazeh, McMillan, Messina, Mestre, Michalik, Millar, Miranda, Molina,
  Molinaro, Molinaro, Moln{\'a}r, Moniez, Montegriffo, Monteiro, Mor, Mora,
  Morbidelli, Morel, Morgenthaler, Morley, Morris, Mulone, Muraveva, Musella,
  Narbonne, Nelemans, Nicastro, Noval, Ord{\'e}novic, Ordieres-Mer{\'e},
  Osborne, Pagani, Pagano, Pailler, Palacin, Palaversa, Parsons, Paulsen,
  Pecoraro, Pedrosa, Pentik{\"a}inen, Pereira, Pichon, Piersimoni, Pineau,
  Plachy, Plum, Poujoulet, Pr{\v s}a, Pulone, Ragaini, Rago, Rambaux,
  Ramos-Lerate, Ranalli, Rauw, Read, Regibo, Renk, Reyl{\'e}, Ribeiro,
  Rimoldini, Ripepi, Riva, Rixon, Roelens, Romero-G{\'o}mez, Rowell, Royer,
  Rudolph, Ruiz-Dern, Sadowski, Sagrist{\`a Sell{\'e}s}, Sahlmann, Salgado,
  Salguero, Sarasso, Savietto, Schnorhk, Schultheis, Sciacca, Segol, Segovia,
  Segransan, Serpell, Shih, Smareglia, Smart, Smith, Solano, Solitro, Sordo,
  Soria~Nieto, Souchay, Spagna, Spoto, Stampa, Steele, Steidelm{\"u}ller,
  Stephenson, Stoev, Suess, S{\"u}veges, Surdej, Szabados, Szegedi-Elek,
  Tapiador, Taris, Tauran, Taylor, Teixeira, Terrett, Tingley, Trager, Turon,
  Ulla, Utrilla, Valentini, van Elteren, Van~Hemelryck, van Leeuwen, Varadi,
  Vecchiato, Veljanoski, Via, Vicente, Vogt, Voss, Votruba, Voutsinas,
  Walmsley, Weiler, Weingrill, Werner, Wevers, Whitehead, Wyrzykowski, Yoldas,
  {\v Z}erjal, Zucker, Zurbach, Zwitter, Alecu, Allen, Allende~Prieto, Amorim,
  Anglada-Escud{\'e}, Arsenijevic, Azaz, Balm, Beck, Bernstein, Bigot, Bijaoui,
  Blasco, Bonfigli, Bono, Boudreault, Bressan, Brown, Brunet, Bunclark,
  Buonanno, Butkevich, Carret, Carrion, Chemin, Ch{\'e}reau, Corcione,
  Darmigny, de~Boer, de~Teodoro, de~Zeeuw, Delle~Luche, Domingues, Dubath,
  Fodor, Fr{\'e}zouls, Fries, Fustes, Fyfe, Gallardo, Gallegos, Gardiol,
  Gebran, Gomboc, G{\'o}mez, Grux, Gueguen, Heyrovsky, Hoar, Iannicola,
  Isasi~Parache, Janotto, Joliet, Jonckheere, Keil, Kim, Klagyivik, Klar,
  Knude, Kochukhov, Kolka, Kos, Kutka, Lainey, LeBouquin, Liu, Loreggia,
  Makarov, Marseille, Martayan, Martinez-Rubi, Massart, Meynadier, Mignot,
  Munari, Nguyen, Nordlander, Ocvirk, O'Flaherty, Olias~Sanz, Ortiz, Osorio,
  Oszkiewicz, Ouzounis, Palmer, Park, Pasquato, Peltzer, Peralta, P{\'e}turaud,
  Pieniluoma, Pigozzi, Poels, Prat, Prod'homme, Raison, Rebordao, Risquez,
  Rocca-Volmerange, Rosen, Ruiz-Fuertes, Russo, Sembay, Serraller~Vizcaino,
  Short, Siebert, Silva, Sinachopoulos, Slezak, Soffel, Sosnowska, Strai{\v
  z}ys, ter Linden, Terrell, Theil, Tiede, Troisi, Tsalmantza, Tur, Vaccari,
  Vachier, Valles, Van~Hamme, Veltz, Virtanen, Wallut, Wichmann, Wilkinson,
  Ziaeepour, \& Zschocke}]{GaiaCollaboration2016}
{Gaia Collaboration}, Prusti, T., de~Bruijne, J. H.~J., {et~al.} 2016, \aap,
  595, A1, \dodoi{10.1051/0004-6361/201629272}

\bibitem[{{Gaia Collaboration} {et~al.}(2018){Gaia Collaboration}, {Brown},
  {Vallenari}, {Prusti}, {de Bruijne}, {Babusiaux}, {Bailer-Jones}, {Biermann},
  {Evans}, {Eyer}, {Jansen}, {Jordi}, {Klioner}, {Lammers}, {Lindegren},
  {Luri}, {Mignard}, {Panem}, {Pourbaix}, {Randich}, {Sartoretti}, {Siddiqui},
  {Soubiran}, {van Leeuwen}, {Walton}, {Arenou}, {Bastian}, {Cropper},
  {Drimmel}, {Katz}, {Lattanzi}, {Bakker}, {Cacciari}, {Casta{\~n}eda},
  {Chaoul}, {Cheek}, {De Angeli}, {Fabricius}, {Guerra}, {Holl}, {Masana},
  {Messineo}, {Mowlavi}, {Nienartowicz}, {Panuzzo}, {Portell}, {Riello},
  {Seabroke}, {Tanga}, {Th{\'e}venin}, {Gracia-Abril}, {Comoretto},
  {Garcia-Reinaldos}, {Teyssier}, {Altmann}, {Andrae}, {Audard},
  {Bellas-Velidis}, {Benson}, {Berthier}, {Blomme}, {Burgess}, {Busso},
  {Carry}, {Cellino}, {Clementini}, {Clotet}, {Creevey}, {Davidson}, {De
  Ridder}, {Delchambre}, {Dell'Oro}, {Ducourant},
  {Fern{\'a}ndez-Hern{\'a}ndez}, {Fouesneau}, {Fr{\'e}mat}, {Galluccio},
  {Garc{\'\i}a-Torres}, {Gonz{\'a}lez-N{\'u}{\~n}ez}, {Gonz{\'a}lez-Vidal},
  {Gosset}, {Guy}, {Halbwachs}, {Hambly}, {Harrison}, {Hern{\'a}ndez},
  {Hestroffer}, {Hodgkin}, {Hutton}, {Jasniewicz}, {Jean-Antoine-Piccolo},
  {Jordan}, {Korn}, {Krone-Martins}, {Lanzafame}, {Lebzelter}, {L{\"o}ffler},
  {Manteiga}, {Marrese}, {Mart{\'\i}n-Fleitas}, {Moitinho}, {Mora}, {Muinonen},
  {Osinde}, {Pancino}, {Pauwels}, {Petit}, {Recio-Blanco}, {Richards},
  {Rimoldini}, {Robin}, {Sarro}, {Siopis}, {Smith}, {Sozzetti}, {S{\"u}veges},
  {Torra}, {van Reeven}, {Abbas}, {Abreu Aramburu}, {Accart}, {Aerts},
  {Altavilla}, {{\'A}lvarez}, {Alvarez}, {Alves}, {Anderson}, {Andrei},
  {Anglada Varela}, {Antiche}, {Antoja}, {Arcay}, {Astraatmadja}, {Bach},
  {Baker}, {Balaguer-N{\'u}{\~n}ez}, {Balm}, {Barache}, {Barata}, {Barbato},
  {Barblan}, {Barklem}, {Barrado}, {Barros}, {Barstow}, {Bartholom{\'e}
  Mu{\~n}oz}, {Bassilana}, {Becciani}, {Bellazzini}, {Berihuete}, {Bertone},
  {Bianchi}, {Bienaym{\'e}}, {Blanco-Cuaresma}, {Boch}, {Boeche}, {Bombrun},
  {Borrachero}, {Bossini}, {Bouquillon}, {Bourda}, {Bragaglia}, {Bramante},
  {Breddels}, {Bressan}, {Brouillet}, {Br{\"u}semeister}, {Brugaletta},
  {Bucciarelli}, {Burlacu}, {Busonero}, {Butkevich}, {Buzzi}, {Caffau},
  {Cancelliere}, {Cannizzaro}, {Cantat-Gaudin}, {Carballo}, {Carlucci},
  {Carrasco}, {Casamiquela}, {Castellani}, {Castro-Ginard}, {Charlot},
  {Chemin}, {Chiavassa}, {Cocozza}, {Costigan}, {Cowell}, {Crifo}, {Crosta},
  {Crowley}, {Cuypers}, {Dafonte}, {Damerdji}, {Dapergolas}, {David}, {David},
  {de Laverny}, {De Luise}, {De March}, {de Martino}, {de Souza}, {de Torres},
  {Debosscher}, {del Pozo}, {Delbo}, {Delgado}, {Delgado}, {Di Matteo},
  {Diakite}, {Diener}, {Distefano}, {Dolding}, {Drazinos}, {Dur{\'a}n},
  {Edvardsson}, {Enke}, {Eriksson}, {Esquej}, {Eynard Bontemps}, {Fabre},
  {Fabrizio}, {Faigler}, {Falc{\~a}o}, {Farr{\`a}s Casas}, {Federici},
  {Fedorets}, {Fernique}, {Figueras}, {Filippi}, {Findeisen}, {Fonti},
  {Fraile}, {Fraser}, {Fr{\'e}zouls}, {Gai}, {Galleti}, {Garabato},
  {Garc{\'\i}a-Sedano}, {Garofalo}, {Garralda}, {Gavel}, {Gavras}, {Gerssen},
  {Geyer}, {Giacobbe}, {Gilmore}, {Girona}, {Giuffrida}, {Glass}, {Gomes},
  {Granvik}, {Gueguen}, {Guerrier}, {Guiraud}, {Guti{\'e}rrez-S{\'a}nchez},
  {Haigron}, {Hatzidimitriou}, {Hauser}, {Haywood}, {Heiter}, {Helmi}, {Heu},
  {Hilger}, {Hobbs}, {Hofmann}, {Holland}, {Huckle}, {Hypki}, {Icardi},
  {Jan{\ss}en}, {Jevardat de Fombelle}, {Jonker}, {Juh{\'a}sz}, {Julbe},
  {Karampelas}, {Kewley}, {Klar}, {Kochoska}, {Kohley}, {Kolenberg},
  {Kontizas}, {Kontizas}, {Koposov}, {Kordopatis}, {Kostrzewa-Rutkowska},
  {Koubsky}, {Lambert}, {Lanza}, {Lasne}, {Lavigne}, {Le Fustec}, {Le
  Poncin-Lafitte}, {Lebreton}, {Leccia}, {Leclerc}, {Lecoeur-Taibi},
  {Lenhardt}, {Leroux}, {Liao}, {Licata}, {Lindstr{\o}m}, {Lister}, {Livanou},
  {Lobel}, {L{\'o}pez}, {Managau}, {Mann}, {Mantelet}, {Marchal}, {Marchant},
  {Marconi}, {Marinoni}, {Marschalk{\'o}}, {Marshall}, {Martino}, {Marton},
  {Mary}, {Massari}, {Matijevi{\v{c}}}, {Mazeh}, {McMillan}, {Messina},
  {Michalik}, {Millar}, {Molina}, {Molinaro}, {Moln{\'a}r}, {Montegriffo},
  {Mor}, {Morbidelli}, {Morel}, {Morris}, {Mulone}, {Muraveva}, {Musella},
  {Nelemans}, {Nicastro}, {Noval}, {O'Mullane}, {Ord{\'e}novic},
  {Ord{\'o}{\~n}ez-Blanco}, {Osborne}, {Pagani}, {Pagano}, {Pailler},
  {Palacin}, {Palaversa}, {Panahi}, {Pawlak}, {Piersimoni}, {Pineau}, {Plachy},
  {Plum}, {Poggio}, {Poujoulet}, {Pr{\v{s}}a}, {Pulone}, {Racero}, {Ragaini},
  {Rambaux}, {Ramos-Lerate}, {Regibo}, {Reyl{\'e}}, {Riclet}, {Ripepi}, {Riva},
  {Rivard}, {Rixon}, {Roegiers}, {Roelens}, {Romero-G{\'o}mez}, {Rowell},
  {Royer}, {Ruiz-Dern}, {Sadowski}, {Sagrist{\`a} Sell{\'e}s}, {Sahlmann},
  {Salgado}, {Salguero}, {Sanna}, {Santana-Ros}, {Sarasso}, {Savietto},
  {Schultheis}, {Sciacca}, {Segol}, {Segovia}, {S{\'e}gransan}, {Shih},
  {Siltala}, {Silva}, {Smart}, {Smith}, {Solano}, {Solitro}, {Sordo}, {Soria
  Nieto}, {Souchay}, {Spagna}, {Spoto}, {Stampa}, {Steele},
  {Steidelm{\"u}ller}, {Stephenson}, {Stoev}, {Suess}, {Surdej}, {Szabados},
  {Szegedi-Elek}, {Tapiador}, {Taris}, {Tauran}, {Taylor}, {Teixeira},
  {Terrett}, {Teyssand ier}, {Thuillot}, {Titarenko}, {Torra Clotet}, {Turon},
  {Ulla}, {Utrilla}, {Uzzi}, {Vaillant}, {Valentini}, {Valette}, {van Elteren},
  {Van Hemelryck}, {van Leeuwen}, {Vaschetto}, {Vecchiato}, {Veljanoski},
  {Viala}, {Vicente}, {Vogt}, {von Essen}, {Voss}, {Votruba}, {Voutsinas},
  {Walmsley}, {Weiler}, {Wertz}, {Wevers}, {Wyrzykowski}, {Yoldas},
  {{\v{Z}}erjal}, {Ziaeepour}, {Zorec}, {Zschocke}, {Zucker}, {Zurbach}, \&
  {Zwitter}}]{GaiaCollaboration2018}
{Gaia Collaboration}, {Brown}, A.~G.~A., {Vallenari}, A., {et~al.} 2018, \aap,
  616, A1, \dodoi{10.1051/0004-6361/201833051}

\bibitem[{{Gaidos} {et~al.}(2014){Gaidos}, {Mann}, {L{\'e}pine}, {Buccino},
  {James}, {Ansdell}, {Petrucci}, {Mauas}, \& {Hilton}}]{Gaidos2014}
{Gaidos}, E., {Mann}, A.~W., {L{\'e}pine}, S., {et~al.} 2014, \mnras, 443,
  2561, \dodoi{10.1093/mnras/stu1313}

\bibitem[{{Galli} {et~al.}(2019){Galli}, {Loinard}, {Bouy}, {Sarro},
  {Ortiz-Le{\'o}n}, {Dzib}, {Olivares}, {Heyer}, {Hernandez},
  {Rom{\'a}n-Z{\'u}{\~n}iga}, {Kounkel}, \& {Covey}}]{Galli2019}
{Galli}, P.~A.~B., {Loinard}, L., {Bouy}, H., {et~al.} 2019, \aap, 630, A137,
  \dodoi{10.1051/0004-6361/201935928}

\bibitem[{{Geiler} {et~al.}(2019){Geiler}, {Krivov}, {Booth}, \&
  {L{\"o}hne}}]{Geiler2019}
{Geiler}, F., {Krivov}, A.~V., {Booth}, M., \& {L{\"o}hne}, T. 2019, \mnras,
  483, 332, \dodoi{10.1093/mnras/sty3160}

\bibitem[{{Gerard} {et~al.}(2016){Gerard}, {Lawler}, {Marois}, {Tannock},
  {Matthews}, \& {Venn}}]{Gerard2016}
{Gerard}, B., {Lawler}, S., {Marois}, C., {et~al.} 2016, \apj, 823, 149,
  \dodoi{10.3847/0004-637X/823/2/149}

\bibitem[{{Gontcharov}(2006)}]{Gontcharov2006}
{Gontcharov}, G.~A. 2006, Astronomy Letters, 32, 759,
  \dodoi{10.1134/S1063773706110065}

\bibitem[{Go{\'z}dziewski \& Migaszewski(2018)}]{Gozdziewski2018}
Go{\'z}dziewski, K., \& Migaszewski, C. 2018, \apjs, 238, 6,
  \dodoi{10.3847/1538-4365/aad3d3}

\bibitem[{{Go{\'z}dziewski} \& {Migaszewski}(2020)}]{Gozdziewski2020}
{Go{\'z}dziewski}, K., \& {Migaszewski}, C. 2020, \apjl, 902, L40,
  \dodoi{10.3847/2041-8213/abb881}

\bibitem[{{Gray} \& {Corbally}(2014)}]{Gray2014}
{Gray}, R.~O., \& {Corbally}, C.~J. 2014, \aj, 147, 80,
  \dodoi{10.1088/0004-6256/147/4/80}

\bibitem[{{Hinz} {et~al.}(2010){Hinz}, {Rodigas}, {Kenworthy}, {Sivanandam},
  {Heinze}, {Mamajek}, \& {Meyer}}]{Hinz2010}
{Hinz}, P.~M., {Rodigas}, T.~J., {Kenworthy}, M.~A., {et~al.} 2010, \apj, 716,
  417, \dodoi{10.1088/0004-637X/716/1/417}

\bibitem[{{Holland} {et~al.}(2017){Holland}, {Matthews}, {Kennedy}, {Greaves},
  {Wyatt}, {Booth}, {Bastien}, {Bryden}, {Butner}, {Chen}, {Chrysostomou},
  {Davies}, {Dent}, {Di Francesco}, {Duch{\^e}ne}, {Gibb}, {Friberg}, {Ivison},
  {Jenness}, {Kavelaars}, {Lawler}, {Lestrade}, {Marshall}, {Moro-Mart{\'i}n},
  {Pani{\'c}}, {Phillips}, {Serjeant}, {Schieven}, {Sibthorpe}, {Vican},
  {Ward-Thompson}, {van der Werf}, {White}, {Wilner}, \&
  {Zuckerman}}]{Holland2017}
{Holland}, W.~S., {Matthews}, B.~C., {Kennedy}, G.~M., {et~al.} 2017, \mnras,
  470, 3606, \dodoi{10.1093/mnras/stx1378}

\bibitem[{{Hughes} {et~al.}(2011){Hughes}, {Wilner}, {Andrews}, {Williams},
  {Su}, {Murray-Clay}, \& {Qi}}]{Hughes2011}
{Hughes}, A.~M., {Wilner}, D.~J., {Andrews}, S.~M., {et~al.} 2011, \apj, 740,
  38, \dodoi{10.1088/0004-637X/740/1/38}

\bibitem[{{Jiang} \& {Tremaine}(2010)}]{Jiang2010}
{Jiang}, Y.-F., \& {Tremaine}, S. 2010, \mnras, 401, 977,
  \dodoi{10.1111/j.1365-2966.2009.15744.x}

\bibitem[{{Joncour} {et~al.}(2018){Joncour}, {Duch{\^e}ne}, {Moraux}, \&
  {Motte}}]{Joncour2018}
{Joncour}, I., {Duch{\^e}ne}, G., {Moraux}, E., \& {Motte}, F. 2018, \aap, 620,
  A27, \dodoi{10.1051/0004-6361/201833042}

\bibitem[{{Kalas} {et~al.}(2005){Kalas}, {Graham}, \& {Clampin}}]{Kalas2005}
{Kalas}, P., {Graham}, J.~R., \& {Clampin}, M. 2005, \nat, 435, 1067,
  \dodoi{10.1038/nature03601}

\bibitem[{{Karim} \& {Mamajek}(2017)}]{Karim2017}
{Karim}, M.~T., \& {Mamajek}, E.~E. 2017, \mnras, 465, 472,
  \dodoi{10.1093/mnras/stw2772}

\bibitem[{{Krivov}(2010)}]{Krivov2010}
{Krivov}, A.~V. 2010, Research in Astronomy and Astrophysics, 10, 383,
  \dodoi{10.1088/1674-4527/10/5/001}

\bibitem[{{Lallement} {et~al.}(2019){Lallement}, {Babusiaux}, {Vergely},
  {Katz}, {Arenou}, {Valette}, {Hottier}, \& {Capitanio}}]{Lallement2019}
{Lallement}, R., {Babusiaux}, C., {Vergely}, J.~L., {et~al.} 2019, \aap, 625,
  A135, \dodoi{10.1051/0004-6361/201834695}

\bibitem[{{Larson}(1981)}]{Larson1981}
{Larson}, R.~B. 1981, \mnras, 194, 809, \dodoi{10.1093/mnras/194.4.809}

\bibitem[{{Lee} \& {Song}(2019)}]{Lee2019}
{Lee}, J., \& {Song}, I. 2019, \mnras, 486, 3434, \dodoi{10.1093/mnras/stz1044}

\bibitem[{{Levison} {et~al.}(2008){Levison}, {Morbidelli}, {Van Laerhoven},
  {Gomes}, \& {Tsiganis}}]{Levison2008}
{Levison}, H.~F., {Morbidelli}, A., {Van Laerhoven}, C., {Gomes}, R., \&
  {Tsiganis}, K. 2008, \icarus, 196, 258, \dodoi{10.1016/j.icarus.2007.11.035}

\bibitem[{{Li} \& {Greenberg}(1998)}]{Li1998}
{Li}, A., \& {Greenberg}, J.~M. 1998, \aap, 331, 291

\bibitem[{{Li} {et~al.}(2019){Li}, {Zhao}, \& {Yang}}]{Li2019}
{Li}, C., {Zhao}, G., \& {Yang}, C. 2019, \apj, 872, 205,
  \dodoi{10.3847/1538-4357/ab0104}

\bibitem[{{Lindegren} {et~al.}(2020){Lindegren}, {Bastian}, {Biermann},
  {Bombrun}, {de Torres}, {Gerlach}, {Geyer}, {Hern{\'a}ndez}, {Hilger},
  {Hobbs}, {Klioner}, {Lammers}, {McMillan}, {Ramos-Lerate},
  {Steidelm{\"u}ller}, {Stephenson}, \& {van Leeuwen}}]{Lindegren2020}
{Lindegren}, L., {Bastian}, U., {Biermann}, M., {et~al.} 2020, arXiv e-prints,
  arXiv:2012.01742.
\newblock \doarXiv{2012.01742}

\bibitem[{{Liseau} {et~al.}(2015){Liseau}, {Vlemmings}, {Bayo}, {Bertone},
  {Black}, {del Burgo}, {Chavez}, {Danchi}, {De la Luz}, {Eiroa}, {Ertel},
  {Fridlund}, {Justtanont}, {Krivov}, {Marshall}, {Mora}, {Montesinos},
  {Nyman}, {Olofsson}, {Sanz-Forcada}, {Th{\'e}bault}, \& {White}}]{Liseau2015}
{Liseau}, R., {Vlemmings}, W., {Bayo}, A., {et~al.} 2015, \aap, 573, L4,
  \dodoi{10.1051/0004-6361/201425189}

\bibitem[{{MacGregor} {et~al.}(2016){MacGregor}, {Wilner}, {Chandler}, {Ricci},
  {Maddison}, {Cranmer}, {Andrews}, {Hughes}, \& {Steele}}]{MacGregor2016}
{MacGregor}, M.~A., {Wilner}, D.~J., {Chandler}, C., {et~al.} 2016, \apj, 823,
  79, \dodoi{10.3847/0004-637X/823/2/79}

\bibitem[{{Maire} {et~al.}(2015){Maire}, {Skemer}, {Hinz}, {Desidera},
  {Esposito}, {Gratton}, {Marzari}, {Skrutskie}, {Biller}, {Defr{\`e}re},
  {Bailey}, {Leisenring}, {Apai}, {Bonnefoy}, {Brandner}, {Buenzli}, {Claudi},
  {Close}, {Crepp}, {De Rosa}, {Eisner}, {Fortney}, {Henning}, {Hofmann},
  {Kopytova}, {Males}, {Mesa}, {Morzinski}, {Oza}, {Patience}, {Pinna},
  {Rajan}, {Schertl}, {Schlieder}, {Su}, {Vaz}, {Ward-Duong}, {Weigelt}, \&
  {Woodward}}]{Maire2015}
{Maire}, A.~L., {Skemer}, A.~J., {Hinz}, P.~M., {et~al.} 2015, \aap, 576, A133,
  \dodoi{10.1051/0004-6361/201425185}

\bibitem[{{Malhotra}(1995)}]{Malhotra1995}
{Malhotra}, R. 1995, \aj, 110, 420, \dodoi{10.1086/117532}

\bibitem[{{Mamajek} \& {Bell}(2014)}]{Mamajek2014}
{Mamajek}, E.~E., \& {Bell}, C. P.~M. 2014, \mnras, 445, 2169,
  \dodoi{10.1093/mnras/stu1894}

\bibitem[{{Mann} {et~al.}(2019){Mann}, {Dupuy}, {Kraus}, {Gaidos}, {Ansdell},
  {Ireland}, {Rizzuto}, {Hung}, {Dittmann}, {Factor}, {Feiden}, {Martinez},
  {Ru{\'\i}z-Rodr{\'\i}guez}, \& {Thao}}]{Mann2019}
{Mann}, A.~W., {Dupuy}, T., {Kraus}, A.~L., {et~al.} 2019, \apj, 871, 63,
  \dodoi{10.3847/1538-4357/aaf3bc}

\bibitem[{{Marino}(2021)}]{Marino2021}
{Marino}, S. 2021, \mnras, \dodoi{10.1093/mnras/stab771}

\bibitem[{{Marino} {et~al.}(2019){Marino}, {Yelverton}, {Booth}, {Faramaz},
  {Kennedy}, {Matr{\`a}}, \& {Wyatt}}]{Marino2019}
{Marino}, S., {Yelverton}, B., {Booth}, M., {et~al.} 2019, \mnras, 484, 1257,
  \dodoi{10.1093/mnras/stz049}

\bibitem[{{Marino} {et~al.}(2018){Marino}, {Carpenter}, {Wyatt}, {Booth},
  {Casassus}, {Faramaz}, {Guzman}, {Hughes}, {Isella}, {Kennedy}, {Matr{\`a}},
  {Ricci}, \& {Corder}}]{Marino2018}
{Marino}, S., {Carpenter}, J., {Wyatt}, M.~C., {et~al.} 2018, \mnras, 479,
  5423, \dodoi{10.1093/mnras/sty1790}

\bibitem[{{Marino} {et~al.}(2020){Marino}, {Zurlo}, {Faramaz}, {Milli},
  {Henning}, {Kennedy}, {Matr{\`a}}, {P{\'e}rez}, {Delorme}, {Cieza}, \&
  {Hughes}}]{Marino2020}
{Marino}, S., {Zurlo}, A., {Faramaz}, V., {et~al.} 2020, \mnras, 498, 1319,
  \dodoi{10.1093/mnras/staa2386}

\bibitem[{Marois {et~al.}(2008)Marois, Macintosh, Barman, Zuckerman, Song,
  Patience, Lafreni{\`e}re, \& Doyon}]{Marois2008}
Marois, C., Macintosh, B., Barman, T., {et~al.} 2008, Science, 322, 1348,
  \dodoi{10.1126/science.1166585}

\bibitem[{Marois {et~al.}(2010)Marois, Zuckerman, Konopacky, Macintosh, \&
  Barman}]{Marois2010}
Marois, C., Zuckerman, B., Konopacky, Q.~M., Macintosh, B., \& Barman, T. 2010,
  \nat, 468, 1080, \dodoi{10.1038/nature09684}

\bibitem[{{Marshall} {et~al.}(2014){Marshall}, {Moro-Mart{\'\i}n}, {Eiroa},
  {Kennedy}, {Mora}, {Sibthorpe}, {Lestrade}, {Maldonado}, {Sanz-Forcada},
  {Wyatt}, {Matthews}, {Horner}, {Montesinos}, {Bryden}, {del Burgo},
  {Greaves}, {Ivison}, {Meeus}, {Olofsson}, {Pilbratt}, \&
  {White}}]{Marshall2014}
{Marshall}, J.~P., {Moro-Mart{\'\i}n}, A., {Eiroa}, C., {et~al.} 2014, \aap,
  565, A15, \dodoi{10.1051/0004-6361/201323058}

\bibitem[{{Matr{\`a}} {et~al.}(2019){Matr{\`a}}, {Wyatt}, {Wilner}, {Dent},
  {Marino}, {Kennedy}, \& {Milli}}]{Matra2019}
{Matr{\`a}}, L., {Wyatt}, M.~C., {Wilner}, D.~J., {et~al.} 2019, \aj, 157, 135,
  \dodoi{10.3847/1538-3881/ab06c0}

\bibitem[{{Matthews} {et~al.}(2014){Matthews}, {Kennedy}, {Sibthorpe}, {Booth},
  {Wyatt}, {Broekhoven-Fiene}, {Macintosh}, \& {Marois}}]{Matthews2014}
{Matthews}, B., {Kennedy}, G., {Sibthorpe}, B., {et~al.} 2014, \apj, 780, 97,
  \dodoi{10.1088/0004-637X/780/1/97}

\bibitem[{{McMullin} {et~al.}(2007){McMullin}, {Waters}, {Schiebel}, {Young},
  \& {Golap}}]{McMullin2007}
{McMullin}, J.~P., {Waters}, B., {Schiebel}, D., {Young}, W., \& {Golap}, K.
  2007, in Astronomical Society of the Pacific Conference Series, Vol. 376,
  Astronomical Data Analysis Software and Systems XVI, ed. R.~A. {Shaw},
  F.~{Hill}, \& D.~J. {Bell}, 127

\bibitem[{{Miret-Roig} {et~al.}(2020){Miret-Roig}, {Galli}, {Brandner}, {Bouy},
  {Barrado}, {Olivares}, {Antoja}, {Romero-G{\'o}mez}, {Figueras}, \&
  {Lillo-Box}}]{Miret-Roig2020}
{Miret-Roig}, N., {Galli}, P.~A.~B., {Brandner}, W., {et~al.} 2020, \aap, 642,
  A179, \dodoi{10.1051/0004-6361/202038765}

\bibitem[{{Morbidelli} \& {Nesvorn{\'y}}(2020)}]{Morbidelli2020}
{Morbidelli}, A., \& {Nesvorn{\'y}}, D. 2020, {Kuiper belt: formation and
  evolution}, ed. D.~{Prialnik}, M.~A. {Barucci}, \& L.~{Young}, 25--59,
  \dodoi{10.1016/B978-0-12-816490-7.00002-3}

\bibitem[{{Moro-Martin}(2013)}]{MoroMartin2013}
{Moro-Martin}, A. 2013, {Dusty Planetary Systems}, ed. T.~D. {Oswalt}, L.~M.
  {French}, \& P.~{Kalas}, 431, \dodoi{10.1007/978-94-007-5606-9_9}

\bibitem[{{Morrison} \& {Malhotra}(2015)}]{Morrison2015}
{Morrison}, S., \& {Malhotra}, R. 2015, \apj, 799, 41,
  \dodoi{10.1088/0004-637X/799/1/41}

\bibitem[{{Mouillet} {et~al.}(1997){Mouillet}, {Larwood}, {Papaloizou}, \&
  {Lagrange}}]{Mouillet1997}
{Mouillet}, D., {Larwood}, J.~D., {Papaloizou}, J.~C.~B., \& {Lagrange}, A.~M.
  1997, \mnras, 292, 896, \dodoi{10.1093/mnras/292.4.896}

\bibitem[{{Murphy} {et~al.}(2020){Murphy}, {Joyce}, {Bedding}, {White}, \&
  {Kama}}]{Murphy2020}
{Murphy}, S.~J., {Joyce}, M., {Bedding}, T.~R., {White}, T.~R., \& {Kama}, M.
  2020, arXiv e-prints, arXiv:2011.11821.
\newblock \doarXiv{2011.11821}

\bibitem[{{Nguyen} {et~al.}(2021){Nguyen}, {De Rosa}, \& {Kalas}}]{Nguyen2021}
{Nguyen}, M.~M., {De Rosa}, R.~J., \& {Kalas}, P. 2021, \aj, 161, 22,
  \dodoi{10.3847/1538-3881/abc012}

\bibitem[{{Oelkers} {et~al.}(2018){Oelkers}, {Rodriguez}, {Stassun}, {Pepper},
  {Somers}, {Kafka}, {Stevens}, {Beatty}, {Siverd}, {Lund}, {Kuhn}, {James}, \&
  {Gaudi}}]{Oelkers2018}
{Oelkers}, R.~J., {Rodriguez}, J.~E., {Stassun}, K.~G., {et~al.} 2018, \aj,
  155, 39, \dodoi{10.3847/1538-3881/aa9bf4}

\bibitem[{{Patience} {et~al.}(2011){Patience}, {Bulger}, {King}, {Ayliffe},
  {Bate}, {Song}, {Pinte}, {Koda}, {Dowell}, \& {Kov{\'a}cs}}]{Patience2011}
{Patience}, J., {Bulger}, J., {King}, R.~R., {et~al.} 2011, \aap, 531, L17,
  \dodoi{10.1051/0004-6361/201117395}

\bibitem[{{Pearce} \& {Wyatt}(2014)}]{Pearce2014}
{Pearce}, T.~D., \& {Wyatt}, M.~C. 2014, \mnras, 443, 2541,
  \dodoi{10.1093/mnras/stu1302}

\bibitem[{{Petit dit de la Roche} {et~al.}(2020){Petit dit de la Roche}, {van
  den Ancker}, {Kissler-Patig}, {Ivanov}, \& {Fedele}}]{Petit2020}
{Petit dit de la Roche}, D.~J.~M., {van den Ancker}, M.~E., {Kissler-Patig},
  M., {Ivanov}, V.~D., \& {Fedele}, D. 2020, \mnras, 491, 1795,
  \dodoi{10.1093/mnras/stz3117}

\bibitem[{{Randich} {et~al.}(2018){Randich}, {Tognelli}, {Jackson}, {Jeffries},
  {Degl'Innocenti}, {Pancino}, {Re Fiorentin}, {Spagna}, {Sacco}, {Bragaglia},
  {Magrini}, {Prada Moroni}, {Alfaro}, {Franciosini}, {Morbidelli},
  {Roccatagliata}, {Bouy}, {Bravi}, {Jim{\'e}nez-Esteban}, {Jordi}, {Zari},
  {Tautvai{\v{s}}iene}, {Drazdauskas}, {Mikolaitis}, {Gilmore}, {Feltzing},
  {Vallenari}, {Bensby}, {Koposov}, {Korn}, {Lanzafame}, {Smiljanic}, {Bayo},
  {Carraro}, {Costado}, {Heiter}, {Hourihane}, {Jofr{\'e}}, {Lewis}, {Monaco},
  {Prisinzano}, {Sbordone}, {Sousa}, {Worley}, \& {Zaggia}}]{Randich2018}
{Randich}, S., {Tognelli}, E., {Jackson}, R., {et~al.} 2018, \aap, 612, A99,
  \dodoi{10.1051/0004-6361/201731738}

\bibitem[{Read {et~al.}(2018)Read, Wyatt, Marino, \& Kennedy}]{Read2018}
Read, M.~J., Wyatt, M.~C., Marino, S., \& Kennedy, G.~M. 2018, \mnras, 475,
  4953, \dodoi{10.1093/mnras/sty141}

\bibitem[{{Ruffio} {et~al.}(2019){Ruffio}, {Macintosh}, {Konopacky}, {Barman},
  {De Rosa}, {Wang}, {Wilcomb}, {Czekala}, \& {Marois}}]{Ruffio2019}
{Ruffio}, J.-B., {Macintosh}, B., {Konopacky}, Q.~M., {et~al.} 2019, \aj, 158,
  200, \dodoi{10.3847/1538-3881/ab4594}

\bibitem[{{Sadakane} \& {Nishida}(1986)}]{Sadakane1986}
{Sadakane}, K., \& {Nishida}, M. 1986, \pasp, 98, 685, \dodoi{10.1086/131813}

\bibitem[{{Schneider} {et~al.}(2019){Schneider}, {Shkolnik}, {Allers}, {Kraus},
  {Liu}, {Weinberger}, \& {Flagg}}]{Schneider2019}
{Schneider}, A.~C., {Shkolnik}, E.~L., {Allers}, K.~N., {et~al.} 2019, \aj,
  157, 234, \dodoi{10.3847/1538-3881/ab1a26}

\bibitem[{{Shkolnik} {et~al.}(2017){Shkolnik}, {Allers}, {Kraus}, {Liu}, \&
  {Flagg}}]{Shkolnik2017}
{Shkolnik}, E.~L., {Allers}, K.~N., {Kraus}, A.~L., {Liu}, M.~C., \& {Flagg},
  L. 2017, \aj, 154, 69, \dodoi{10.3847/1538-3881/aa77fa}

\bibitem[{{Soummer} {et~al.}(2011){Soummer}, {Hagan}, {Pueyo}, {Thormann},
  {Rajan}, \& {Marois}}]{Soummer2011}
{Soummer}, R., {Hagan}, J.~B., {Pueyo}, L., {et~al.} 2011, \apj, 741, 55,
  \dodoi{10.1088/0004-637X/741/1/55}

\bibitem[{{Su} {et~al.}(2009){Su}, {Rieke}, {Stapelfeldt}, {Malhotra},
  {Bryden}, {Smith}, {Misselt}, {Moro-Mart{\'i}n}, \& {Williams}}]{Su2009}
{Su}, K.~Y.~L., {Rieke}, G.~H., {Stapelfeldt}, K.~R., {et~al.} 2009, \apj, 705,
  314, \dodoi{10.1088/0004-637X/705/1/314}

\bibitem[{{Su} {et~al.}(2017){Su}, {MacGregor}, {Booth}, {Wilner}, {Flaherty},
  {Hughes}, {Phillips}, {Malhotra}, {Hales}, {Morrison}, {Ertel}, {Matthews},
  {Dent}, \& {Casassus}}]{Su2017}
{Su}, K. Y.~L., {MacGregor}, M.~A., {Booth}, M., {et~al.} 2017, \aj, 154, 225,
  \dodoi{10.3847/1538-3881/aa906b}

\bibitem[{{van Leeuwen}(2007)}]{vanLeeuwen2007}
{van Leeuwen}, F. 2007, \aap, 474, 653, \dodoi{10.1051/0004-6361:20078357}

\bibitem[{{Wang} {et~al.}(2018{\natexlab{a}}){Wang}, {Mawet}, {Fortney},
  {Hood}, {Morley}, \& {Benneke}}]{WangJi2018}
{Wang}, J., {Mawet}, D., {Fortney}, J.~J., {et~al.} 2018{\natexlab{a}}, \aj,
  156, 272, \dodoi{10.3847/1538-3881/aae47b}

\bibitem[{{Wang} {et~al.}(2018{\natexlab{b}}){Wang}, {Graham}, {Dawson},
  {Fabrycky}, {De Rosa}, {Pueyo}, {Konopacky}, {Macintosh}, {Marois}, {Chiang},
  {Ammons}, {Arriaga}, {Bailey}, {Barman}, {Bulger}, {Chilcote}, {Cotten},
  {Doyon}, {Duch{\^e}ne}, {Esposito}, {Fitzgerald}, {Follette}, {Gerard},
  {Goodsell}, {Greenbaum}, {Hibon}, {Hung}, {Ingraham}, {Kalas}, {Larkin},
  {Maire}, {Marchis}, {Marley}, {Metchev}, {Millar-Blanchaer}, {Nielsen},
  {Oppenheimer}, {Palmer}, {Patience}, {Perrin}, {Poyneer}, {Rajan}, {Rameau},
  {Rantakyr{\"o}}, {Ruffio}, {Savransky}, {Schneider}, {Sivaramakrishnan},
  {Song}, {Soummer}, {Thomas}, {Wallace}, {Ward-Duong}, {Wiktorowicz}, \&
  {Wolff}}]{WangJason2018}
{Wang}, J.~J., {Graham}, J.~R., {Dawson}, R., {et~al.} 2018{\natexlab{b}}, \aj,
  156, 192, \dodoi{10.3847/1538-3881/aae150}

\bibitem[{{Wertz} {et~al.}(2017){Wertz}, {Absil}, {G{\'o}mez Gonz{\'a}lez},
  {Milli}, {Girard}, {Mawet}, \& {Pueyo}}]{Wertz2017}
{Wertz}, O., {Absil}, O., {G{\'o}mez Gonz{\'a}lez}, C.~A., {et~al.} 2017, \aap,
  598, A83, \dodoi{10.1051/0004-6361/201628730}

\bibitem[{{White} {et~al.}(2019){White}, {Aufdenberg}, {Boley}, {Devlin},
  {Dicker}, {Hauschildt}, {Hughes}, {Hughes}, {Mason}, {Matthews}, {Mo{\'o}r},
  {Mroczkowski}, {Romero}, {Sievers}, {Stanchfield}, {Tapia}, \&
  {Wilner}}]{White2019}
{White}, J.~A., {Aufdenberg}, J., {Boley}, A.~C., {et~al.} 2019, \apj, 875, 55,
  \dodoi{10.3847/1538-4357/ab0e7f}

\bibitem[{{White} {et~al.}(2020){White}, {Tapia-V{\'a}zquez}, {Hughes},
  {Mo{\'o}r}, {Matthews}, {Wilner}, {Aufdenberg}, {Hughes}, {De la Luz}, \&
  {Boley}}]{White2020}
{White}, J.~A., {Tapia-V{\'a}zquez}, F., {Hughes}, A.~G., {et~al.} 2020, \apj,
  894, 76, \dodoi{10.3847/1538-4357/ab8467}

\bibitem[{{Williams} \& {Andrews}(2006)}]{Williams2006}
{Williams}, J.~P., \& {Andrews}, S.~M. 2006, \apj, 653, 1480,
  \dodoi{10.1086/508919}

\bibitem[{{Wilner} {et~al.}(2018){Wilner}, {MacGregor}, {Andrews}, {Hughes},
  {Matthews}, \& {Su}}]{Wilner2018}
{Wilner}, D.~J., {MacGregor}, M.~A., {Andrews}, S.~M., {et~al.} 2018, \apj,
  855, 56, \dodoi{10.3847/1538-4357/aaacd7}

\bibitem[{{Wright}(2020)}]{Wright2020}
{Wright}, N.~J. 2020, \nar, 90, 101549, \dodoi{10.1016/j.newar.2020.101549}

\bibitem[{{Wyatt}(2008)}]{Wyatt2008}
{Wyatt}, M.~C. 2008, \araa, 46, 339,
  \dodoi{10.1146/annurev.astro.45.051806.110525}

\bibitem[{{Wyatt} {et~al.}(1999){Wyatt}, {Dermott}, {Telesco}, {Fisher},
  {Grogan}, {Holmes}, \& {Pi{\~n}a}}]{Wyatt1999}
{Wyatt}, M.~C., {Dermott}, S.~F., {Telesco}, C.~M., {et~al.} 1999, \apj, 527,
  918, \dodoi{10.1086/308093}

\bibitem[{{Xuan} {et~al.}(2020){Xuan}, {Kennedy}, {Wyatt}, \&
  {Yelverton}}]{Xuan2020}
{Xuan}, J.~W., {Kennedy}, G.~M., {Wyatt}, M.~C., \& {Yelverton}, B. 2020,
  \mnras, 499, 5059, \dodoi{10.1093/mnras/staa3155}

\bibitem[{{Yamamoto} {et~al.}(2003){Yamamoto}, {Onishi}, {Mizuno}, \&
  {Fukui}}]{Yamamoto2003}
{Yamamoto}, H., {Onishi}, T., {Mizuno}, A., \& {Fukui}, Y. 2003, \apj, 592,
  217, \dodoi{10.1086/375128}

\bibitem[{{Yan} {et~al.}(2019){Yan}, {Yang}, {Sun}, {Su}, \& {Xu}}]{Yan2019}
{Yan}, Q.-Z., {Yang}, J., {Sun}, Y., {Su}, Y., \& {Xu}, Y. 2019, \apj, 885, 19,
  \dodoi{10.3847/1538-4357/ab458e}

\bibitem[{{Yelverton} {et~al.}(2020){Yelverton}, {Kennedy}, \&
  {Su}}]{Yelverton2020}
{Yelverton}, B., {Kennedy}, G.~M., \& {Su}, K. Y.~L. 2020, \mnras, 495, 1943,
  \dodoi{10.1093/mnras/staa1316}

\bibitem[{{Yelverton} {et~al.}(2019){Yelverton}, {Kennedy}, {Su}, \&
  {Wyatt}}]{Yelverton2019}
{Yelverton}, B., {Kennedy}, G.~M., {Su}, K. Y.~L., \& {Wyatt}, M.~C. 2019,
  \mnras, 488, 3588, \dodoi{10.1093/mnras/stz1927}

\bibitem[{{Zucker} {et~al.}(2019){Zucker}, {Speagle}, {Schlafly}, {Green},
  {Finkbeiner}, {Goodman}, \& {Alves}}]{Zucker2019}
{Zucker}, C., {Speagle}, J.~S., {Schlafly}, E.~F., {et~al.} 2019, \apj, 879,
  125, \dodoi{10.3847/1538-4357/ab2388}

\bibitem[{{Zuckerman} {et~al.}(2011){Zuckerman}, {Rhee}, {Song}, \&
  {Bessell}}]{Zuckerman2011}
{Zuckerman}, B., {Rhee}, J.~H., {Song}, I., \& {Bessell}, M.~S. 2011, \apj,
  732, 61, \dodoi{10.1088/0004-637X/732/2/61}

\bibitem[{{Zurlo} {et~al.}(2016){Zurlo}, {Vigan}, {Galicher}, {Maire}, {Mesa},
  {Gratton}, {Chauvin}, {Kasper}, {Moutou}, {Bonnefoy}, {Desidera}, {Abe},
  {Apai}, {Baruffolo}, {Baudoz}, {Baudrand}, {Beuzit}, {Blancard},
  {Boccaletti}, {Cantalloube}, {Carle}, {Cascone}, {Charton}, {Claudi},
  {Costille}, {de Caprio}, {Dohlen}, {Dominik}, {Fantinel}, {Feautrier},
  {Feldt}, {Fusco}, {Gigan}, {Girard}, {Gisler}, {Gluck}, {Gry}, {Henning},
  {Hugot}, {Janson}, {Jaquet}, {Lagrange}, {Langlois}, {Llored}, {Madec},
  {Magnard}, {Martinez}, {Maurel}, {Mawet}, {Meyer}, {Milli},
  {Moeller-Nilsson}, {Mouillet}, {Orign{\'e}}, {Pavlov}, {Petit}, {Puget},
  {Quanz}, {Rabou}, {Ramos}, {Rousset}, {Roux}, {Salasnich}, {Salter},
  {Sauvage}, {Schmid}, {Soenke}, {Stadler}, {Suarez}, {Turatto}, {Udry},
  {Vakili}, {Wahhaj}, {Wildi}, \& {Antichi}}]{Zurlo2016}
{Zurlo}, A., {Vigan}, A., {Galicher}, R., {et~al.} 2016, \aap, 587, A57,
  \dodoi{10.1051/0004-6361/201526835}

\end{thebibliography}

\end{document}